\documentclass[12pt,3p]{llncs}

\usepackage[sectionbib]{natbib}

\usepackage{color}
\usepackage{eufrak}
\usepackage{bm}
\usepackage{comment}
\usepackage{amsmath}
\usepackage{mathrsfs,amsmath} 
\usepackage{amsfonts}
\usepackage{graphicx} 
\usepackage{epstopdf}
\usepackage{enumitem}
\usepackage{xcolor,colortbl}
\usepackage{subfigure}
\usepackage{setspace}
\usepackage{bm}
\definecolor{Gray}{gray}{0.85}
\usepackage{appendix}
\usepackage[top=25mm, bottom=25mm, left=20mm, right=20mm]{geometry}
\allowdisplaybreaks

\begin{document}

\title{Complex frequency band structure of periodic thermo-diffusive materials by Floquet-Bloch theory}

\author{Andrea Bacigalupo, Maria Laura De Bellis, Giorgio Gnecco}
\institute{-
\email{\-}}

\maketitle

\begin{abstract}

{\color{black} This work deals with the micromechanical study of periodic thermo-diffusive elastic multi-layered
materials, which are of interest for the fabrication of solid oxide fuel cells (SOFCs). The focus is on  the dynamic regime, that is investigating the dispersive wave propagation within the periodic material. In this framework, a generalization of the Floquet-Bloch theory is adopted, able to determine the complex band structure of such materials. The infinite algebraic linear system, obtained by exploiting both bilateral Laplace transform and Fourier transform, is replaced by its finite counterpart, resulting from a proper truncation at a finite number of considered equations.       
A regularization technique is herein useful to get rid of the Gibbs phenomenon. The solution of the problem is, finally, found in terms of complex angular frequencies, corresponding to a finite sequence of eigenvalue problems for given values of the wave vector.
The paper is complemented by numerical examples taking into account thermo-mechanical coupling. The overall behaviour is found to be strongly influenced by the interaction between thermal and mechanical  phenomena. } 
\end{abstract}


\section{Introduction}


The advances in different engineering fields, involving, among others, structural monitoring, nano-medicine, military-aerospace industries and wearable electronics call for the need of composite materials with increasingly improved performances. Optimized responses in terms of toughness, strength, mass diffusivity, thermal conductivity or also electric permittivity can be, indeed, obtained by properly combining materials characterized by different physical properties.
In the framework of renewable energy applications, multiphase materials have been widely adopted especially in the manufacturing of battery devices, such as Solid Oxide Fuel Cells 
 (SOFCs) \citep{Hajetal2011,Naketal2012,Devetal2014} and lithium ions batteries \citep{Ellisetal2012,DIOUF2015375,Ricetal2012}. In particular, 
SOFCs are electrochemical devices able to convert the chemical energy originally stored in a fuel into electrical power. Such fuel cells are characterized by very high energy conversion performances and are extremely eligible for realizing efficient small-scale power generation systems, with efficiency up to 60$\%$. The typical SOFC building block, schematized in Figure \ref{fig:figura1SOFC}, consists of a layer of doped ceramic
material, used as  electrolyte, that is sandwiched between two heat-resistant electrodes, i.e. cathode and anode, made of porous materials \citep{BoveUbertini2008}. The working principle of SOFC devices can be described as follows. The cathode is supplied with both
oxygen, as oxidant material, and also electrons that are provided by an external circuit. At this point, a chemical reaction takes place and oxygen ions are 
released.  Such ions migrate into the electrolyte towards the anode, on which they combine with the H$_2$ fuel, forming water and electrons. These electrons travel
along the circuit, generating an electrical current. The single
cell composed by anode, electrolyte and cathode is provided 
with an interconnect separator and flow channels on
each side. A number of cells are typically stacked together, working 
in series, in order to build up
the voltage for the power generating capacity required  in practical uses. The resulting composite material is, thus, periodic.\\
The strengths of SOFCs are the great flexibility in sizing and nearly zero emissions \citep{KAKAC2007761}, while the weaknesses concern both the very high operating temperature involved, usually in the range
$60^\circ$-$1000^\circ$ C, and the intense particle flows. 
As a result, these devices typically undergo severe stresses induced by coupled chemo-thermo-mechanical phenomena, thus also influencing the choice of the materials adopted in their manufacturing \citep{Olesiak1994,PitBus2005}.\\
A comprehensive understanding of the SOFC functioning cannot disregard the strong mutual influence between the 
macroscopic behaviour of the thermodiffusive composite
materials and the multiphysics phenomena taking place at the scale  of the constituents \citep{BoveUbertini2008,Hajetal2011}. In this context, multi-scale techniques are powerful tools to gather global constitutive information  starting from the detailed description of the microscopic material scale. In particular, various homogenization approaches, including computational and asymptotic strategies, have been proposed for
investigating the overall elastic properties of composite materials with periodic microstructures, both in the static \citep{Addessi2013,Addessi2016,Bacca2013a,Bacca2013b,Bacigalupo2013,
BacigalupoDeBellis2015,
BACIGALUPO2017,Bigoni2007,FREUND201456,DeBellis-Addessi11,Geers2010}  and in the dynamic regimes \citep{chen2001dispersive,Bacigalupo2014,SRIDHAR2018414}. Examples of multi-physics homogenization approaches range from thermo-elastic \citep{francfort1983homogenization,johnson2008effective,
aboudi2001linear,kanoute2009multiscale,zhang2007thermo}, to piezoelectric and thermo-piezoelectric problems \citep{galka1996some,DERAEMAEKER20103272,debellis2017auxetic,fantoni2017multi}, up  to  thermo-diffusive phenomena \citep{BACIGALUPO201415017,
BacigalupoMorini2016,bacigalupo2016overall,salvadori2014computational}.\\
Standard homogenization approaches, nevertheless, show some inner limitations in the study of high frequency dispersive waves characterized by short wavelengths. In such case, it can be useful resorting to micromechanical approaches, as in
\citep{hawwa1995general,aboudi2001micromechanical,sharma2009modelling,dev2014mechanical,
ignaczak2010thermoelasticity,suiker2001micro,aaberg2000micromechanical}.\\
In this work, a micromechanical dynamical study of SOFCs-like cells is carried out, taking into account the coupling between thermal, mechanical and diffusive phenomena. In this framework, we are interested in investigating the possible 
influence that the interaction between such phenomena might have on the dispersive waves propagation within the periodic material. With this in mind, we consider the governing equations of the coupled problem \citep{lord1967generalized,nowacki1974dynamical1,nowacki1974dynamical2,
nowacki1974dynamical3,NOWACKI1976261,
SHERIEF2004591} and apply the Floquet-Bloch theory. In particular, by exploiting both bilateral Laplace transforms in time and  Fourier transforms in space, we obtain an infinite algebraic linear system in terms of  complex frequencies and  wave vectors. Such resulting system admits eigenvalues representing the complex spectrum of the periodic material. An accurate numerical solution can be found by truncating the infinite system at a given finite number of considered equations, determined on the basis of a properly conceived convergence analysis. In this context, regularization  techniques are useful to smooth the results in terms of constitutive components, that can be affected by the Gibbs phenomenon, that is oscillations particularly pronounced near the discontinuities between materials phases in the periodic cells. Finally, after performing a  discretization of the wave vectors space, a finite sequence of eigenvalue problems is obtained, which solutions are the complex angular frequencies corresponding to a given wave vector.\\
The paper is organized as follows. In Section \ref{S1} the governing equations of the periodic thermodiffusive material, considered as a first order continuum, are recalled. By exploiting the periodicity of the medium, the Christoffel equations are complemented by the Floquet-Bloch boundary conditions. Section \ref{Damped} is devoted to the determination of the infinite algebraic  system, characterizing the propagation of damped Bloch-waves. Moreover, in Section \ref{sec:4} the generalized infinite eigenvalue problem is truncated and the approximate solution of the Floquet-Bloch spectrum is found. Section \ref{sec:numerical_results} presents numerical examples focused on the study of damped wave propagation in SOFC-like devices. Finally, some concluding remarks are summarized in Section \ref{S:FR}. \\

\section{Governing equations of periodic elastic thermodiffusive first order continuum}\label{S1}
\begin{figure}[ht!]
\centering
\includegraphics[scale=1]{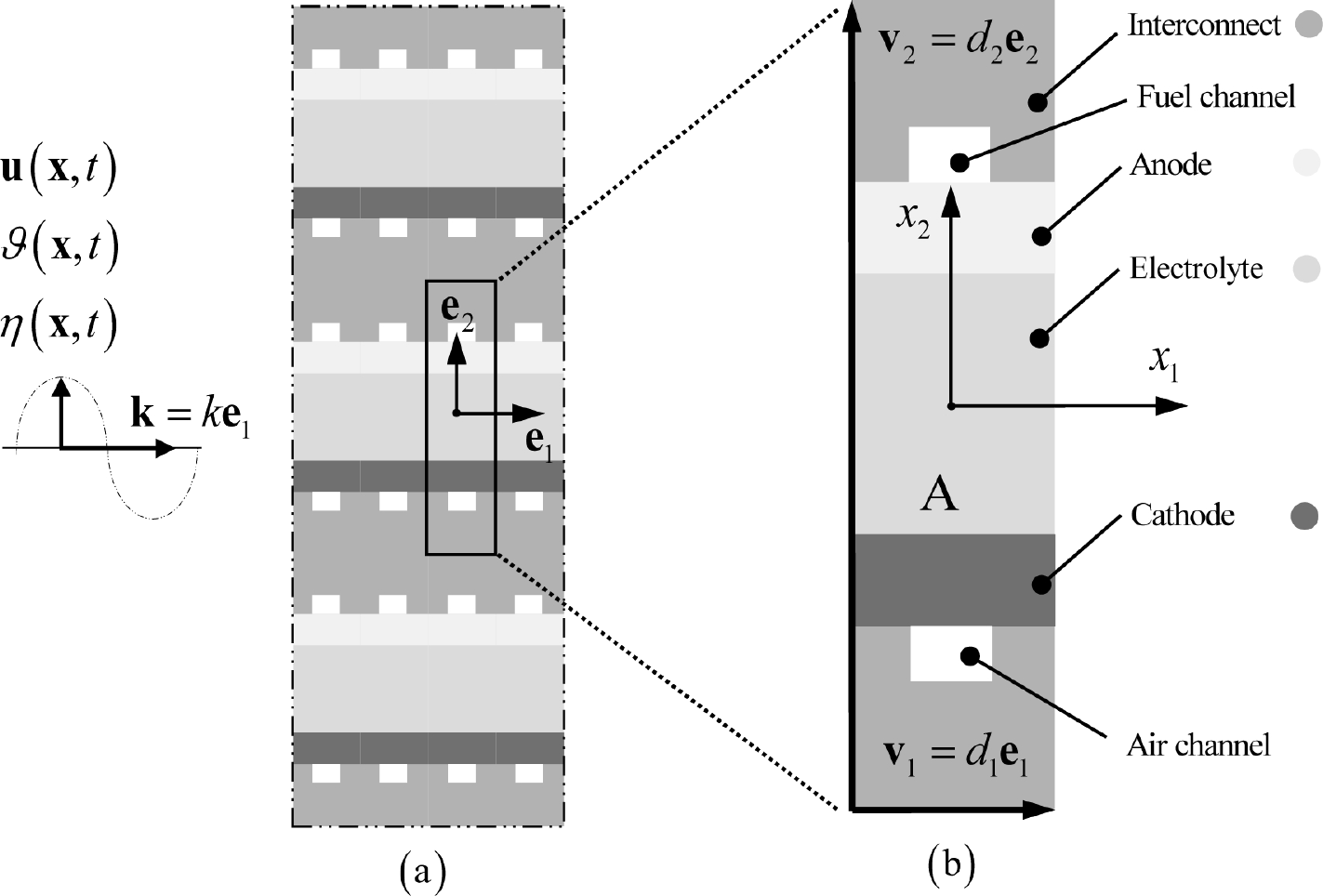}
\caption{ (a) SOFC-like device; (b) zoomed view of a periodic cell.}\label{fig:figura1SOFC}
\end{figure}

Let us consider a periodic elastic heterogeneous material undergoing thermo-diffusive phenomena. Restricting our analysis to the two-dimensional case, we adopt a linear elastic thermodiffusive first order continuum to model each material phase of the heterogeneous medium schematically shown in Figure \ref{fig:figura1SOFC}(a). Given the medium periodicity,
it is possible to identify the periodic cell ${\bf A}=[0,d_1]\times[0,d_2]$ in Figure \ref{fig:figura1SOFC}(b),  characterized by two orthogonal vectors of periodicity $\textbf{v}_1=d_1 \textbf{e}_1$, $\textbf{v}_2=d_2 \textbf{e}_2$, with $\textbf{e}_1$ and $\textbf{e}_2$ being a given orthogonal base. \\
\noindent Each material point $\textbf{x}=x_1 \textbf{e}_1+x_2\textbf{e}_2$, is endowed with a displacement field $\textbf{u}(\textbf{x},t)=u_i\textbf{e}_i$, a temperature field $\theta(\textbf{x},t)$, and a chemical potential field $\eta(\textbf{x},t)$. The coupled constitutive relations link the stress tensor $ \boldsymbol{\sigma}(\textbf{x},t)=\sigma_{ij}\textbf{e}_i\otimes\textbf{e}_j$, the heat and mass fluxes ${\mathbf{ q}}(\textbf{x},t)=q_i\textbf{e}_i$ and ${\mathbf{ j}}(\textbf{x},t)=j_i\textbf{e}_i$, respectively, to the aforementioned relevant fields $\textbf{u}(\textbf{x})$, $\theta(\textbf{x})$ and $\eta(\textbf{x})$, that is 
\begin{align}
 \boldsymbol{\sigma}(\textbf{x},t)  &= {\mathbb{C}}(\textbf{x})\nabla {\mathbf{ u}}(\textbf{x},t) - {{ \boldsymbol{\alpha }}}(\textbf{x}) \vartheta(\textbf{x},t)  - {{\boldsymbol{\beta }}}(\textbf{x}) \eta(\textbf{x},t) \,,\label{sigma}\\
{\mathbf{ q}}(\textbf{x},t) &=  - {{\mathbf{K}}}(\textbf{x}) \nabla  \vartheta(\textbf{x},t) \,,\label{q}\\
{\mathbf{ j}}(\textbf{x},t) &=  - {{\mathbf{D}}}(\textbf{x}) \nabla  \eta(\textbf{x},t) \,,\label{j}
\end{align}
where ${\mathbb{C}}=C_{ijhk}\textbf{e}_i \otimes \textbf{e}_j \otimes\textbf{e}_h \otimes\textbf{e}_k$ is the fourth-order elastic tensor, ${{\mathbf{K}}}=K_{ij}\textbf{e}_i \otimes \textbf{e}_j$ is the second-order heat conduction tensor, ${{\mathbf{D}}}=D_{ij}\textbf{e}_i \otimes \textbf{e}_j$ is the second-order mass diffusion tensor, ${{ \boldsymbol{\alpha }}}=\alpha_{ij}\textbf{e}_i \otimes \textbf{e}_j$ is the second-order thermal dilatation tensor, and ${{\boldsymbol{\beta }}}=\beta_{ij}\textbf{e}_i \otimes \textbf{e}_j$ is the diffusive expansion tensor.\\
The following balance equations hold
\begin{align}
 \nabla \cdot \boldsymbol{\sigma}(\textbf{x},t)  + \textbf{f}(\textbf{x},t)&=\rho(\textbf{x}) \ddot{{\mathbf{ u}}}(\textbf{x},t),\label{bal1}\\
-\nabla \cdot {\mathbf{ q}}(\textbf{x},t) + r(\textbf{x},t)&= p(\textbf{x}) \ \dot{\theta}(\textbf{x},t) +\boldsymbol{\alpha}(\textbf{x}) : \nabla \dot{\textbf{u}}(\textbf{x},t)+ \Psi(\textbf{x}) \dot{\eta}(\textbf{x},t),\label{bal2}\\
-\nabla \cdot {\mathbf{ j}}(\textbf{x},t) + s(\textbf{x},t)&=  c(\textbf{x}) \, \dot{\eta}(\textbf{x},t)  + \boldsymbol{\beta}(\textbf{x}) : \nabla \dot{\textbf{u}}(\textbf{x},t)+ \Psi(\textbf{x}) \dot{\theta}(\textbf{x},t)\,,\label{bal3}
\end{align}
with $\textbf{f}=f_i\textbf{e}_i$ being the body forces, ${r}$ the heat source and ${s}$ the mass source, $\rho$ is the mass density,  $p$ and $c$ are thermodiffusive constants, and $\Psi$ is a thermodiffusive coupling constant.\\
The governing equations are finally obtained by plugging the constitutive equations into the balance equations, that is 
\begin{align}
\nabla  \cdot \left( {{\mathbb{C}}\left(\mathbf{x}\right)\nabla {\mathbf{u}}}(\textbf{x},t) - {{\boldsymbol{\alpha}}\left(\mathbf{x}\right)\vartheta(\textbf{x},t) } -{{\boldsymbol{\beta}}\left(\mathbf{x}\right) \eta(\textbf{x},t) }
 \right) + {\mathbf{f}}(\textbf{x},t) &= \rho \left(\mathbf{x}\right){\mathbf{\ddot u}(\textbf{x},t)}\,,\label{gov1}\\
 \nabla  \cdot \left( {{{\mathbf{K}}}(\textbf{x})\nabla \vartheta(\textbf{x},t) } \right) - {{\boldsymbol {\alpha }}} (\textbf{x}):\nabla {\mathbf{\dot u}}(\textbf{x},t) - {\Psi } (\textbf{x})\dot \eta(\textbf{x},t)  + r\left( {\mathbf{x}},t \right) &= {p}(\textbf{x})\dot \vartheta(\textbf{x},t)\,,\label{gov2}\\
 \nabla  \cdot \left( {{{\mathbf{D}}}(\textbf{x}) \nabla \eta(\textbf{x},t) } \right) - {{\boldsymbol{\beta }}}(\textbf{x}):\nabla {\mathbf{\dot u}}(\textbf{x},t) - {\Psi }(\textbf{x})\dot \vartheta(\textbf{x},t)  + s (\textbf{x},t) &= {c}(\textbf{x})\dot \eta(\textbf{x},t)\,.\label{gov3}
\end{align}
\noindent Since we focus on periodic materials, the constitutive tensors involved in Equations (\ref{gov1}),(\ref{gov2}) and (\ref{gov3}) satisfy periodicity conditions within the periodic cell, i.e. $\forall \textbf{x} \in \textbf{A}$. This means that the elastic tensor, the heat conduction and the mass diffusion tensors are characterized by $A$-periodicity, i.e.
\begin{align}
{\mathbb{C}}\left( {{\mathbf{x}} + {{\mathbf{v}}_i}} \right) &= {\mathbb{C}}\left( {\mathbf{x}} \right)\,,\label{PerC}\\
{{\mathbf{K}}}\left( {{\mathbf{x}} + {{\mathbf{v}}_i}} \right) &= {{\mathbf{K}}}\left( {\mathbf{x}} \right)\,,\label{PerK}\\
{{\mathbf{D}}}\left( {{\mathbf{x}} + {{\mathbf{v}}_i}} \right) &= {{\mathbf{D}}}\left( {\mathbf{x}} \right).\label{PerD}
\end{align}
Similarly, the inertial terms result as
\begin{align}{\rho }\left( {{\mathbf{x}} + {{\mathbf{v}}_i}} \right) &= {\rho }\left( {\mathbf{x}} \right)\,,\label{Perro}\\	
{p}\left( {{\mathbf{x}} + {{\mathbf{v}}_i}} \right) &= {p}\left( {\mathbf{x}} \right)\,,\label{Perp}\\
{c}\left( {{\mathbf{x}} + {{\mathbf{v}}_i}} \right) &= {c}\left( {\mathbf{x}} \right)\,,\label{Perc}
\end{align}
and, finally, the thermal dilatation tensor, the diffusive expansion tensor and the thermodiffusive coupling constant comply with the following relations
\begin{align}
{{\boldsymbol{\alpha }}}\left( {{\mathbf{x}} + {{\mathbf{v}}_i}} \right) &= {{\boldsymbol{\alpha }}}\left( {\mathbf{x}} \right)\,,\label{Peralf}\\
{{\boldsymbol{\beta }}}\left( {{\mathbf{x}} + {{\mathbf{v}}_i}} \right) &= {{\boldsymbol{\beta }}}\left( {\mathbf{x}} \right)\,,\label{Perbeta}\\
{\Psi }\left( {{\mathbf{x}} + {{\mathbf{v}}_i}} \right) &= {\Psi}\left( {\mathbf{x}} \right).\label{PerPsi}
\end{align}
\noindent We focus the attention on the study of wave propagation  in the thermodiffusive periodic material. Note that, due to the structure of the governing equations (\ref{gov1})-(\ref{gov3}), the waves propagating in the medium are in general dispersive and characterized by  temporal damping. 
To this aim, we apply the time bilateral Laplace transform to Equations (\ref{bal1})-(\ref{bal3})
 in the case of
zero source terms, i.e. ${\mathbf{f}} = {\mathbf{0}}$, $r =0, s = 0$.
We recall that the time bilateral Laplace transform (or Fourier transform with complex argument) \citep{VanBre1947,Wei1965} 
of a given a function 
$\mathbf{g}\left( {{\mathbf{x}}{\text{, }}t} \right)$, is defined as 
 \begin{equation}
\mathcal{L}_{t} \left[ {{\mathbf{g}}\left( {{\mathbf{x}}{\text{, }}t} \right)} \right] = \int_{-\infty}^{ + \infty } {{\mathbf{g}}\left( {{\mathbf{x}}{\text{, }}t} \right){e^{ - I\omega t}}dt} = {\mathbf{\hat g}}\left( {\mathbf{x}} \right)\,,
\end{equation}
where $\omega$ is the 
complex angular frequency ($\omega \in \mathcal{C}$), $I$ is the imaginary unit, with $I^2=-1$.
It follows that the governing equations in the transformed
space, also referred to as generalized Christoffel equations, are
\begin{align}
\nabla  \cdot \left( {{\mathbb{C}}\nabla {\mathbf{\hat u}}} -  {{{\boldsymbol{\alpha }}}\hat \vartheta }  - \nabla  {{{\boldsymbol{\beta }}}\hat \eta } \right) + \rho {\omega ^2}{\mathbf{\hat u}} &= {\mathbf{0}}\,,\label{eq:Chr1}\\
\nabla  \cdot \left( {{{\mathbf{K}}}\nabla \hat \vartheta } \right) - I\omega \left( {{\boldsymbol{\alpha }}}:\nabla {\mathbf{\hat u}} + {\Psi }\hat \eta + {p}\hat \vartheta \right) &= 0\,,\label{eq:Chr2}\\
\nabla  \cdot \left( {{{\mathbf{D}}}\nabla \hat \eta } \right) - I\omega  \left( {{\boldsymbol{\beta }}}:\nabla {\mathbf{\hat u}} + {\Psi }\hat \vartheta + {c}\hat \eta \right)&= 0\,,\label{eq:Chr3}
\end{align}
where $\mathbf{\hat u}$, $\hat \vartheta$, $\hat \eta$ are the
bilateral Laplace transforms of the displacement field, the temperature field and the chemical potential field, respectively.\\
By exploiting the periodicity of the medium, it is possible to analyse the generalized Christoffel equations (\ref{eq:Chr1}), (\ref{eq:Chr2}), (\ref{eq:Chr3}) in the periodic cell $\textbf{A}$ undergoing the Floquet-Bloch boundary conditions. 
Let  $\textbf{k}=k_1 \textbf{e}_1+k_2 \textbf{e}_2 \in \textbf{B}$ be the wave vector, being $k_1$, $k_2$ the wave numbers and $\textbf{B}=[-\pi/d_1,\pi/d_1]\times[-\pi/d_2,\pi/d_2]$ the first Brillouin zone   related to the periodic cell with orthogonal periodicity vectors. Moreover, it is possible to define the unit vector
of propagation
$\textbf{m} = \textbf{k} / ||\textbf{k} ||$. The Floquet-Bloch boundary conditions read
\begin{align}
  {{{\mathbf{\hat u}}} }(\textbf{x}+\textbf{v}_p) &= {e^{I\left( {{\mathbf{k}} \cdot {{\mathbf{v}}_p}} \right)}}{{{\mathbf{\hat u}}} }(\textbf{x}),\label{eq:FB1}\\
  {\text{ }}{{\hat \vartheta } }(\textbf{x}+\textbf{v}_p) &= {e^{I\left( {{\mathbf{k}} \cdot {{\mathbf{v}}_p}} \right)}}{{\hat \vartheta } }(\textbf{x}),\\
  {\text{ }}{{\hat \eta } }(\textbf{x}+\textbf{v}_p) &= {e^{I\left( {{\mathbf{k}} \cdot {{\mathbf{v}}_p}} \right)}}{{\hat \eta } }(\textbf{x})\,, \hfill \\
  \boldsymbol{\hat \sigma}(\textbf{x}+\textbf{v}_p){\mathbf{n}}_h(\textbf{x}+\textbf{v}_p)  &=  - {e^{I\left( {{\mathbf{k}} \cdot {{\mathbf{v}}_p}} \right)}} \boldsymbol{\hat \sigma} (\textbf{x}){\mathbf{n}}_h(\textbf{x}) \,, \hfill \\
  {{{\mathbf{\hat q}}} }(\textbf{x}+\textbf{v}_p){\mathbf{n}}_h(\textbf{x}+\textbf{v}_p) +  &=  - {e^{I\left( {{\mathbf{k}} \cdot {{\mathbf{v}}_p}} \right)}}{{{\mathbf{\hat q}}} }(\textbf{x}){\mathbf{n}}_h(\textbf{x}) \,, \hfill \\
  {{{\mathbf{\hat j}}} }(\textbf{x}+\textbf{v}_p){\mathbf{n}}_h(\textbf{x}+\textbf{v}_p)  &=  - {e^{I\left( {{\mathbf{k}} \cdot {{\mathbf{v}}_p}} \right)}}{{{\mathbf{\hat j}}}}(\textbf{x}){\mathbf{n}}_h(\textbf{x}) \,. \label{eq:FB6}
\end{align}
where  $\textbf{v}_p$ is the periodicity vector ($p=1,2$), $\textbf{n}_h$ is the outward normal unit vector ($h=1,2$), and $\boldsymbol{\hat \sigma}$, $\mathbf{\hat q}$, $\mathbf{\hat j}$ are the bilateral Laplace transforms of the stress tensor, heat and mass fluxes, respectively. It is worth noting that by adopting the constitutive relations, the following expressions hold  
\begin{align}
\boldsymbol{\hat \sigma}  &= {\mathbb{C}}\nabla {\mathbf{\hat u}} - {{\boldsymbol{\alpha }}}\hat \vartheta  - {{\boldsymbol{\beta }}}\hat \eta \,,\\
{\mathbf{\hat q}} &=  - {{\mathbf{K}}}\nabla \hat \vartheta \,,\\
{\mathbf{\hat j}} &=  - {{\mathbf{D}}}\nabla \hat \eta \,.
\end{align}

\section{ Damped Bloch wave propagation and  Floquet Bloch spectrum }\label{Damped}
In accordance with the Floquet-Bloch theory, here generalized to the case of an elastic thermodiffusive medium, we resort to the 
Floquet-Bloch decomposition, that is 
\begin{align}
{\mathbf{\hat u}}\left( {\mathbf{x}} \right) = {\mathbf{\tilde u}}\left( {\mathbf{x}} \right){e^{I\left( {{\mathbf{k}} \cdot {\mathbf{x}}} \right)}},\label{eq:FBdec1}\\
{\text{       }}\hat \vartheta \left( {\mathbf{x}} \right) = \tilde \vartheta \left( {\mathbf{x}} \right){e^{I\left( {{\mathbf{k}} \cdot {\mathbf{x}}} \right)}},\label{eq:FBdec2}\\
{\text{       }}\hat \eta \left( {\mathbf{x}} \right) = \tilde \eta \left( {\mathbf{x}} \right){e^{I\left( {{\mathbf{k}} \cdot {\mathbf{x}}} \right)}}\,,\label{eq:FBdec3}
\end{align}
with ${\mathbf{\tilde u}}\left( {\mathbf{x}} \right), \tilde \vartheta \left( {\mathbf{x}} \right), \tilde \eta \left( {\mathbf{x}} \right)$ being the $A$-periodic Bloch amplitudes of the displacement, temperature, and chemical potential fields, in the transformed space. It is worth noting that Equations (\ref{eq:FBdec1})-(\ref{eq:FBdec3}) 
automatically satisfy the Floquet-Bloch boundary conditions (\ref{eq:FB1})-(\ref{eq:FB6}). \\
By plugging in Equations (\ref{eq:FBdec1})-(\ref{eq:FBdec3}) into Equations (\ref{eq:Chr1})-(\ref{eq:Chr3}), the generalized Christoffel equations take the following form
\begin{eqnarray}
&& \nabla  \cdot \left( {{\mathbb{C}}\left( \mathbf{x} \right)\nabla \left( {\mathbf{\tilde u}}\left( {\mathbf{x}} \right){e^{I\left( {{\mathbf{k}} \cdot {\mathbf{x}}} \right)}} \right)} \right) - \nabla  \cdot \left( {{{\boldsymbol{\alpha }}}\left( \mathbf{x} \right) \tilde \vartheta \left( {\mathbf{x}} \right){e^{I\left( {{\mathbf{k}} \cdot {\mathbf{x}}} \right)}}} \right) \nonumber \\
&& - \nabla  \cdot \left( {{{\boldsymbol{\beta }}}\left( \mathbf{x} \right) \tilde \eta \left( {\mathbf{x}} \right){e^{I\left( {{\mathbf{k}} \cdot {\mathbf{x}}} \right)}}} \right) + \rho \left( \mathbf{x} \right){\omega ^2}{ {\mathbf{\tilde u}}\left( {\mathbf{x}} \right){e^{I\left( {{\mathbf{k}} \cdot {\mathbf{x}}} \right)}} } = {\mathbf{0}}\,,\label{eq:PDE1}\\[5pt]	
&& \nabla  \cdot \left( {{{\mathbf{K}}}\left( \mathbf{x} \right)\nabla \left( \tilde \vartheta \left( {\mathbf{x}} \right){e^{I\left( {{\mathbf{k}} \cdot {\mathbf{x}}} \right)}} \right)} \right) - I\omega {\text{ }}{{\boldsymbol{\alpha }}}\left( \mathbf{x} \right):\nabla {\left( {\mathbf{\tilde u}}\left( {\mathbf{x}} \right){e^{I\left( {{\mathbf{k}} \cdot {\mathbf{x}}} \right)}} \right)} \nonumber \\
&&- I\omega {\Psi }\left( \mathbf{x} \right) \tilde \eta \left( {\mathbf{x}} \right){e^{I\left( {{\mathbf{k}} \cdot {\mathbf{x}}} \right)}} -I\omega {p}\left( \mathbf{x} \right) \tilde \vartheta \left( {\mathbf{x}} \right){e^{I\left( {{\mathbf{k}} \cdot {\mathbf{x}}} \right)}}=0\,,\label{eq:PDE2}\\[5pt]
&& \nabla  \cdot \left( {{{\mathbf{D}}}\left( \mathbf{x} \right)\nabla \left( \tilde \eta \left( {\mathbf{x}} \right){e^{I\left( {{\mathbf{k}} \cdot {\mathbf{x}}} \right)}} \right) } \right) - I\omega {\text{ }}{{\boldsymbol{\beta }}}\left( \mathbf{x} \right):\nabla {\left( {\mathbf{\tilde u}}\left( {\mathbf{x}} \right){e^{I\left( {{\mathbf{k}} \cdot {\mathbf{x}}} \right)}} \right)} \nonumber \\
&& - I\omega {\Psi }\left( \mathbf{x} \right) \tilde \vartheta \left( {\mathbf{x}} \right){e^{I\left( {{\mathbf{k}} \cdot {\mathbf{x}}} \right)}}  -I\omega {c}\left( \mathbf{x} \right) \tilde \eta \left( {\mathbf{x}} \right){e^{I\left( {{\mathbf{k}} \cdot {\mathbf{x}}} \right)}}=0\,.\label{eq:PDE3}
\end{eqnarray}
The governing equations (\ref{eq:PDE1})-(\ref{eq:PDE3}) can be properly manipulated by performing the space Fourier transform, which for a generic function $\textbf{g}(\mathbf{x})$ is defined as
\begin{align}
\mathcal{F}_{\mathbf{x}} \left[ {{\mathbf{g}}\left( {\mathbf{x}} \right)} \right] = \int_{{\mathbb{R}^2}}^{} {{\mathbf{g}}\left( {\mathbf{x}} \right){e^{ - I{\mathbf{r}} \cdot {\mathbf{x}}}}d{\mathbf{x}}},
\end{align}
where $\textbf{r}$ is a vector in the transformed space of wave vectors, with $\textbf{r} \in \textbf{B}$. 
It follows that in transformed space and frequency domain, Equations (\ref{eq:PDE1})-(\ref{eq:PDE3}), in the component form, become
\begin{eqnarray}
  && I {r_j} \left(\mathcal{F}_{\mathbf{x}} \left[ {C_{ijhk}} \right] * \left(I {r_k} \left(\mathcal{F}_{\mathbf{x}} \left[ {{{\tilde u}_h}} \right] * \mathcal{F}_{\mathbf{x}} \left[ e^{I\left( {{\mathbf{k}} \cdot {\mathbf{x}}} \right)} \right] \right) \right) \right) - I{r_j} \left( \mathcal{F}_{\mathbf{x}} \left[ {\alpha _{ij}} \right] * \mathcal{F}_{\mathbf{x}} \left[ {\tilde \vartheta } \right] * \mathcal{F}_{\mathbf{x}} \left[ e^{I\left( {{\mathbf{k}} \cdot {\mathbf{x}}} \right)} \right] \right) \nonumber \\
	&& - I{r_j} \left( \mathcal{F}_{\mathbf{x}} \left[ {\beta _{ij}} \right] * \mathcal{F}_{\mathbf{x}} \left[ {\tilde \eta } \right] * \mathcal{F}_{\mathbf{x}} \left[ e^{I\left( {{\mathbf{k}} \cdot {\mathbf{x}}} \right)} \right] \right) + {\omega ^2}\mathcal{F}_{\mathbf{x}} \left[ {{\rho }} \right] * \mathcal{F}_{\mathbf{x}} \left[ {{{\tilde u}_i}} \right] * \mathcal{F}_{\mathbf{x}} \left[ e^{I\left( {{\mathbf{k}} \cdot {\mathbf{x}}} \right)} \right] = 0\,,\label{eq:ChrFourier1}\\[5pt]
	&& I {r_i} \left( \mathcal{F}_{\mathbf{x}} \left[ {K_{ij}} \right] * \left( I r_j \left( \mathcal{F}_{\mathbf{x}} \left[ {\tilde \vartheta } \right] * \mathcal{F}_{\mathbf{x}} \left[ e^{I\left( {{\mathbf{k}} \cdot {\mathbf{x}}} \right)} \right] \right) \right) \right) - I \omega \left( \mathcal{F}_{\mathbf{x}} \left[ {\alpha _{ij}} \right] * \left( I {r_j} \left( \mathcal{F}_{\mathbf{x}} \left[ {{{\tilde u}_i}} \right] * \mathcal{F}_{\mathbf{x}} \left[ e^{I\left( {{\mathbf{k}} \cdot {\mathbf{x}}} \right)} \right] \right) \right) \right) \nonumber \\
&& - I\omega \mathcal{F}_{\mathbf{x}} \left[ {{\Psi}} \right] * \mathcal{F}_{\mathbf{x}} \left[ {\tilde \eta } \right] * \mathcal{F}_{\mathbf{x}} \left[ e^{I\left( {{\mathbf{k}} \cdot {\mathbf{x}}} \right)} \right] - I \omega \mathcal{F}_{\mathbf{x}} \left[ {{p}} \right] * \mathcal{F}_{\mathbf{x}} \left[ {\tilde \vartheta } \right] * \mathcal{F}_{\mathbf{x}} \left[ e^{I\left( {{\mathbf{k}} \cdot {\mathbf{x}}} \right)} \right] = 0\,, \label{eq:ChrFourier2}\\[5pt]
	&& I {r_i} \left( \mathcal{F}_{\mathbf{x}} \left[ {D_{ij}} \right] * \left( I r_j \left( \mathcal{F}_{\mathbf{x}} \left[ {\tilde \eta } \right] * \mathcal{F}_{\mathbf{x}} \left[ e^{I\left( {{\mathbf{k}} \cdot {\mathbf{x}}} \right)} \right] \right) \right) \right) - I \omega \left( \mathcal{F}_{\mathbf{x}} \left[ {\beta _{ij}} \right] * \left( I {r_j} \left( \mathcal{F}_{\mathbf{x}} \left[ {{{\tilde u}_i}} \right] * \mathcal{F}_{\mathbf{x}} \left[ e^{I\left( {{\mathbf{k}} \cdot {\mathbf{x}}} \right)} \right] \right) \right) \right) \nonumber \\
&&-  I\omega \mathcal{F}_{\mathbf{x}} \left[ {{\Psi }} \right] * \mathcal{F}_{\mathbf{x}} \left[ {\tilde \vartheta } \right] * \mathcal{F}_{\mathbf{x}} \left[ e^{I\left( {{\mathbf{k}} \cdot {\mathbf{x}}} \right)} \right] - I \omega \mathcal{F}_{\mathbf{x}} \left[ {{c}} \right] * \mathcal{F}_{\mathbf{x}} \left[ {\tilde \eta } \right] * \mathcal{F}_{\mathbf{x}} \left[ e^{I\left( {{\mathbf{k}} \cdot {\mathbf{x}}} \right)} \right] = 0\,, \label{eq:ChrFourier3}
  \end{eqnarray}
where $*$ denotes the convolution product symbol.\\
By performing the convolution products in Equations (\ref{eq:ChrFourier1})-(\ref{eq:ChrFourier3}), after some manipulations described in  Appendix A, we obtain 
\begin{eqnarray}
  && -4 \pi^2 {r_j} \int_{\mathbb{R}^2} {q_k} \mathcal{F}_{\mathbf{x}} \left[ {C_{ijhk}} \right](\mathbf{r}-\mathbf{q})\, \mathcal{F}_{\mathbf{x}} \left[ {{{\tilde u}_h}} \right](\mathbf{q}-\mathbf{k})\, d{\mathbf{q}}- 4 \pi^2 I{r_j} \int_{\mathbb{R}^2} \mathcal{F}_{\mathbf{x}} \left[ {\alpha _{ij}} \right](\mathbf{r}-\mathbf{q})\, \mathcal{F}_{\mathbf{x}} \left[ {\tilde \vartheta } \right] (\mathbf{q}-\mathbf{k})\, d{\mathbf{q}} \nonumber \\
	&& - 4 \pi^2 I{r_j} \int_{\mathbb{R}^2} \mathcal{F}_{\mathbf{x}} \left[ {\beta _{ij}} \right](\mathbf{r}-\mathbf{q})\, \mathcal{F}_{\mathbf{x}} \left[ {\tilde \eta } \right](\mathbf{q}-\mathbf{k})\, d{\mathbf{q}} + 4 \pi^2 {\omega ^2} \int_{\mathbb{R}^2} \mathcal{F}_{\mathbf{x}} \left[ {{\rho }} \right](\mathbf{r}-\mathbf{q})\, \mathcal{F}_{\mathbf{x}} \left[ {{{\tilde u}_i}} \right] (\mathbf{q}-\mathbf{k})\, d{\mathbf{q}} = 0\,, \nonumber\\ \label{eq:conv1}\\
	&&  -4 \pi^2 {r_i} \int_{\mathbb{R}^2} { q_j} \mathcal{F}_{\mathbf{x}} \left[ {K_{ij}} \right](\mathbf{r}-\mathbf{q})\,\mathcal{F}_{\mathbf{x}} \left[ {\tilde \vartheta } \right](\mathbf{q}-\mathbf{k})\, d{\mathbf{q}} + 4 \pi^2 \omega \int_{\mathbb{R}^2} q_j \mathcal{F}_{\mathbf{x}} \left[ {\alpha _{ij}} \right](\mathbf{r}-\mathbf{q})\, \mathcal{F}_{\mathbf{x}} \left[ {{{\tilde u}_i}} \right](\mathbf{q}-\mathbf{k})\, d{\mathbf{q}} \nonumber \\
  && -  4 \pi^2 I\omega \int_{\mathbb{R}^2} \mathcal{F}_{\mathbf{x}} \left[ {{\Psi }} \right](\mathbf{r}-\mathbf{q})\, \mathcal{F}_{\mathbf{x}} \left[ {\tilde \eta } \right](\mathbf{q}-\mathbf{k})\, d{\mathbf{q}} - 4 \pi^2 I \omega \int_{\mathbb{R}^2} \mathcal{F}_{\mathbf{x}} \left[ {{p}} \right](\mathbf{r}-\mathbf{q})\, \mathcal{F}_{\mathbf{x}} \left[ {\tilde \vartheta } \right](\mathbf{q}-\mathbf{k})\, d{\mathbf{q}} = 0\,, \nonumber\\ \label{eq:conv1var}\\
	&& -4 \pi^2 {r_i} \int_{\mathbb{R}^2} { q_j} \mathcal{F}_{\mathbf{x}} \left[ {D_{ij}} \right](\mathbf{r}-\mathbf{q})\,\mathcal{F}_{\mathbf{x}} \left[ {\tilde \eta } \right](\mathbf{q}-\mathbf{k})\, d{\mathbf{q}} + 4 \pi^2 \omega \int_{\mathbb{R}^2} q_j  \mathcal{F}_{\mathbf{x}} \left[ {\beta _{ij}} \right](\mathbf{q})\, \mathcal{F}_{\mathbf{x}} \left[ {{{\tilde u}_i}} \right](\mathbf{q}-\mathbf{k})\, d{\mathbf{q}} \nonumber \\
  && -  4 \pi^2 I\omega \int_{\mathbb{R}^2} \mathcal{F}_{\mathbf{x}} \left[ {{\Psi }} \right](\mathbf{r}-\mathbf{q})\, \mathcal{F}_{\mathbf{x}} \left[ {\tilde \vartheta } \right](\mathbf{q}-\mathbf{k})\, d{\mathbf{q}} - 4 \pi^2 I \omega \int_{\mathbb{R}^2} \mathcal{F}_{\mathbf{x}} \left[ {{c}} \right](\mathbf{r}-\mathbf{q})\, \mathcal{F}_{\mathbf{x}} \left[ {\tilde \eta } \right](\mathbf{q}-\mathbf{k})\, d{\mathbf{q}} = 0\,. \nonumber \label{eq:conv1var2}\\
  \end{eqnarray} 
with $\textbf{q} \in \textbf{B}$. Moreover, by exploiting the $A$-periodicity of the constitutive tensors, (\ref{PerC})-(\ref{PerD}), (\ref{Peralf})-(\ref{PerPsi}), of the inertial terms (\ref{Perro})-(\ref{Perc}), and of the unknown Bloch amplitudes ${\mathbf{\tilde u}} \left( {\mathbf{x}} \right)$, $\tilde \vartheta \left( {\mathbf{x}} \right)$, and $\tilde \eta \left( {\mathbf{x}} \right)$,  their  Fourier transforms take the form of weighted Dirac combs, whose weights are the related Fourier coefficients.
 It is worth noting that while the Fourier coefficients associated with the constitutive tensors and inertial terms are fully determined, those associated with the functions ${\mathbf{\tilde u}} \left( {\mathbf{x}} \right)$, $\tilde \vartheta \left( {\mathbf{x}} \right)$, and $\tilde \eta \left( {\mathbf{x}} \right)$ are unknowns. In particular, we first introduce 
  the following vector 
   \begin{align}\label{eq:pid}
{\bm{\pi}}_d &= \left(\frac{2 \pi}{d_1}, \frac{2 \pi}{d_2} \right) \in \mathbb{R}^2\,,
\end{align}
together with the two generic vectors ${\mathbf{m}}$ and ${\mathbf{n}}$ with integer components
\begin{align}\label{eq:indices}
{\mathbf{m}} &= \left(m_1, m_2 \right) \in \mathbb{Z}^2\,,\\
{\mathbf{n}} &= \left(n_1, n_2 \right) \in \mathbb{Z}^2\,,
\end{align}
 and we exploit the Hadamard product as
 \begin{align}\label{eq:Hp}
{\bm{\pi}}_d \circ {\mathbf{m}} & = \left( \frac{2 \pi}{d_1} m_1, \frac{2 \pi}{d_2} m_2\right) \,,\\
{\bm{\pi}}_d \circ {\mathbf{n}} &= \left( \frac{2 \pi}{d_1} n_1, \frac{2 \pi}{d_2} n_2\right)\,.
\end{align}
It follows that Equations (\ref{eq:conv1})-(\ref{eq:conv1var2}) can be specialized as
\begin{eqnarray}
  && -4 \pi^2 {r_j} \int_{\mathbb{R}^2} {q_k} \sum_{\mathbf{m} \in \mathbb{Z}^2} C_{i\,j\,h\,k}^{m_1\,m_2} \delta \left(\mathbf{r}-\mathbf{q} - {\bm{\pi}}_d \circ {\mathbf{m}}  \right) \sum_{\mathbf{n} \in \mathbb{Z}^2} \tilde{u}_h^{n_1\,n_2} \delta\left( \mathbf{q}-\mathbf{k} -{\bm{\pi}}_d \circ {\mathbf{n}} \right) \, d{\mathbf{q}} \nonumber \\
&& -4 \pi^2 I {r_j} \int_{\mathbb{R}^2} \sum_{\mathbf{m} \in \mathbb{Z}^2} \alpha_{i\,j}^{m_1\,m_2} \delta \left(\mathbf{r}-\mathbf{q} - {\bm{\pi}}_d \circ {\mathbf{m}}  \right) \sum_{\mathbf{n} \in \mathbb{Z}^2} \tilde{\vartheta}^{n_1\,n_2} \delta\left( \mathbf{q}-\mathbf{k} -{\bm{\pi}}_d \circ {\mathbf{n}} \right) \, d{\mathbf{q}} \nonumber \\
&& -4 \pi^2 I {r_j} \int_{\mathbb{R}^2} \sum_{\mathbf{m} \in \mathbb{Z}^2} \beta_{i\,j}^{m_1\,m_2} \delta \left(\mathbf{r}-\mathbf{q} - {\bm{\pi}}_d \circ {\mathbf{m}}  \right) \sum_{\mathbf{n} \in \mathbb{Z}^2} \tilde{\eta}^{n_1\,n_2} \delta\left( \mathbf{q}-\mathbf{k} -{\bm{\pi}}_d \circ {\mathbf{n}} \right) \, d{\mathbf{q}} \nonumber \\
&& +4 \pi^2 \omega^2 \int_{\mathbb{R}^2} \sum_{\mathbf{m} \in \mathbb{Z}^2} \rho^{m_1\,m_2} \delta \left(\mathbf{r}-\mathbf{q} - {\bm{\pi}}_d \circ {\mathbf{m}}  \right) \sum_{\mathbf{n} \in \mathbb{Z}^2} \tilde{u}_i^{n_1\,n_2} \delta\left( \mathbf{q}-\mathbf{k} -{\bm{\pi}}_d \circ {\mathbf{n}} \right) \, d{\mathbf{q}} = 0\,, \label{eq:Pattabi1}\\[5pt]
  && -4 \pi^2 {r_i} \int_{\mathbb{R}^2} {q_j} \sum_{\mathbf{m} \in \mathbb{Z}^2} K_{i\,j}^{m_1\,m_2} \delta \left(\mathbf{r}-\mathbf{q} - {\bm{\pi}}_d \circ {\mathbf{m}}  \right) \sum_{\mathbf{n} \in \mathbb{Z}^2} \tilde{\vartheta}^{n_1\,n_2} \delta\left( \mathbf{q}-\mathbf{k} -{\bm{\pi}}_d \circ {\mathbf{n}} \right) \, d{\mathbf{q}} \nonumber \\
&& +4 \pi^2 \omega \int_{\mathbb{R}^2} q_j  \sum_{\mathbf{m} \in \mathbb{Z}^2} \alpha_{i\,j}^{m_1\,m_2} \delta \left(\mathbf{r}-\mathbf{q} - {\bm{\pi}}_d \circ {\mathbf{m}}  \right) \sum_{\mathbf{n} \in \mathbb{Z}^2} \tilde{u}_i^{n_1\,n_2} \delta\left( \mathbf{q}-\mathbf{k} -{\bm{\pi}}_d \circ {\mathbf{n}} \right) \, d{\mathbf{q}} \nonumber \\
&& -4 \pi^2 I {\omega} \int_{\mathbb{R}^2} \sum_{\mathbf{m} \in \mathbb{Z}^2} \Psi^{m_1\,m_2} \delta \left(\mathbf{r}-\mathbf{q} - {\bm{\pi}}_d \circ {\mathbf{m}}  \right) \sum_{\mathbf{n} \in \mathbb{Z}^2} \tilde{\eta}^{n_1\,n_2} \delta\left( \mathbf{q}-\mathbf{k} -{\bm{\pi}}_d \circ {\mathbf{n}} \right) \, d{\mathbf{q}} \nonumber \\
&& - 4 \pi^2 I \omega \int_{\mathbb{R}^2} \sum_{\mathbf{m} \in \mathbb{Z}^2} p^{m_1\,m_2} \delta \left(\mathbf{r}-\mathbf{q} - {\bm{\pi}}_d \circ {\mathbf{m}}  \right) \sum_{\mathbf{n} \in \mathbb{Z}^2} \tilde{\vartheta}^{n_1\,n_2} \delta\left( \mathbf{q}-\mathbf{k} -{\bm{\pi}}_d \circ {\mathbf{n}} \right) \, d{\mathbf{q}} = 0\,, \label{eq:Pattabi2}\\[5pt]
  && -4 \pi^2 {r_i} \int_{\mathbb{R}^2} {q_j} \sum_{\mathbf{m} \in \mathbb{Z}^2} D_{i\,j}^{m_1\,m_2} \delta \left(\mathbf{r}-\mathbf{q} - {\bm{\pi}}_d \circ {\mathbf{m}}  \right) \sum_{\mathbf{n} \in \mathbb{Z}^2} \tilde{\eta}^{n_1\,n_2} \delta\left( \mathbf{q}-\mathbf{k} -{\bm{\pi}}_d \circ {\mathbf{n}} \right) \, d{\mathbf{q}} \nonumber \\
&& + 4 \pi^2 \omega \int_{\mathbb{R}^2} q_j \sum_{\mathbf{m} \in \mathbb{Z}^2} \beta_{i\,j}^{m_1\,m_2} \delta \left(\mathbf{r}-\mathbf{q} - {\bm{\pi}}_d \circ {\mathbf{m}}  \right) \sum_{\mathbf{n} \in \mathbb{Z}^2} \tilde{u}_i^{n_1\,n_2} \delta\left( \mathbf{q}-\mathbf{k} -{\bm{\pi}}_d \circ {\mathbf{n}} \right) \, d{\mathbf{q}} \nonumber \\
&& -4 \pi^2 I {\omega} \int_{\mathbb{R}^2} \sum_{\mathbf{m} \in \mathbb{Z}^2} \Psi^{m_1\,m_2} \delta \left(\mathbf{r}-\mathbf{q} - {\bm{\pi}}_d \circ {\mathbf{m}}  \right) \sum_{\mathbf{n} \in \mathbb{Z}^2} \tilde{\vartheta}^{n_1\,n_2} \delta\left( \mathbf{q}-\mathbf{k} -{\bm{\pi}}_d \circ {\mathbf{n}} \right) \, d{\mathbf{q}} \nonumber \\
&& - 4 \pi^2 I \omega \int_{\mathbb{R}^2} \sum_{\mathbf{m} \in \mathbb{Z}^2} c^{m_1\,m_2} \delta \left(\mathbf{r}-\mathbf{q} - {\bm{\pi}}_d \circ {\mathbf{m}}  \right) \sum_{\mathbf{n} \in \mathbb{Z}^2} \tilde{\eta}^{n_1\,n_2} \delta\left( \mathbf{q}-\mathbf{k} -{\bm{\pi}}_d \circ {\mathbf{n}} \right) \, d{\mathbf{q}} = 0\,. \label{eq:Pattabi3}
  \end{eqnarray}  
where  $\delta(\cdot)$ denotes the Dirac-delta generalized function,  $C_{i\,j\,h\,k}^{m_1\,m_2}$, $K_{i\,j}^{m_1\,m_2}$, $D_{i\,j}^{m_1\,m_2}$, $\alpha_{i\,j}^{m_1\,m_2}$, $\beta_{i\,j}^{m_1\,m_2}$, $\Psi^{m_1\,m_2}$, $\rho^{m_1\,m_2}$, $p^{m_1\,m_2}$, $c^{m_1\,m_2}$ are the weights of the Dirac combs associated with the constitutive tensors and the inertial terms, and $\tilde{u}_{h}^{n_1\,n_2}$, $\tilde{\vartheta}^{n_1\,n_2}$, $\tilde{\eta}^{n_1\,n_2}$ are the weights of the Dirac combs associated with the unknown Bloch amplitudes.\\ 
By recalling the property  $g(\mathbf{r}) \delta(\mathbf{r}-\mathbf{r}_0)=g(\mathbf{r}_0) \delta(\mathbf{r}-\mathbf{r}_0)$ which applies to a generic smooth function $g(\mathbf{r})$ and a constant vector $\mathbf{r}_0$, and by computing the integrals in (\ref{eq:Pattabi1})-(\ref{eq:Pattabi3}), we obtain
\begin{eqnarray}
  && -4 \pi^2 \sum_{\mathbf{m}, \mathbf{n}\in \mathbb{Z}^2}\left(k_j +  \frac{2 \pi (m_j + n_j)}{d_j} \right)   \left(k_k+  \frac{2 \pi n_k}{d_k} \right)   C_{i\,j\,h\,k}^{m_1\,m_2} \tilde{u}_h^{n_1\,n_2} \delta\left( \mathbf{r}-\mathbf{k} -{\bm{\pi}}_d \circ {(\mathbf{m}+\mathbf{n})} \right) \nonumber \\
&& -4 \pi^2 I \sum_{\mathbf{m}, \mathbf{n}\in \mathbb{Z}^2} \left(k_j +  \frac{2 \pi (m_j + n_j)}{d_j} \right) \alpha_{i\,j}^{m_1\,m_2} \tilde{\vartheta}^{n_1\,n_2}  \delta\left( \mathbf{r}-\mathbf{k} -{\bm{\pi}}_d \circ {(\mathbf{m}+\mathbf{n})} \right) \nonumber \\
&& -4 \pi^2 I \sum_{\mathbf{m}, \mathbf{n}\in \mathbb{Z}^2} \left(k_j +  \frac{2 \pi (m_j + n_j)}{d_j} \right) \beta_{i\,j}^{m_1\,m_2} \tilde{\eta}^{n_1\,n_2}  \delta\left( \mathbf{r}-\mathbf{k} -{\bm{\pi}}_d \circ {(\mathbf{m}+\mathbf{n})} \right) \nonumber \\
&& +4 \pi^2 \omega^2 \sum_{\mathbf{m}, \mathbf{n}\in \mathbb{Z}^2} \rho^{m_1\,m_2}  \tilde{u}_i^{n_1\,n_2}  \delta\left( \mathbf{r}-\mathbf{k} -{\bm{\pi}}_d \circ {(\mathbf{m}+\mathbf{n})} \right) = 0\,, \label{eq:Pattabi1bis}\\[5pt]
  && -4 \pi^2 \sum_{\mathbf{m}, \mathbf{n}\in \mathbb{Z}^2} \left(k_i +  \frac{2 \pi (m_i + n_i)}{d_i} \right) \left(k_j +  \frac{2 \pi n_j}{d_j} \right)  K_{i\,j}^{m_1\,m_2}  \tilde{\vartheta}^{n_1\,n_2}  \delta\left( \mathbf{r}-\mathbf{k} -{\bm{\pi}}_d \circ {(\mathbf{m}+\mathbf{n})} \right) \nonumber \\
&& +4 \pi^2 \omega \sum_{\mathbf{m}, \mathbf{n}\in \mathbb{Z}^2} \left(k_j +  \frac{2 \pi n_j}{d_j} \right) \alpha_{i\,j}^{m_1\,m_2}  \tilde{u}_i^{n_1\,n_2}  \delta\left( \mathbf{r}-\mathbf{k} -{\bm{\pi}}_d \circ {(\mathbf{m}+\mathbf{n})} \right) \nonumber \\
&& -4 \pi^2 I {\omega} \sum_{\mathbf{m}, \mathbf{n}\in \mathbb{Z}^2} \Psi^{m_1\,m_2} \tilde{\eta}^{n_1\,n_2}  \delta\left( \mathbf{r}-\mathbf{k} -{\bm{\pi}}_d \circ {(\mathbf{m}+\mathbf{n})} \right) \nonumber \\
&& - 4 \pi^2 I \omega \sum_{\mathbf{m}, \mathbf{n}\in \mathbb{Z}^2} p^{m_1\,m_2}  \tilde{\vartheta}^{n_1\,n_2}  \delta\left( \mathbf{r}-\mathbf{k} -{\bm{\pi}}_d \circ {(\mathbf{m}+\mathbf{n})} \right) = 0\,, \label{eq:Pattabi2bis}\\[5pt]
  && -4 \pi^2 \sum_{\mathbf{m}, \mathbf{n}\in \mathbb{Z}^2} \left(k_i +  \frac{2 \pi (m_i + n_i)}{d_i} \right) \left(k_j +  \frac{2 \pi n_j}{d_j} \right) D_{i\,j}^{m_1\,m_2}  \tilde{\eta}^{n_1\,n_2}  \delta\left( \mathbf{r}-\mathbf{k} -{\bm{\pi}}_d \circ {(\mathbf{m}+\mathbf{n})} \right) \nonumber \\
&& + 4 \pi^2 \omega \sum_{\mathbf{m}, \mathbf{n}\in \mathbb{Z}^2} \left(k_j +  \frac{2 \pi n_j}{d_j} \right) \beta_{i\,j}^{m_1\,m_2}  \tilde{u}_i^{n_1\,n_2}  \delta\left( \mathbf{r}-\mathbf{k} -{\bm{\pi}}_d \circ {(\mathbf{m}+\mathbf{n})} \right) \nonumber \\
&& -4 \pi^2 I {\omega} \sum_{\mathbf{m}, \mathbf{n}\in \mathbb{Z}^2} \Psi^{m_1\,m_2}  \tilde{\vartheta}^{n_1\,n_2}  \delta\left( \mathbf{r}-\mathbf{k} -{\bm{\pi}}_d \circ {(\mathbf{m}+\mathbf{n})} \right) \nonumber \\
&& - 4 \pi^2 I \omega \sum_{\mathbf{m}, \mathbf{n}\in \mathbb{Z}^2} c^{m_1\,m_2} \tilde{\eta}^{n_1\,n_2}  \delta\left( \mathbf{r}-\mathbf{k} -{\bm{\pi}}_d \circ {(\mathbf{m}+\mathbf{n})} \right) = 0\,. \label{eq:Pattabi3bis}
  \end{eqnarray} 
Moreover, by defining the following vectors with integer components
\begin{align}\label{eq:rq}
{\bar{\mathbf{r}}} & = {\mathbf{m}} +{\mathbf{n}}\in \mathbb{Z}^2 \,,\\
{\bar{\mathbf{q}}} & = {\mathbf{m}} \in \mathbb{Z}^2\,,
\end{align}
Equations (\ref{eq:Pattabi1bis})-(\ref{eq:Pattabi3bis}) become  
\begin{eqnarray}
  && -4 \pi^2 \sum_{\bar{\mathbf{r}} \in \mathbb{Z}^2}\left[\left(k_j +  \frac{2 \pi \bar{r}_j}{d_j} \right) \sum_{\bar{\mathbf{q}} \in \mathbb{Z}^2} \left(k_k+  \frac{2 \pi \bar{q}_k}{d_k} \right)   C_{i\,j\,h\,k}^{\bar{r}_1-\bar{q}_1\,\bar{r}_2-\bar{q}_2} \tilde{u}_h^{\bar{q}_1\,\bar{q}_2} \right] \delta\left( \mathbf{r}-\mathbf{k} -{\bm{\pi}}_d \circ \bar{\mathbf{r}} \right) \nonumber \\
&& -4 \pi^2 I \sum_{\bar{\mathbf{r}} \in \mathbb{Z}^2} \left[\left(k_j +  \frac{2 \pi \bar{r}_j}{d_j} \right) \sum_{\bar{\mathbf{q}} \in \mathbb{Z}^2} \alpha_{i\,j}^{\bar{r}_1-\bar{q}_1\,\bar{r}_2-\bar{q}_2} \tilde{\vartheta}^{\bar{q}_1\,\bar{q}_2}\right]  \delta\left( \mathbf{r}-\mathbf{k} -{\bm{\pi}}_d \circ \bar{\mathbf{r}} \right) \nonumber \\
&& -4 \pi^2 I \sum_{\bar{\mathbf{r}} \in \mathbb{Z}^2} \left[ \left(k_j +  \frac{2 \pi \bar{r}_j}{d_j} \right) \sum_{\bar{\mathbf{q}} \in \mathbb{Z}^2} \beta_{i\,j}^{\bar{r}_1-\bar{q}_1\,\bar{r}_2-\bar{q}_2} \tilde{\eta}^{\bar{q}_1\,\bar{q}_2}\right]  \delta\left( \mathbf{r}-\mathbf{k} -{\bm{\pi}}_d \circ \bar{\mathbf{r}} \right) \nonumber \\
&& +4 \pi^2 \omega^2 \sum_{\bar{\mathbf{r}} \in \mathbb{Z}^2} \left[\sum_{\bar{\mathbf{q}} \in \mathbb{Z}^2} \rho^{\bar{r}_1-\bar{q}_1\,\bar{r}_2-\bar{q}_2}  \tilde{u}_i^{\bar{q}_1\,\bar{q}_2}\right]  \delta\left( \mathbf{r}-\mathbf{k} -{\bm{\pi}}_d \circ \bar{\mathbf{r}} \right) = 0\,, \label{eq:Pattabi1ter}\\[5pt]
  && -4 \pi^2 \sum_{\bar{\mathbf{r}} \in \mathbb{Z}^2} \left[ \left(k_i +  \frac{2 \pi \bar{r}_i}{d_i} \right) \sum_{\bar{\mathbf{q}} \in \mathbb{Z}^2} \left(k_j +  \frac{2 \pi \bar{q}_j}{d_j} \right)  K_{i\,j}^{\bar{r}_1-\bar{q}_1\,\bar{r}_2-\bar{q}_2}  \tilde{\vartheta}^{\bar{q}_1\,\bar{q}_2} \right] \delta\left( \mathbf{r}-\mathbf{k} -{\bm{\pi}}_d \circ \bar{\mathbf{r}} \right) \nonumber \\
&& +4 \pi^2 \omega \sum_{\bar{\mathbf{r}} \in \mathbb{Z}^2} \left[ \sum_{\bar{\mathbf{q}} \in \mathbb{Z}^2} \left(k_j +  \frac{2 \pi \bar{q}_j}{d_j} \right) \alpha_{i\,j}^{\bar{r}_1-\bar{q}_1\,\bar{r}_2-\bar{q}_2}  \tilde{u}_i^{\bar{q}_1\,\bar{q}_2}\right]  \delta\left( \mathbf{r}-\mathbf{k} -{\bm{\pi}}_d \circ \bar{\mathbf{r}} \right) \nonumber \\
&& -4 \pi^2 I {\omega} \sum_{\bar{\mathbf{r}} \in \mathbb{Z}^2} \left[ \sum_{\bar{\mathbf{q}} \in \mathbb{Z}^2} \Psi^{\bar{r}_1-\bar{q}_1\,\bar{r}_2-\bar{q}_2} \tilde{\eta}^{\bar{q}_1\,\bar{q}_2}\right]  \delta\left( \mathbf{r}-\mathbf{k} -{\bm{\pi}}_d \circ \bar{\mathbf{r}} \right) \nonumber \\
&& - 4 \pi^2 I \omega \sum_{\bar{\mathbf{r}} \in \mathbb{Z}^2} \left[ \sum_{\bar{\mathbf{q}} \in \mathbb{Z}^2} p^{\bar{r}_1-\bar{q}_1\,\bar{r}_2-\bar{q}_2}  \tilde{\vartheta}^{\bar{q}_1\,\bar{q}_2} \right]  \delta\left( \mathbf{r}-\mathbf{k} -{\bm{\pi}}_d \circ \bar{\mathbf{r}} \right) = 0\,, \label{eq:Pattabi2ter}\\[5pt]
  && -4 \pi^2 \sum_{\bar{\mathbf{r}} \in \mathbb{Z}^2} \left[ \left(k_i +  \frac{2 \pi \bar{r}_i}{d_i} \right) \sum_{\bar{\mathbf{q}} \in \mathbb{Z}^2}\left(k_j +  \frac{2 \pi \bar{q}_j}{d_j} \right) D_{i\,j}^{\bar{r}_1-\bar{q}_1\,\bar{r}_2-\bar{q}_2}  \tilde{\eta}^{\bar{q}_1\,\bar{q}_2} \right]  \delta\left( \mathbf{r}-\mathbf{k} -{\bm{\pi}}_d \circ \bar{\mathbf{r}} \right) \nonumber \\
&& + 4 \pi^2 \omega \sum_{\bar{\mathbf{r}} \in \mathbb{Z}^2} \left[ \sum_{\bar{\mathbf{q}} \in \mathbb{Z}^2} \left(k_j +  \frac{2 \pi \bar{q}_j}{d_j} \right) \beta_{i\,j}^{\bar{r}_1-\bar{q}_1\,\bar{r}_2-\bar{q}_2}  \tilde{u}_i^{\bar{q}_1\,\bar{q}_2} \right] \delta\left( \mathbf{r}-\mathbf{k} -{\bm{\pi}}_d \circ \bar{\mathbf{r}} \right) \nonumber \\
&& -4 \pi^2 I {\omega} \sum_{\bar{\mathbf{r}} \in \mathbb{Z}^2} \left[ \sum_{\bar{\mathbf{q}} \in \mathbb{Z}^2} \Psi^{\bar{r}_1-\bar{q}_1\,\bar{r}_2-\bar{q}_2}  \tilde{\vartheta}^{\bar{q}_1\,\bar{q}_2} \right] \delta\left( \mathbf{r}-\mathbf{k} -{\bm{\pi}}_d \circ \bar{\mathbf{r}} \right) \nonumber \\
&& - 4 \pi^2 I \omega \sum_{\bar{\mathbf{r}} \in \mathbb{Z}^2} \left[ \sum_{\bar{\mathbf{q}} \in \mathbb{Z}^2} c^{\bar{r}_1-\bar{q}_1\,\bar{r}_2-\bar{q}_2} \tilde{\eta}^{\bar{q}_1\,\bar{q}_2} \right]  \delta\left( \mathbf{r}-\mathbf{k} -{\bm{\pi}}_d \circ \bar{\mathbf{r}} \right) = 0\,. \label{eq:Pattabi3ter}
  \end{eqnarray}
\noindent Note that Equations  (\ref{eq:Pattabi1ter})-(\ref{eq:Pattabi3ter}) are weighted Dirac combs set equal to zero. Therefore,   the summation of all the weights associated with the Dirac-delta generalized function $\delta\left( \mathbf{r}-\mathbf{k} -{\bm{\pi}}_d \circ \bar{\mathbf{r}} \right)$ must vanish, for each $\bar{\mathbf{r}} \in \mathbb{Z}^2$.
This produces the following generalized eigenvalue problem, expressed as an infinite algebraic linear system
\begin{eqnarray}
  && - \left(\frac{2 \pi \bar{r}_j}{d_j}+k_j\right) \sum_{\bar{\mathbf{q}} \in {\mathbb{Z}^2}} \left(\frac{2 \pi { \bar{q}_k}}{d_k} +k_k \right) C_{ijhk}^{\bar{r}_1-\bar{q}_1\,\,\bar{r}_2-\bar{q}_2} {\tilde u}_h^{\bar{q}_1 \bar{q}_2}- I \left(\frac{2 \pi \bar{r}_j}{d_j} +k_j \right) \sum_{\bar{\mathbf{q}} \in {\mathbb{Z}^2}} \alpha_{ij}^{\bar{r}_1-\bar{q}_1\,\,\bar{r}_2-\bar{q}_2} {\tilde \vartheta}^{\bar{q}_1 \bar{q}_2}\nonumber \\
	&& - I \left(\frac{2 \pi \bar{r}_j}{d_j} +k_j \right) \sum_{\bar{\mathbf{q}} \in {\mathbb{Z}^2}} \beta_{ij}^{\bar{r}_1-\bar{q}_1\,\,\bar{r}_2-\bar{q}_2} {\tilde \eta}^{\bar{q}_1 \bar{q}_2} + \omega^2 \sum_{\bar{\mathbf{q}} \in {\mathbb{Z}^2}} {\rho}^{\bar{r}_1-\bar{q}_1\,\,\bar{r}_2-\bar{q}_2} {\tilde u}_i^{\bar{q}_1 \bar{q}_2} = 0\,,  \label{eq:conv2}\\[5pt]
	&& - \left(\frac{2 \pi \bar{r}_i}{d_i}+k_i\right) \sum_{\bar{\mathbf{q}} \in {\mathbb{Z}^2}} \left(\frac{2 \pi { \bar{q}_j}}{d_j}+k_j\right) K_{ij}^{\bar{r}_1-\bar{q}_1\,\,\bar{r}_2-\bar{q}_2} {\tilde \vartheta}^{\bar{q}_1 \bar{q}_2} + \omega \sum_{\bar{\mathbf{q}} \in {\mathbb{Z}^2}} \left(\frac{2 \pi \bar{q}_j}{d_j}+k_j\right) \alpha_{ij}^{\bar{r}_1-\bar{q}_1\,\,\bar{r}_2-\bar{q}_2} {\tilde u}_i^{\bar{q}_1 \bar{q}_2}\nonumber \label{eq:conv2var}\\
	&& -  I \omega \sum_{\bar{\mathbf{q}} \in {\mathbb{Z}^2}} \Psi^{\bar{r}_1-\bar{q}_1\,\,\bar{r}_2-\bar{q}_2} {\tilde \eta}^{\bar{q}_1 \bar{q}_2} - I \omega \sum_{\bar{\mathbf{q}} \in {\mathbb{Z}^2}} {p}^{\bar{r}_1-\bar{q}_1\,\,\bar{r}_2-\bar{q}_2} {\tilde \vartheta}^{\bar{q}_1 \bar{q}_2} = 0\,,  \\[5pt]
	&& - \left(\frac{2 \pi \bar{r}_i}{d_i}+k_i\right) \sum_{\bar{\mathbf{q}} \in {\mathbb{Z}^2}} \left(\frac{2 \pi { \bar{q}_j}}{d_j}+k_j\right)  D_{ij}^{\bar{r}_1-\bar{q}_1\,\,\bar{r}_2-\bar{q}_2} {\tilde \eta}^{\bar{q}_1 \bar{q}_2} + \omega  \sum_{\bar{\mathbf{q}} \in {\mathbb{Z}^2}} \left(\frac{2 \pi \bar{q}_j}{d_j}+k_j\right) \beta_{ij}^{\bar{r}_1-\bar{q}_1\,\,\bar{r}_2-\bar{q}_2} {\tilde u}_i^{\bar{q}_1 \bar{q}_2}\nonumber \\
	&& -  I \omega \sum_{\bar{\mathbf{q}} \in {\mathbb{Z}^2}} \Psi^{\bar{r}_1-\bar{q}_1\,\,\bar{r}_2-\bar{q}_2} {\tilde \vartheta}^{\bar{q}_1 \bar{q}_2} - I \omega \sum_{\bar{\mathbf{q}} \in {\mathbb{Z}^2}} {c}^{\bar{r}_1-\bar{q}_1\,\,\bar{r}_2-\bar{q}_2} {\tilde \eta}^{\bar{q}_1 \bar{q}_2} = 0\,, \label{eq:conv2var2}
  \end{eqnarray}
where $k_j$ is the $j$-th component of the wave vector ${\mathbf{k}} \in {\mathbb{R}^2}$, and  $\omega  = {\omega _r}+ I{\omega _i}$ is the complex angular frequency, whose real  and complex  parts characterize the propagation and the attenuation modes of dispersive Bloch waves, respectively.\\
Equations (\ref{eq:conv2})-(\ref{eq:conv2var2}) can be expressed in compact form, by introducing the linear operators $\textbf{\textit{A}}$, $\textbf{\textit{B}}$, and $\textbf{\textit{C}}$ applied to the vector $\mathbf{z}$ that collects the Fourier coefficients of the Bloch amplitudes of $\mathbf{\tilde u}_1$, $\mathbf{\tilde u}_2$, $\boldsymbol{\tilde \vartheta}$, $\boldsymbol{\tilde \eta}$,  for the details see Appendix B. Therefore, it results
\begin{equation}\label{eq:generalized}
\left( {{\omega^2} \textbf{\textit{A}} + \omega \textbf{\textit{B}} + \textbf{\textit{C}}} \right)\mathbf{z} = {\mathbf{0}}\,,
\end{equation}
taking the form of a quadratic generalized eigenvalue problem, where $\omega$ and $\mathbf{z}$ are the generalized eigenvalue and  eigenvector, respectively. It is worth noting that $\mathbf{z}$ is the polarization vector of the damped Bloch wave.  \\
The quadratic generalized eigenvalue problem (\ref{eq:generalized}) can be transformed into the following equivalent linear generalized eigenvalue problem
\begin{equation}\label{eq:generalized2}
(\omega \textbf{\textit{A}}'+\textbf{\textit{B}}') \, \mathbf{z}'=\mathbf{0}'\,,
\end{equation}
 where the operators $\textbf{\textit{A}}'$ and $\textbf{\textit{B}}'$ are applied to the generalized 
 eigenvector $\mathbf{z}'$ collecting the vectors  $\omega \mathbf{z}$ and $\mathbf{z}$ appearing in 
 (\ref{eq:generalized}), see Appendix B. Note that the linear generalized eigenvalue problem (\ref{eq:generalized2}) has a nontrivial solution $\mathbf{z}'$ if and only if the linear operator $\textbf{\textit{B}}'+\omega \textbf{\textit{A}}'$ is not invertible. In general, an infinite number of generalized eigenvalues $\omega$ is obtained.

\section{Truncation of the generalized eigenevualue problem and approximate solution of the Floquet-Bloch spectrum}\label{sec:4}

\begin{figure}[ht]
\centering
\subfigure[]
{\includegraphics[scale=0.56,trim={3.8cm 9.5cm 2.5cm 9.5cm}]{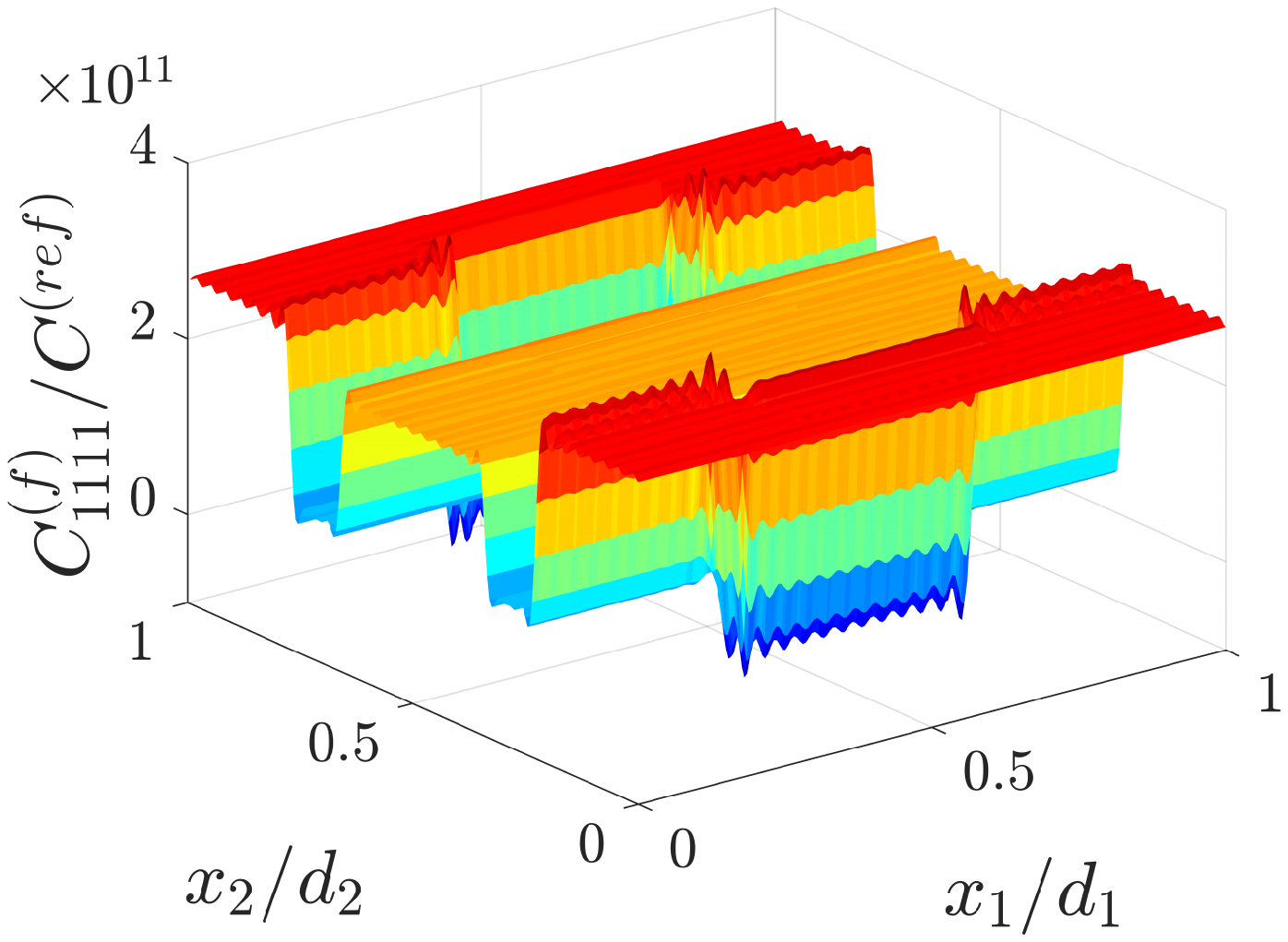}}
\subfigure[]
{\includegraphics[scale=0.56,trim={3.8cm 9.5cm 2.5cm 9.5cm}]{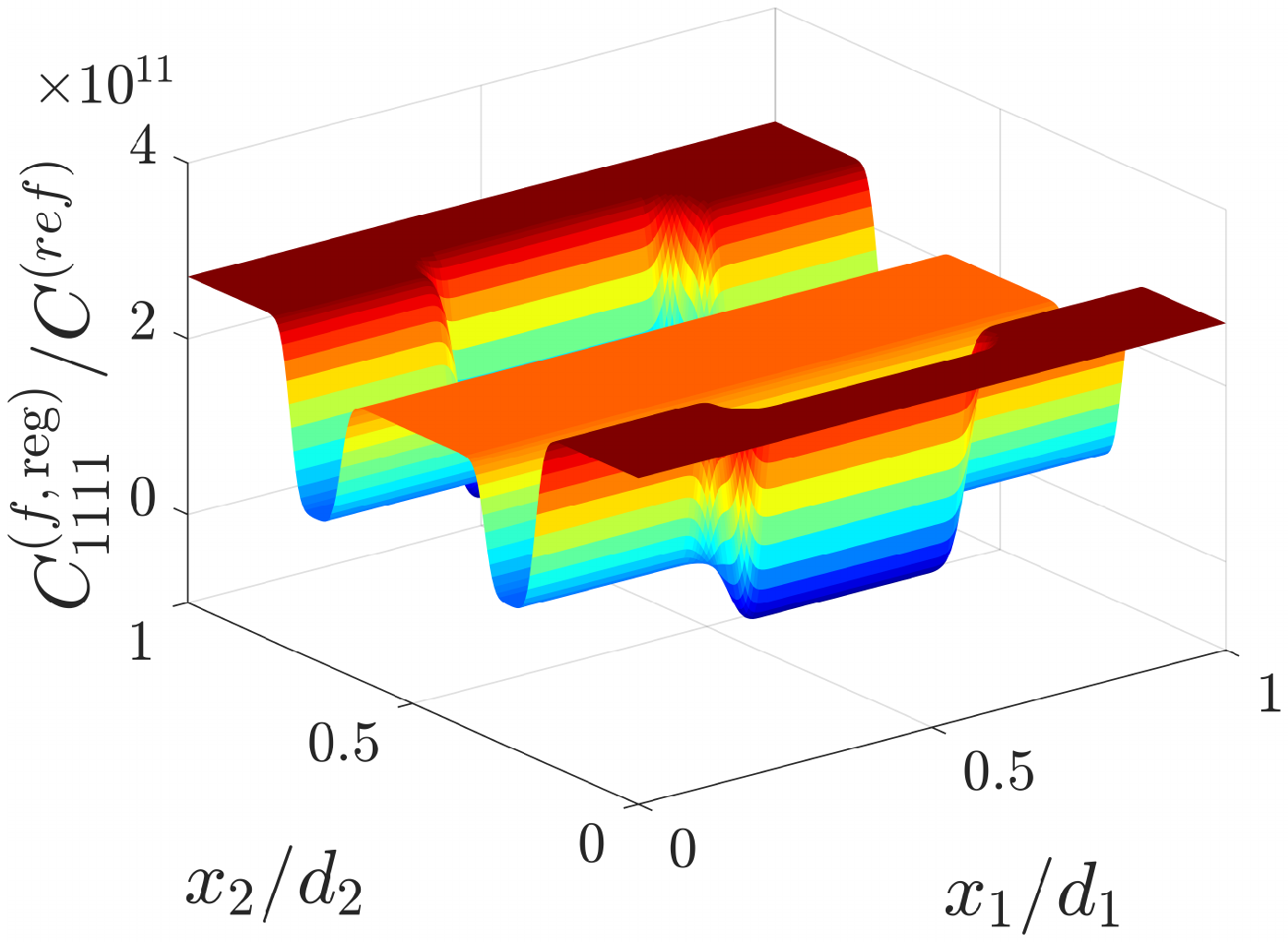}}
\caption{(a) Component $C_{1111}^{(f)}/C^{(ref)}$, with 961 terms kept in the Fourier expansion; (b) Component  $C_{1111}^{(f,\,{\rm reg})}/C^{(ref)}$ with $\sigma=0.005$.}\label{fig:regularization_new}
\end{figure}

The eigenvalue problem (\ref{eq:generalized2}) corresponds to the compact form of the infinite-dimensional algebraic system of Equations (\ref{eq:conv2})-(\ref{eq:conv2var2}). Therefore, 
 in order to obtain an approximate solution for the generalized eigenvalue $\omega$, the aforementioned algebraic system is truncated by restricting the discrete variables $\bar{\mathbf{r}},\bar{\mathbf{q}}$ 
 to the set $ \{-Q,\ldots,Q\}^2$,  with $Q \in \mathbb{Z}^+$. It follows that the resulting finite algebraic linear system is characterized by the same number of equations and unknowns, thus resulting in
 
\begin{eqnarray}\label{eq:conv3}
  && - \left(\frac{2 \pi \bar{r}_j}{d_j}+k_j\right) \sum_{\bar{\mathbf{q}} \in \{-Q,\ldots,Q\}^2} \left(\frac{2 \pi \bar{q}_k}{d_k} +k_k \right) C_{ijhk}^{\bar{r}_1-\bar{q}_1\,\,\bar{r}_2-\bar{q}_2} {\tilde u}_h^{\bar{q}_1 \bar{q}_2}\nonumber \\
	&& - I \left(\frac{2 \pi \bar{r}_j}{d_j} +k_j \right) \sum_{\bar{\mathbf{q}} \in \{-Q,\ldots,Q\}^2} \alpha_{ij}^{\bar{r}_1-\bar{q}_1\,\,\bar{r}_2-\bar{q}_2} {\tilde \vartheta}^{\bar{q}_1 \bar{q}_2}\nonumber \\
	&& - I \left(\frac{2 \pi \bar{r}_j}{d_j} +k_j \right) \sum_{\bar{\mathbf{q}} \in \{-Q,\ldots,Q\}^2} \beta_{ij}^{\bar{r}_1-\bar{q}_1\,\,\bar{r}_2-\bar{q}_2} {\tilde \eta}^{\bar{q}_1 \bar{q}_2} \nonumber \\
	&& + \omega^2 \sum_{\bar{\mathbf{q}} \in \{-Q,\ldots,Q\}^2} {\rho}^{\bar{r}_1-\bar{q}_1\,\,\bar{r}_2-\bar{q}_2} {\tilde u}_i^{\bar{q}_1 \bar{q}_2} = 0\,,
	\end{eqnarray}
	\begin{eqnarray}\label{eq:conv3var}
	&& - \left(\frac{2 \pi \bar{r}_i}{d_i}+k_i\right) \sum_{\bar{\mathbf{q}} \in \{-Q,\ldots,Q\}^2} \left(\frac{2 \pi \bar{q}_j}{d_j}+k_j\right) K_{ij}^{\bar{r}_1-\bar{q}_1\,\,\bar{r}_2-\bar{q}_2} {\tilde \vartheta}^{\bar{q}_1 \bar{q}_2} \nonumber \\
	&& + \omega \sum_{\bar{\mathbf{q}} \in \{-Q,\ldots,Q\}^2} \left(\frac{2 \pi \bar{q}_j}{d_j}+k_j\right) \alpha_{ij}^{\bar{r}_1-\bar{q}_1\,\,\bar{r}_2-\bar{q}_2} {\tilde u}_i^{\bar{q}_1 \bar{q}_2}\nonumber \\
	&& - I \omega \sum_{\bar{\mathbf{q}} \in \{-Q,\ldots,Q\}^2} \psi^{\bar{r}_1-\bar{q}_1\,\,\bar{r}_2-\bar{q}_2} {\tilde \eta}^{\bar{q}_1 \bar{q}_2} \nonumber \\
	&& - I \omega \sum_{\bar{\mathbf{q}} \in \{-Q,\ldots,Q\}^2} {p}^{\bar{r}_1-\bar{q}_1\,\,\bar{r}_2-\bar{q}_2} {\tilde \vartheta}^{\bar{q}_1 \bar{q}_2} = 0\,,
	\end{eqnarray}
	\begin{eqnarray}\label{eq:conv3var2}
	&& - \left(\frac{2 \pi \bar{r}_i}{d_i}+k_i\right)  \sum_{\bar{\mathbf{q}} \in \{-Q,\ldots,Q\}^2} \left(\frac{2 \pi \bar{q}_j}{d_j}+k_j\right) D_{ij}^{\bar{r}_1-\bar{q}_1\,\,\bar{r}_2-\bar{q}_2} {\tilde \eta}^{\bar{q}_1 \bar{q}_2} 	\nonumber \\
	&& + \omega \sum_{\bar{\mathbf{q}} \in \{-Q,\ldots,Q\}^2} \left(\frac{2 \pi \bar{q}_j}{d_j}+k_j\right) \beta_{ij}^{\bar{r}_1-\bar{q}_1\,\,\bar{r}_2-\bar{q}_2} {\tilde u}_i^{\bar{q}_1 \bar{q}_2}\nonumber \\
	&& - I \omega \sum_{\bar{\mathbf{q}} \in \{-Q,\ldots,Q\}^2} \psi^{\bar{r}_1-\bar{q}_1\,\,\bar{r}_2-\bar{q}_2} {\tilde \vartheta}^{\bar{q}_1 \bar{q}_2} \nonumber \\
	&& - I \omega \sum_{\bar{\mathbf{q}} \in \{-Q,\ldots,Q\}^2} {c}^{\bar{r}_1-\bar{q}_1\,\,\bar{r}_2-\bar{q}_2} {\tilde \eta}^{\bar{q}_1 \bar{q}_2} = 0\,.
  \end{eqnarray}
\noindent In (\ref{eq:conv3})-(\ref{eq:conv3var2}), 
the indices of the Fourier coefficients  ${\tilde u}^{\bar{q}_1 \bar{q}_2}, \tilde \vartheta^{\bar{q}_1 \bar{q}_2}, \tilde \eta^{\bar{q}_1 \bar{q}_2}$ result defined  as $\bar{\mathbf{q}}=(\bar{q}_1, \bar{q}_2) \in \{-Q,\ldots,Q\}^2$. Concerning the Fourier coefficients associated with the components of the  
constitutive tensors, instead, the related indices are  $\bar{\mathbf{r}}-\bar{\mathbf{q}}=(\bar{r}_1-\bar{q}_1, \bar{r}_2-\bar{q}_2) \in \{-2Q,\ldots,2Q\}^2$, since it also results that $\bar{\mathbf{r}}=(\bar{r}_1, \bar{r}_2) \in \{-Q,\ldots,Q\}^2$. Properly conceived convergence analyses  will be, thus, required as $Q$, i.e. the number of considered harmonics related to the dimension of the finite-dimensional algebraic system, increases. \\
It is worth noting that we are interested to the study of multi-phase materials, in which the components of the constitutive tensors are discontinuous. In this context, the undesired Gibbs phenomenon arises when an approximate solution is sought, i.e.
when the infinite weighted Dirac combs are replaced by finite weighted Dirac combs. This means that spurious  
oscillations appear in the inverse Fourier transforms of the finite weighted Dirac combs, in the neighbourhood  of the discontinuities between material phases.
A regularization technique is, thus, required in order to significantly reduce such oscillations, see \citep{Jerri1998}.  Moreover,
by adopting the regularization technique an
 additional benefit is found. It is, indeed, possible to account for a reduced number of harmonics $Q$ in order to practically fulfill the convergence.\\
By way of example, we consider a generic component $C_{ijhk}\left(\mathbf{x}\right)$ of the elastic tensor, and compute the inverse Fourier transform of the associated infinite weighted Dirac comb, as
\begin{equation}\label{Cijhk-1}
C_{ijhk}\left(\mathbf{x}\right)=\sum_{\bar{\mathbf{r}} \in \mathbf{Z}^2} \frac{C_{ijhk}^{\bar{r}_1 \bar{r}_2}}{4 \pi^2} e^{I \left(\bar{\mathbf{r}} \cdot \mathbf{x}\right)}\,.
\end{equation} 
\noindent When the infinite weighted Dirac comb is truncated, we obtain its finite-dimensional  approximation, denoted with the superscript $(f)$, as 
\begin{equation}
C_{ijhk}^{(f)}\left(\mathbf{x}\right)=\sum_{\bar{\mathbf{r}} \in \{-2 Q, \ldots, 2 Q\}^2} \frac{C_{ijhk}^{\bar{r}_1 \bar{r}_2}}{4 \pi^2} e^{I \left(\bar{\mathbf{r}} \cdot \mathbf{x}\right)}\,.\label{finite-dim}
\end{equation}
Finally, a regularization of the approximation above is obtained, by replacing each Fourier coefficient $C_{ijhk}^{\bar{r}_1\,\,\bar{r}_2}$ with
\begin{equation}
C_{ijhk}^{\bar{r}_1 \bar{r}_2 \,{ (reg)}}= C_{ijhk}^{\bar{r}_1 \bar{r}_2} e^{-\sigma(\bar{r}_1^2 + \bar{r}_2^2)}\,,
\end{equation}
where the regularization parameter $\sigma>0$ is introduced, involved in the decaying exponential factor
$e^{-\sigma(\bar{r}_1^2 + \bar{r}_2^2)}$.
After introducing the regularization, Equation (\ref{finite-dim}) becomes
\begin{equation}\label{Cijhk-4}
C_{ijhk}^{ (f,\,reg)}\left(\mathbf{x}\right)=\sum_{\bar{\mathbf{r}} \in \{-2 Q, \ldots, 2 Q\}^2} \frac{C_{ijhk}^{\bar{r}_1 \bar{r}_2 \,(reg)}}{4 \pi^2} e^{I \left(\bar{\mathbf{r}} \cdot \mathbf{x}\right)}.
\end{equation}
In order to achieve a good trade-off between the reduction of the oscillations near the discontinuities, and the accuracy of the approximation in the periodic cell $\textbf{A}$, the value of the parameter 
 $\sigma$ is chosen depending on the considered number of terms in the truncated Fourier expansion.\\
The effectiveness of such a regularization is demonstrated in Figure \ref{fig:regularization_new} where the component  $C_{1111}$, related to a specific  periodic cell with geometry in Figure \ref{fig:figura1SOFC}(b), is shown. In Figure \ref{fig:regularization_new}(a) the dimensionless component $C_{1111}^{(f)}/C^{(ref)}$, while in Figure \ref{fig:regularization_new}(b) the dimensionless component $C_{1111}^{(f,\,reg)}/C^{(ref)}$ (where $C^{(ref)}$ is a reference value) are reported versus the dimensionless coordinates $x_1/ d_1$, $x_2/ d_2$.
The same procedure (\ref{Cijhk-1})-(\ref{Cijhk-4}) analogously apply to $K_{ij}$, $D_{ij}$, $\alpha_{ij}$, $\beta_{ij}$, $\Psi$, $\rho$, $p$, $c$. \\
It follows that the infinite-dimensional operators $\textbf{\textit{A}}$, $\textbf{\textit{B}}$, $\textbf{\textit{C}}$ are replaced by their finite-dimensional regularized counterparts $\textbf{\textit{A}}^{(f,\,\,reg)}$,
  $\textbf{\textit{B}}^{(f,\,\,reg)}$, $\textbf{\textit{C}}^{(f,\,\,reg)}$, and the  generalized quadratic eigenvalue problem (\ref{eq:generalized}) becomes
\begin{equation}\label{eq:generalized3}
\left( {{\omega ^2} \textbf{\textit{A}}^{(f,\,\,reg)}} + \omega \textbf{\textit{B}}^{(f,\,\,reg)} + \textbf{\textit{C}}^{(f,\,\,reg)} \right)\mathbf{z}^{(f,\,\,reg)} = {\mathbf{0}}^{(f)}\,.
\end{equation}
where
 the vector $\mathbf{z}^{(f,\,\,reg)}$ collects the Fourier coefficients of the finite dimensional Bloch amplitudes of ${\mathbf{\tilde u}_1}^{(f,\,\,reg)}$, ${\mathbf{\tilde u}_2}^{(f,\,\,reg)}$, $\boldsymbol{\tilde \vartheta}^{(f,\,\,reg)}$, $\boldsymbol{\tilde \eta}^{(f,\,\,reg)}$,  for the details see Appendix B.\\
The solution $\omega$ of the generalized eigenvualue problem (\ref{eq:generalized3}) is an approximation, including a finite number of eigenvalues, of 
the corresponding generalized eigenvalue  solution of (\ref{eq:generalized}). Note that the dimension of $\mathbf{z}^{(f, \, \, reg)}$ is $4(2Q +1)^2$, since $(2Q+1)^2$ Fourier coefficients are considered for each of the four unknown Bloch amplitudes  ${\mathbf{\tilde u}}\left({\mathbf{x}}\right)$, ${\tilde \vartheta}\left({\mathbf{x}}\right)$, and ${\tilde \eta}\left({\mathbf{x}}\right)$, having ${\mathbf{\tilde u}}\left({\mathbf{x}}\right)$ two components.
Consistently to what done in Equation (\ref{eq:generalized2}), the  linear generalized eigenvalue problem equivalent to (\ref{eq:generalized3}) is
\begin{equation}\label{eq:generalized4}
(\omega \textbf{\textit{A}}'^{(f,\,\,reg)}+\textbf{\textit{B}}'^{(f,\,\,reg)}) \, \mathbf{z}'^{(f,\, reg)}=\mathbf{0}'^{(f)}\,,
\end{equation}
where, in this case, the operators $ \textbf{\textit{A}}'^{(f,\,\,reg)}$ and $ \textbf{\textit{B}}'^{(f,\,\,reg)}$ are applied to the generalized 
 eigenvector $\mathbf{z}'^{(f,\, reg)}$ collecting the vectors  $\omega \mathbf{z}^{(f,\, reg)}$ and $\mathbf{z}^{(f,\, reg)}$ appearing in 
 (\ref{eq:generalized3}), see Appendix B.\\ 
  Finally, the generalized eigenvalues $\omega$, i.e. the complex-frequencies, associated with the problem (\ref{eq:generalized4}) are obtained as the roots of the characteristic equation
\begin{equation}\label{eq:characteristic}
{\rm det}\, (\omega  \textbf{\textit{A}}'^{(f,\,\,reg)}+\textbf{\textit{B}}'^{(f,\,\,reg)})=0\,.
\end{equation}
Such a characteristic equation, thus, provides the complex-frequency  band structure of the thermo-diffusive periodic material. \\
As detailed in Appendix B,  
$\textit{\textbf{A}}^{(f,\,\,reg)}$, $\textit{\textbf{B}}^{(f,\,\,reg)}$, and $\textit{\textbf{C}}^{(f,\,\,reg)}$, together with  
 $\textbf{\textit{A}}'^{(f,\,\,reg)}$ and $\textbf{\textit{B}}'^{(f,\,\,reg)}$, that are constructed starting from the former group, are complex operators.
Moreover, the operator $\textit{\textbf{B}}^{(f,\,\,reg)}$ and $\textit{\textbf{C}}^{(f,\,\,reg)}$ depend on the wave vector $\mathbf{k}$ linearly and quadratically, respectively, while $\textit{\textbf{A}}^{(f,\,\,reg)}$ is constant with respect to  $\mathbf{k}$.
 This property can be exploited to speed up their construction as $\mathbf{k}$ varies. Due to this dependence on $\mathbf{k}$, also the generalized eigenvalues $\omega$ depend on $\mathbf{k}$.

\subsection{Specialization of the generalized eigenvalue problem to the case of orthotropic or isotropic material phases }\label{sec:details}
In the case of periodic thermo-diffusive materials, characterized by orthotropic phases, in which the orthotropic directions are parallel to the reference system, Equations (\ref{eq:conv3})-(\ref{eq:conv3var2}) strongly simplify and take the following form, with all the indices made explicit 
\begin{eqnarray}\label{eq:pattaman1}
  && - \left(\frac{2 \pi \bar{r}_1}{d_1}+k_1\right) \sum_{\bar{\mathbf{q}} \in \{-Q,\ldots,Q\}^2} \left(\frac{2 \pi \bar{q}_1}{d_1} +k_1 \right) C_{1111}^{\bar{r}_1-\bar{q}_1\,\,\bar{r}_2-\bar{q}_2\,\,(reg)} {\tilde u}_1^{\bar{q}_1 \bar{q}_2\,\,(reg)}\nonumber \\
	&& - \left(\frac{2 \pi \bar{r}_1}{d_1}+k_1\right) \sum_{\bar{\mathbf{q}} \in \{-Q,\ldots,Q\}^2} \left(\frac{2 \pi \bar{q}_2}{d_2} +k_2 \right) C_{1122}^{\bar{r}_1-\bar{q}_1\,\,\bar{r}_2-\bar{q}_2\,\,(reg)} {\tilde u}_2^{\bar{q}_1 \bar{q}_2\,\,(reg)}\nonumber \\
	&& - \left(\frac{2 \pi \bar{r}_2}{d_2}+k_2\right) \sum_{\bar{\mathbf{q}} \in \{-Q,\ldots,Q\}^2} \left(\frac{2 \pi \bar{q}_1}{d_1} +k_1 \right) C_{1212}^{\bar{r}_1-\bar{q}_1\,\,\bar{r}_2-\bar{q}_2\,\,(reg)} {\tilde u}_2^{\bar{q}_1 \bar{q}_2\,\,(reg)}\nonumber \\
	&& - \left(\frac{2 \pi \bar{r}_2}{d_2}+k_2\right) \sum_{\bar{\mathbf{q}} \in \{-Q,\ldots,Q\}^2} \left(\frac{2 \pi \bar{q}_2}{d_2} +k_2 \right) C_{1212}^{\bar{r}_1-\bar{q}_1\,\,\bar{r}_2-\bar{q}_2\,\,(reg)} {\tilde u}_1^{\bar{q}_1 \bar{q}_2\,\,(reg)}\nonumber \\
	&& - I \left(\frac{2 \pi \bar{r}_1}{d_1} +k_1 \right) \sum_{\bar{\mathbf{q}} \in \{-Q,\ldots,Q\}^2} \alpha_{11}^{\bar{r}_1-\bar{q}_1\,\,\bar{r}_2-\bar{q}_2 \,\,(reg)} {\tilde \vartheta}^{\bar{q}_1 \bar{q}_2\,\,(reg)}\nonumber \\
	&& - I \left(\frac{2 \pi \bar{r}_1}{d_1} +k_1 \right) \sum_{\bar{\mathbf{q}} \in \{-Q,\ldots,Q\}^2} \beta_{11}^{\bar{r}_1-\bar{q}_1\,\,\bar{r}_2-\bar{q}_2\,\,(reg)} {\tilde \eta}^{\bar{q}_1 \bar{q}_2\,\,(reg)} \nonumber \\
	&& + \omega^2 \sum_{\bar{\mathbf{q}} \in \{-Q,\ldots,Q\}^2} {\rho}^{\bar{r}_1-\bar{q}_1\,\,\bar{r}_2-\bar{q}_2\,\,(reg)} {\tilde u}_1^{\bar{q}_1 \bar{q}_2\,\,(reg)} = 0\,,
	\end{eqnarray}

\begin{eqnarray}\label{eq:pattaman2}
  && - \left(\frac{2 \pi \bar{r}_1}{d_1}+k_1\right) \sum_{\bar{\mathbf{q}} \in \{-Q,\ldots,Q\}^2} \left(\frac{2 \pi \bar{q}_2}{d_2} +k_2 \right) C_{1212}^{\bar{r}_1-\bar{q}_1\,\,\bar{r}_2-\bar{q}_2\,\,(reg)} {\tilde u}_1^{\bar{q}_1 \bar{q}_2\,\,(reg)}\nonumber \\
	&& - \left(\frac{2 \pi \bar{r}_1}{d_1}+k_1\right) \sum_{\bar{\mathbf{q}} \in \{-Q,\ldots,Q\}^2} \left(\frac{2 \pi \bar{q}_1}{d_1} +k_1 \right) C_{1212}^{\bar{r}_1-\bar{q}_1\,\,\bar{r}_2-\bar{q}_2\,\,(reg)} {\tilde u}_2^{\bar{q}_1 \bar{q}_2\,\,(reg)}\nonumber \\
	&& - \left(\frac{2 \pi \bar{r}_2}{d_2}+k_2\right) \sum_{\bar{\mathbf{q}} \in \{-Q,\ldots,Q\}^2} \left(\frac{2 \pi \bar{q}_1}{d_1} +k_1 \right) C_{1122}^{\bar{r}_1-\bar{q}_1\,\,\bar{r}_2-\bar{q}_2\,\,(reg)} {\tilde u}_1^{\bar{q}_1 \bar{q}_2\,\,(reg)}\nonumber \\
	&& - \left(\frac{2 \pi \bar{r}_2}{d_2}+k_2\right) \sum_{\bar{\mathbf{q}} \in \{-Q,\ldots,Q\}^2} \left(\frac{2 \pi \bar{q}_2}{d_2} +k_2 \right) C_{2222}^{\bar{r}_1-\bar{q}_1\,\,\bar{r}_2-\bar{q}_2\,\,(reg)} {\tilde u}_2^{\bar{q}_1 \bar{q}_2\,\,(reg)}\nonumber \\
	&& - I \left(\frac{2 \pi \bar{r}_2}{d_2} +k_2 \right) \sum_{\bar{\mathbf{q}} \in \{-Q,\ldots,Q\}^2} \alpha_{22}^{\bar{r}_1-\bar{q}_1\,\,\bar{r}_2-\bar{q}_2 \,\,(reg)} {\tilde \vartheta}^{\bar{q}_1 \bar{q}_2\,\,(reg)}\nonumber \\
	&& - I \left(\frac{2 \pi \bar{r}_2}{d_2} +k_2 \right) \sum_{\bar{\mathbf{q}} \in \{-Q,\ldots,Q\}^2} \beta_{22}^{\bar{r}_1-\bar{q}_1\,\,\bar{r}_2-\bar{q}_2\,\,(reg)} {\tilde \eta}^{\bar{q}_1 \bar{q}_2\,\,(reg)} \nonumber \\
	&& + \omega^2 \sum_{\bar{\mathbf{q}} \in \{-Q,\ldots,Q\}^2} {\rho}^{\bar{r}_1-\bar{q}_1\,\,\bar{r}_2-\bar{q}_2\,\,(reg)} {\tilde u}_2^{\bar{q}_1 \bar{q}_2\,\,(reg)} = 0\,,
	\end{eqnarray}

	\begin{eqnarray}\label{eq:pattaman3}
	&& - \left(\frac{2 \pi \bar{r}_1}{d_1}+k_1\right) \sum_{\bar{\mathbf{q}} \in \{-Q,\ldots,Q\}^2} \left(\frac{2 \pi \bar{q}_1}{d_1}+k_1\right) K_{11}^{\bar{r}_1-\bar{q}_1\,\,\bar{r}_2-\bar{q}_2\,\,(reg)} {\tilde \vartheta}^{\bar{q}_1 \bar{q}_2\,\,(reg)} \nonumber \\
	&& - \left(\frac{2 \pi \bar{r}_2}{d_2}+k_2\right) \sum_{\bar{\mathbf{q}} \in \{-Q,\ldots,Q\}^2} \left(\frac{2 \pi \bar{q}_2}{d_2}+k_2\right) K_{22}^{\bar{r}_1-\bar{q}_1\,\,\bar{r}_2-\bar{q}_2\,\,(reg)} {\tilde \vartheta}^{\bar{q}_1 \bar{q}_2\,\,(reg)} \nonumber \\
	&& + \omega \sum_{\bar{\mathbf{q}} \in \{-Q,\ldots,Q\}^2} \left(\frac{2 \pi \bar{q}_1}{d_1}+k_1\right) \alpha_{11}^{\bar{r}_1-\bar{q}_1\,\,\bar{r}_2-\bar{q}_2\,\,(reg)} {\tilde u}_1^{\bar{q}_1 \bar{q}_2\,\,(reg)}\nonumber \\
	&& + \omega \sum_{\bar{\mathbf{q}} \in \{-Q,\ldots,Q\}^2} \left(\frac{2 \pi \bar{q}_j}{d_2}+k_2\right) \alpha_{22}^{\bar{r}_1-\bar{q}_1\,\,\bar{r}_2-\bar{q}_2\,\,(reg)} {\tilde u}_2^{\bar{q}_1 \bar{q}_2\,\,(reg)}\nonumber \\
	&& - I \omega \sum_{\bar{\mathbf{q}} \in \{-Q,\ldots,Q\}^2} \psi^{\bar{r}_1-\bar{q}_1\,\,\bar{r}_2-\bar{q}_2\,\,(reg)} {\tilde \eta}^{\bar{q}_1 \bar{q}_2\,\,(reg)} \nonumber \\
	&& - I \omega \sum_{\bar{\mathbf{q}} \in \{-Q,\ldots,Q\}^2} {p}^{\bar{r}_1-\bar{q}_1\,\,\bar{r}_2-\bar{q}_2\,\,(reg)} {\tilde \vartheta}^{\bar{q}_1 \bar{q}_2\,\,(reg)} = 0\,,
	\end{eqnarray}
	\begin{eqnarray}\label{eq:pattaman4}
	&& - \left(\frac{2 \pi \bar{r}_1}{d_1}+k_1\right)  \sum_{\bar{\mathbf{q}} \in \{-Q,\ldots,Q\}^2} \left(\frac{2 \pi \bar{q}_1}{d_1}+k_1\right) D_{11}^{\bar{r}_1-\bar{q}_1\,\,\bar{r}_2-\bar{q}_2\,\,(reg)} {\tilde \eta}^{\bar{q}_1 \bar{q}_2\,\,(reg)} 	\nonumber \\
	&& - \left(\frac{2 \pi \bar{r}_2}{d_2}+k_2\right)  \sum_{\bar{\mathbf{q}} \in \{-Q,\ldots,Q\}^2} \left(\frac{2 \pi \bar{q}_2}{d_2}+k_2\right) D_{22}^{\bar{r}_1-\bar{q}_1\,\,\bar{r}_2-\bar{q}_2\,\,(reg)} {\tilde \eta}^{\bar{q}_1 \bar{q}_2\,\,(reg)} 	\nonumber \\
	&& + \omega \sum_{\bar{\mathbf{q}} \in \{-Q,\ldots,Q\}^2} \left(\frac{2 \pi \bar{q}_1}{d_1}+k_1\right) \beta_{11}^{\bar{r}_1-\bar{q}_1\,\,\bar{r}_2-\bar{q}_2\,\,(reg)} {\tilde u}_1^{\bar{q}_1 \bar{q}_2\,\,(reg)}\nonumber \\
	&& + \omega \sum_{\bar{\mathbf{q}} \in \{-Q,\ldots,Q\}^2} \left(\frac{2 \pi \bar{q}_2}{d_2}+k_2\right) \beta_{22}^{\bar{r}_1-\bar{q}_1\,\,\bar{r}_2-\bar{q}_2\,\,(reg)} {\tilde u}_2^{\bar{q}_1 \bar{q}_2\,\,(reg)}\nonumber \\
	&& - I \omega \sum_{\bar{\mathbf{q}} \in \{-Q,\ldots,Q\}^2} \psi^{\bar{r}_1-\bar{q}_1\,\,\bar{r}_2-\bar{q}_2\,\,(reg)} {\tilde \vartheta}^{\bar{q}_1 \bar{q}_2\,\,(reg)} \nonumber \\
	&& - I \omega \sum_{\bar{\mathbf{q}} \in \{-Q,\ldots,Q\}^2} {c}^{\bar{r}_1-\bar{q}_1\,\,\bar{r}_2-\bar{q}_2\,\,(reg)} {\tilde \eta}^{\bar{q}_1 \bar{q}_2\,\,(reg)} = 0\,.
  \end{eqnarray}
It stands to reason that the operators  $\textbf{\textit{A}}^{(f,\,\,reg)}$,
  $\textbf{\textit{B}}^{(f,\,\,reg)}$, $\textbf{\textit{C}}^{(f,\,\,reg)}$, and $\textbf{\textit{A}}'^{(f,\,\,reg)}$,
  $\textbf{\textit{B}}'^{(f,\,\,reg)}$, explicitly defined in Appendix B,
are simplified accordingly.\\
Finally, in the case of isotropic phases the non-vanishing components of the constitutive and coupling tensors become $C_{1111}=C_{2222}=\widetilde{E}/(1-\widetilde{\nu}^2)$, 
$C_{1122}=\widetilde{\nu} \widetilde{E}/(1-\widetilde{\nu}^2)$, $C_{1212}= \widetilde{E}/2(1+\widetilde{\nu})$, $K_{11}=K_{22}=K$, $D_{11}=D_{22}=D$, $\alpha_{11}=\alpha_{22}=\widetilde{\alpha}(1-2\widetilde{\nu})/ (1-\widetilde{\nu})$, $\beta_{11}=\beta_{22}=\widetilde{\beta}(1-2\widetilde{\nu})/ (1-\widetilde{\nu})$. In particular, in the case of plane state characterized by  $sym(\nabla \textbf{u})\textbf{e}_3=\textbf{0}$, $\nabla \vartheta \cdot \textbf{e}_3=0  $, $\nabla \eta \cdot \textbf{e}_3=0  $, with $\textbf{e}_3 = \textbf{e}_1 \times \textbf{e}_2$ being the out-of-plane unit vector, the constants are $\widetilde{E}=E/(1-\nu^2)$, $\widetilde{\nu}=\nu/(1-\nu)$, $\widetilde{\alpha}=(1-2 \nu)/(1-3 \nu)\alpha$, $\widetilde{\beta}=(1-2 \nu)/(1-3 \nu)\beta$, while, in the case with $\boldsymbol{\sigma} \textbf{e}_3=\textbf{0}$, $\textbf{q} \cdot \textbf{e}_3=0$, $\textbf{j} \cdot \textbf{e}_3=0$, the constants are $\widetilde{E}=E$, $\widetilde{\nu}=\nu$, $\widetilde{\alpha}=\alpha$, $\widetilde{\beta}=\beta$. In all the previous definitions $E$ is the Young modulus, $\nu$ is the Poisson's ratio, $\alpha$ is the thermal dilatation constant, and $\beta$ is the diffusive expansion constant.

\section{Damped wave propagation in SOFC-like devices}\label{sec:numerical_results}
The procedure discussed in Sections \ref{Damped} and \ref{sec:4}, proposed to determine the band structure of thermo-diffusive heterogeneous materials with periodic microstructure, is here specialized to the case of a thermo-elastic  periodic material. The focus is, thus, on the thermo-mechanical coupling phenomena, and the ion diffusion is neglected for the sake of simplicity and without loss of generality.\\
\begin{figure}[ht]
\centering
\includegraphics[scale=0.87,trim={0cm 16cm 3cm 4.1cm}]{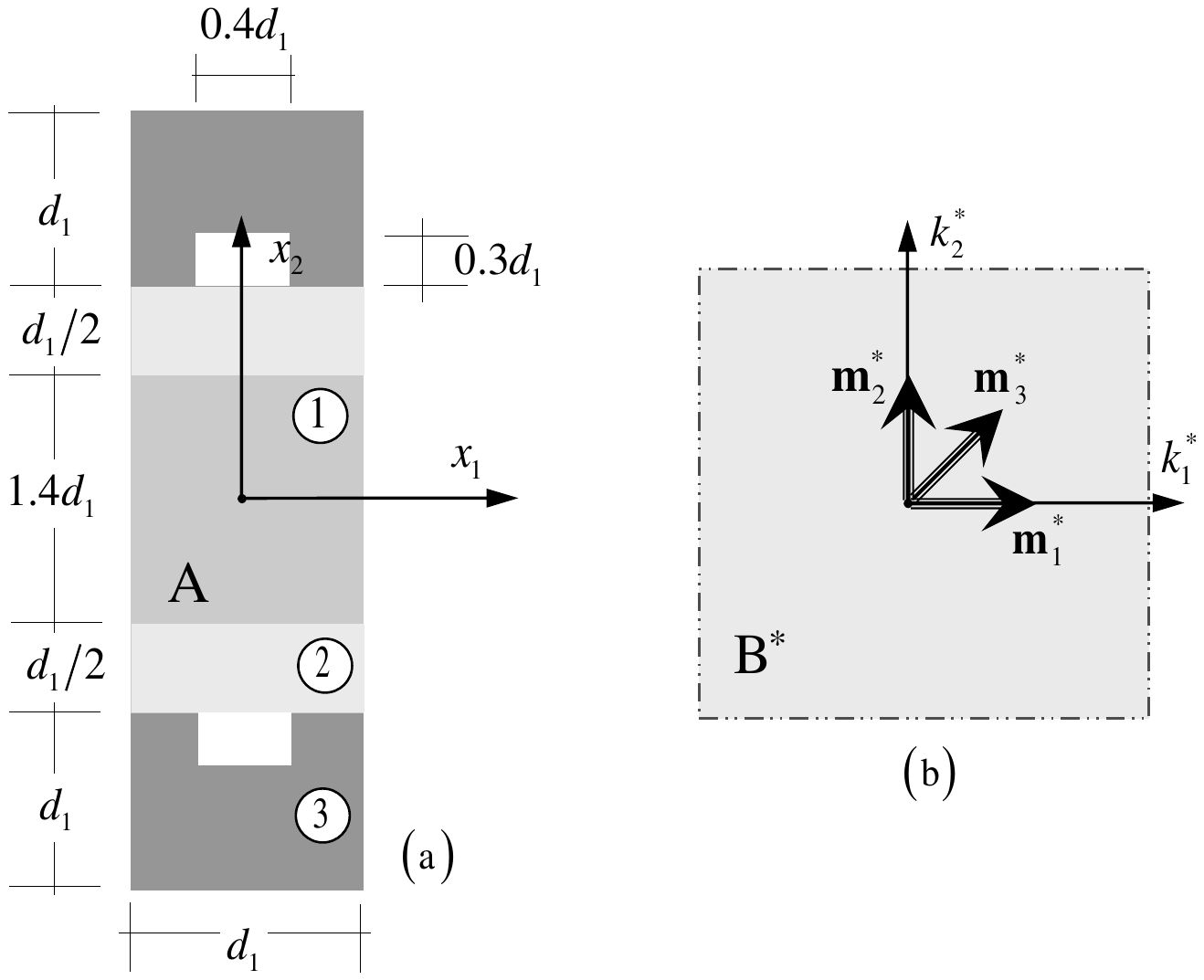}
\caption{(a) Schematic of the SOFC-like periodic cell adopted in the numerical analyses; (b) Dimensionless first Brillouin zone
 and unit vectors of propagation $\textbf{m}^*_i$ (i=1,2,3).}\label{fig:figura2SOFC}
\end{figure}
In this context, we consider a periodic multi-phase laminate, generated by the spatial repetition of solid oxide fuel cells (SOFC-like material). In Figure \ref{fig:figura2SOFC} a generic cell is reported, characterized by dimensions $d_1= 100 \, \mu m$, and $d_2= 440 \,\mu m$.  We assume all the constituents being linear isotropic, perfectly bonded and in plane state characterized by $sym(\nabla \textbf{u})\textbf{e}_3=\textbf{0}$, $\nabla \vartheta \cdot \textbf{e}_3=0  $, $\nabla \eta \cdot \textbf{e}_3=0  $. The phase 1 identifies the ceramic electrolyte of the SOFC cell and it 
 is made of Yttria-stabilized zirconia (YSZ) , with Young modulus $E=155\cdot 10^9 N/m^2$, Poisson ratio $\nu=0.3$, the heat conduction constant $K=0.009 \, W/(m K^2)$, the thermal dilatation constant $\alpha=4223.75 \cdot 10^{3} \, N/(m^2K)$, inertial terms  $\rho=5.532 \cdot 10^3 kg/m^3$, and $p=7548.35 \, N/(m^2 K^2)$. The phase 2, instead, represents both cathode and anode electrodes, and it is made of Nichel oxide (NiO), characterized by $E=50\cdot 10^9 \, N/m^2$, Poisson ratio $\nu=0.25$, the heat conduction constant $K=0.034 \, W/(m K^2)$, the thermal dilatation constant $\alpha=1250 \cdot 10^{3} \, N/(m^2K)$, inertial terms  $\rho=6.67 \cdot 10^3 \, kg/m^3$, and $p=10011.26 \, N/(m^2 K^2)$. Finally, the phase 3 mimics conductive interconnections, able to connect adjacent SOFC cells, is made of steel with $E=2.01\cdot 10^{11} \, N/m^2$, Poisson ratio $\nu=0.3$, the heat conduction constant $K=0.1228 \, W/(m K^2)$, the thermal dilatation constant $\alpha=6030 \cdot 10^{3} \, N/(m^2K)$, inertial terms  $\rho=7.86 \cdot 10^3 \, kg/m^3$, and $p=13459.7 \, N/(m^2 K^2)$.\\
A set of numerical applications is herein reported, aimed at investigating the complex band structure of the periodic material, as the direction of wave propagation varies. In particular, we analyse both the cases with and without thermo-mechanical coupling (the thermal dilatation tensor $\boldsymbol{\alpha}$ is either different or equal to zero, respectively), in order to grasp the role of the coupling phenomena on the overall material behaviour.
We consider three unit vectors of propagation in the first Brillouin zone $\textbf{m}_i \in \textbf{B}$, the first  
$\textbf{m}_1=\textbf{e}_1$ propagating parallel to the material layers, the second $\textbf{m}_2= \textbf{e}_2$ propagating orthogonal to the material layers, while the third $\textbf{m}_3= d_2/\sqrt{d_1^2+d_2^2}\,\textbf{e}_1+ d_1/\sqrt{d_1^2+d_2^2}\,\textbf{e}_2$. For the sake of convenience, let us define the dimensionless wave vector $\textbf{k}^*=k_1^* \textbf{e}_1+k_2^* \textbf{e}_2 \in \textbf{B}^*$, being $k_1^*=k_1 d_1$ and  $k_2^*=k_2 d_2$  the dimensionless wave numbers, and the dimensionless first Brillouin zone  $\textbf{B}^*=[-\pi,\pi]\times[-\pi,\pi]$. It is, consequently, possible to define the unit vector of propagation $\textbf{m}^* = \textbf{k}^* / ||\textbf{k}^* ||$. The three considered unit vectors of propagation, thus, become $\textbf{m}_1^*= \textbf{e}_1$, $\textbf{m}_2^*= \textbf{e}_2$ and $\textbf{m}_3^*= \sqrt{2}/2\textbf{e}_1+\sqrt{2}/2\textbf{e}_2$.\\
It stands to reason that, under the working hypothesis of thermo-mechanical coupling,  the algebraic linear system generated by Equations (\ref{eq:conv3})-(\ref{eq:conv3var2}) assumes a reduced form, in which only the equations of the form (\ref{eq:conv3}) and (\ref{eq:conv3var}) are taken into account. Moreover, it results that the  terms containing $\beta_{ij}^{\bar{r}_1-\bar{q}_1\,\,\bar{r}_2-\bar{q}_2}$, $\psi^{\bar{r}_1-\bar{q}_1\,\,\bar{r}_2-\bar{q}_2}$, and ${\tilde \eta}^{\bar{q}_1 \bar{q}_2}$ are neglected, and the unknowns are reduced to ${\tilde u}_1^{\bar{q}_1 \bar{q}_2}$, ${\tilde u}_2^{\bar{q}_1 \bar{q}_2}$, and ${\tilde \vartheta}^{\bar{q}_1 \bar{q}_2}$. As a consequence, the equations in compact form (\ref{eq:generalized3}) and (\ref{eq:generalized4}) still apply, provided that reduced dimensions of the vectors $\mathbf{z}^{(f,\,\,reg)}$, $\mathbf{z}'^{(f,\, reg)}$ 
and 
of the operators $\textbf{\textit{A}}^{(f,\,\,reg)}$, $\textbf{\textit{B}}^{(f,\,\,reg)}$, $\textbf{\textit{C}}^{(f,\,\,reg)}$, $\textbf{\textit{A}}'^{(f,\,\,reg)}$, $\textbf{\textit{B}}'^{(f,\,\,reg)}$ are considered. Note that the numerical simulations have been repeated for different values of $Q$,
in order to test the convergence of the results as the number of considered harmonics increases.   \\
The equivalent linear generalized eigenvalue problem in Equation (\ref{eq:generalized4}) has been solved using  MATLAB$\textsuperscript{\textregistered}$ together with the Advanpix-Multiprecision Computing Toolbox, that enables the use of higher precision with respect to the standard double precision, up to quad precision. 
Such advanced computational tool is here required because, 
in the specific settings investigated in the numerical simulations, 
 typically the 
 components of the matrices associated with the operators $\textbf{\textit{A}}'^{(f,\,\,reg)}$ and $\textbf{\textit{B}}'^{(f,\,\,reg)}$, related either to the mechanical, to the thermal or to the coupling parts, are characterized by different orders of magnitude between each other. Moreover, in order to compute the generalized eigenvalues, since the matrices associated with the operators $\textbf{\textit{A}}'^{(f,\,\,reg)}$ and $\textbf{\textit{B}}'^{(f,\,\,reg)}$ are in general neither symmetric, nor Hermitian, their preliminary generalized Schur decomposition  has been exploited \citep{GolvanLoa1996}. In addition, in order to reduce as much as possible the computational noise and improve the accuracy, a  preconditioning of such matrices has been performed, by exploiting the algorithm developed in \citep{Ward1981}. More precisely, this preconditioning replaces the operators $\textbf{\textit{A}}'^{(f,\,\,reg)}$ and $\textbf{\textit{B}}'^{(f,\,\,reg)}$, respectively, with $\textbf{\textit{T}}_1 \textbf{\textit{A}}'^{(f,\,\,reg)} \textbf{\textit{T}}_2$ and $\textbf{\textit{T}}_1 \textbf{\textit{B}}'^{(f,\,\,reg)} \textbf{\textit{T}}_2$, where $\textbf{\textit{T}}_1$ and $\textbf{\textit{T}}_2$ are suitable invertible operators, determined in accordance with \citep{Ward1981}. Finally, for the optimization of the numerical code, a sparse representation of the aforementioned matrices has been adopted, together with their suitable reorganization according to a block structure.

 \subsection{Numerical examples}
\begin{figure}[t!]
\centering
\subfigure[]
{\includegraphics[scale=0.6,trim={4cm 9.5cm 2.5cm 9.5cm}]{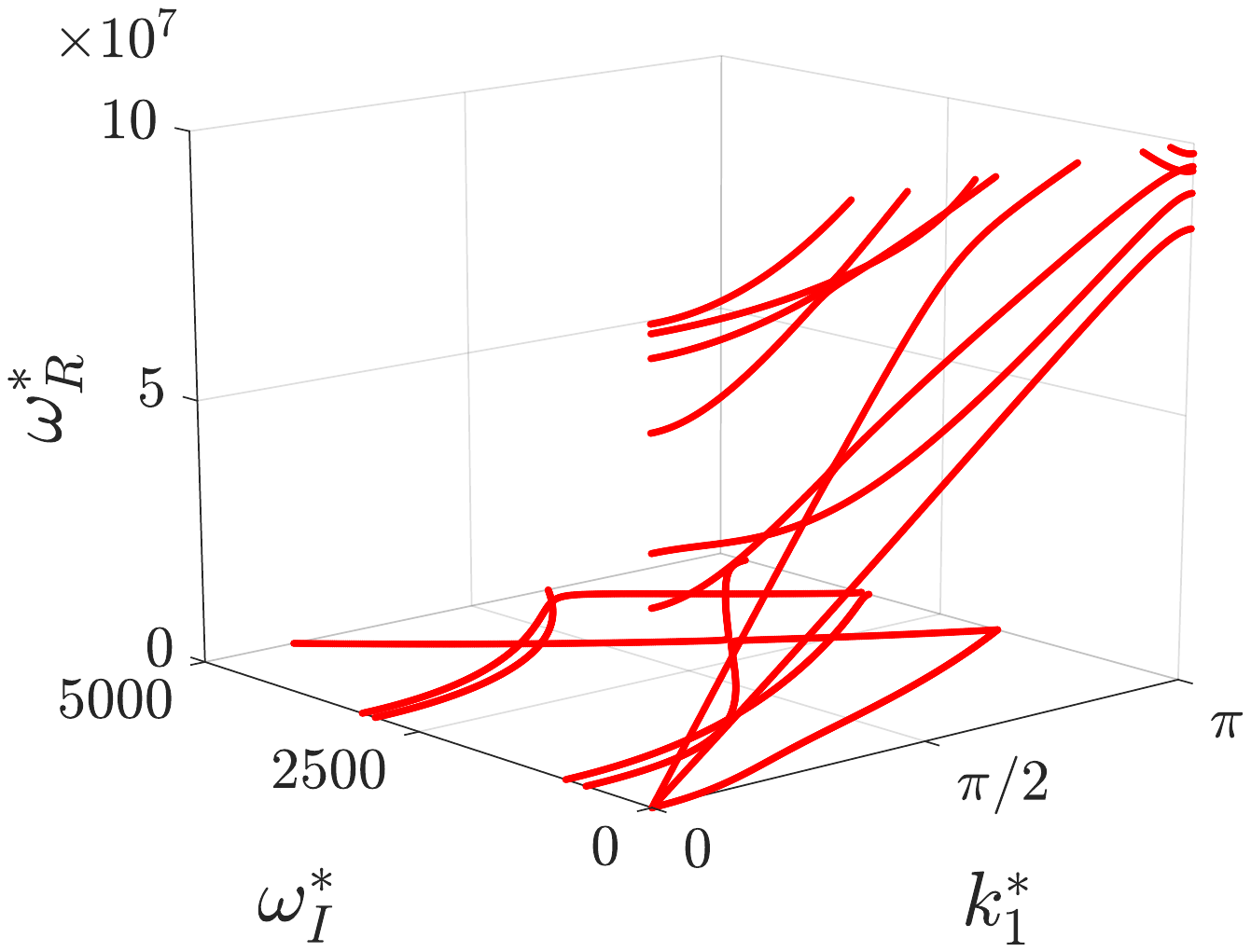}}
\subfigure[]
{\includegraphics[scale=0.6,trim={4cm 9.5cm 2.5cm 9.5cm}]{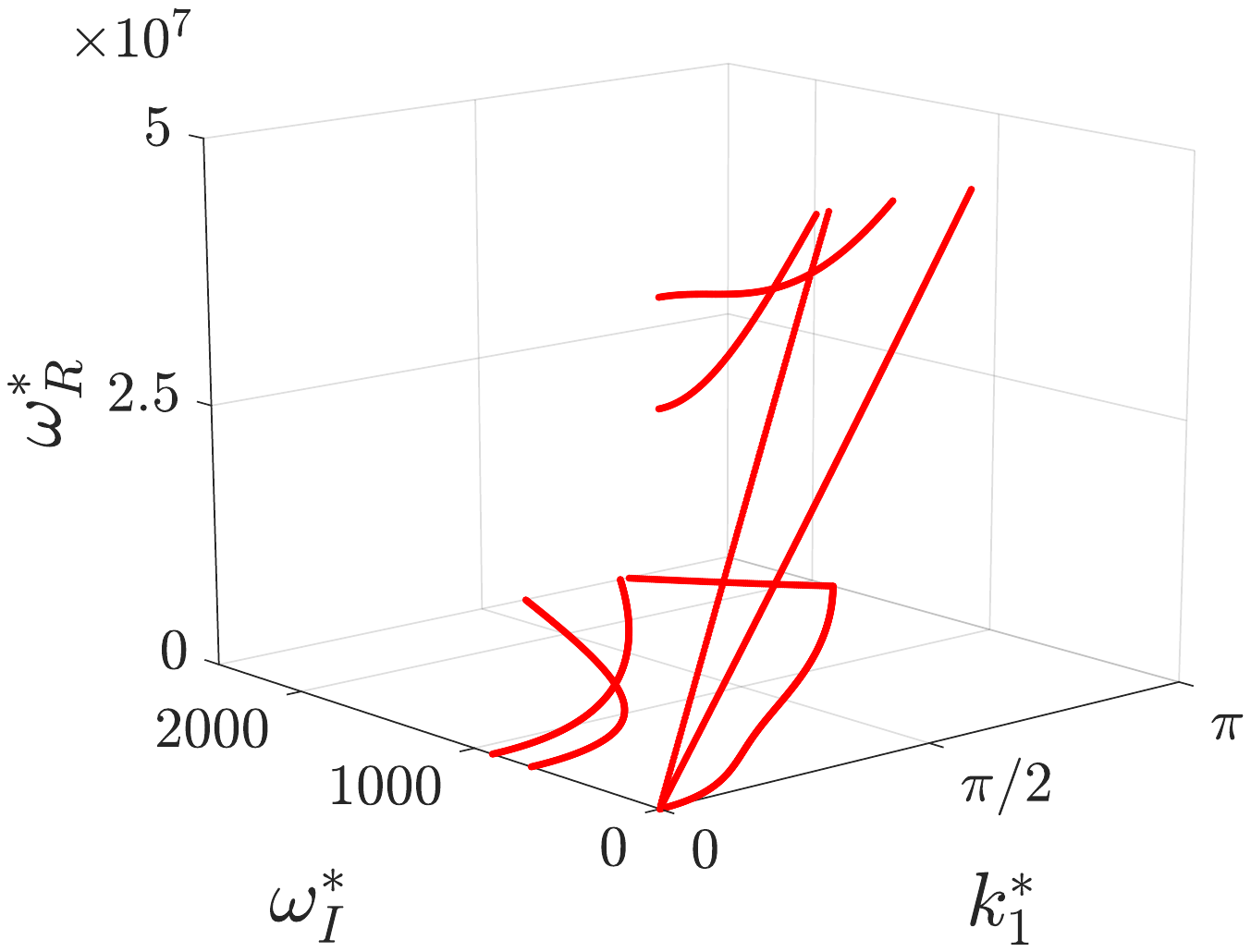}}\\
\subfigure[]
{\includegraphics[scale=0.6,trim={4cm 9.5cm 2.5cm 9.5cm}]{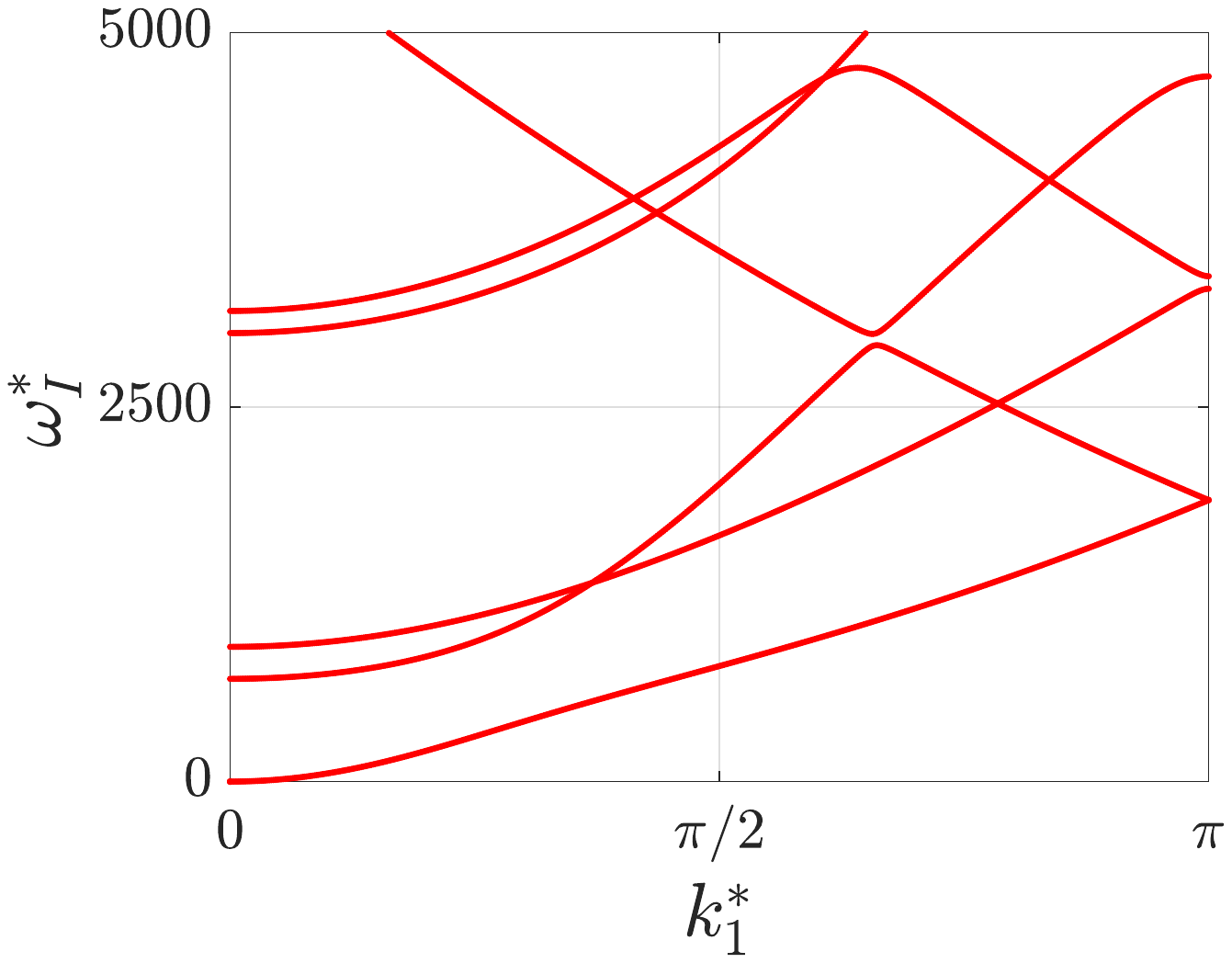}}
\subfigure[]
{\includegraphics[scale=0.6,trim={4cm 9.5cm 2.5cm 9.5cm}]{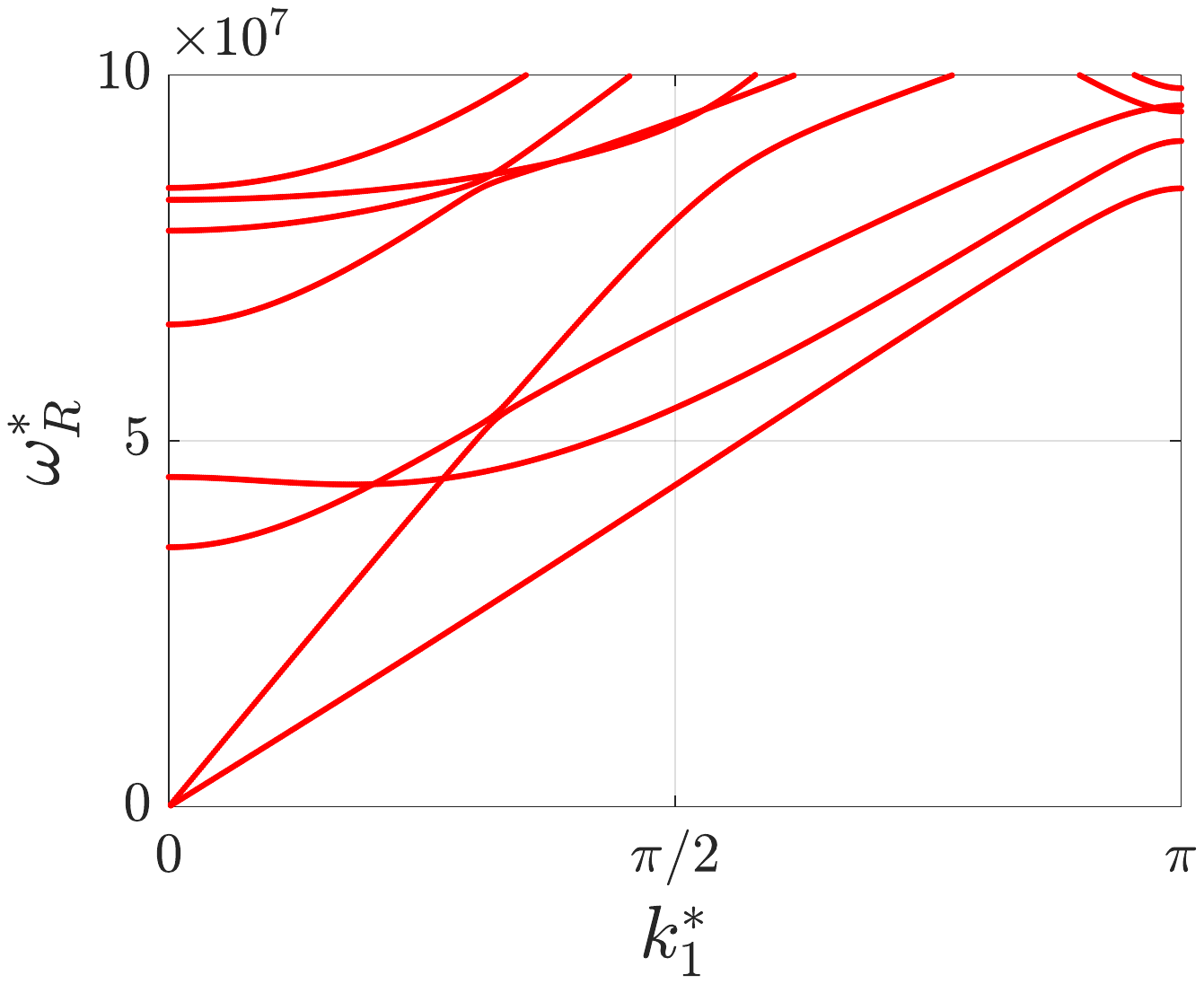}}
\caption{ Complex Floquet-Bloch spectrum of the thermo-elastic SOFC-like material related to waves with unit vector of propagation $\textbf{m}_1^*$. (a) Complex frequencies $\omega$=$\omega_R^*$+$I$ $\omega_I^*$ versus $k_1^*$; (b) zoomed view of the lowest frequency branches of Figure \ref{fig:k1}(a); (c) view of the dispersive curves associated with temporal damping modes; (d) projection view on the plane $\omega_R^*$-$k_1^*$.}\label{fig:k1}
\end{figure}

\begin{figure}[!t]
\centering
\subfigure[]
{\includegraphics[scale=0.6,trim={4cm 9.5cm 2.5cm 9.5cm}]{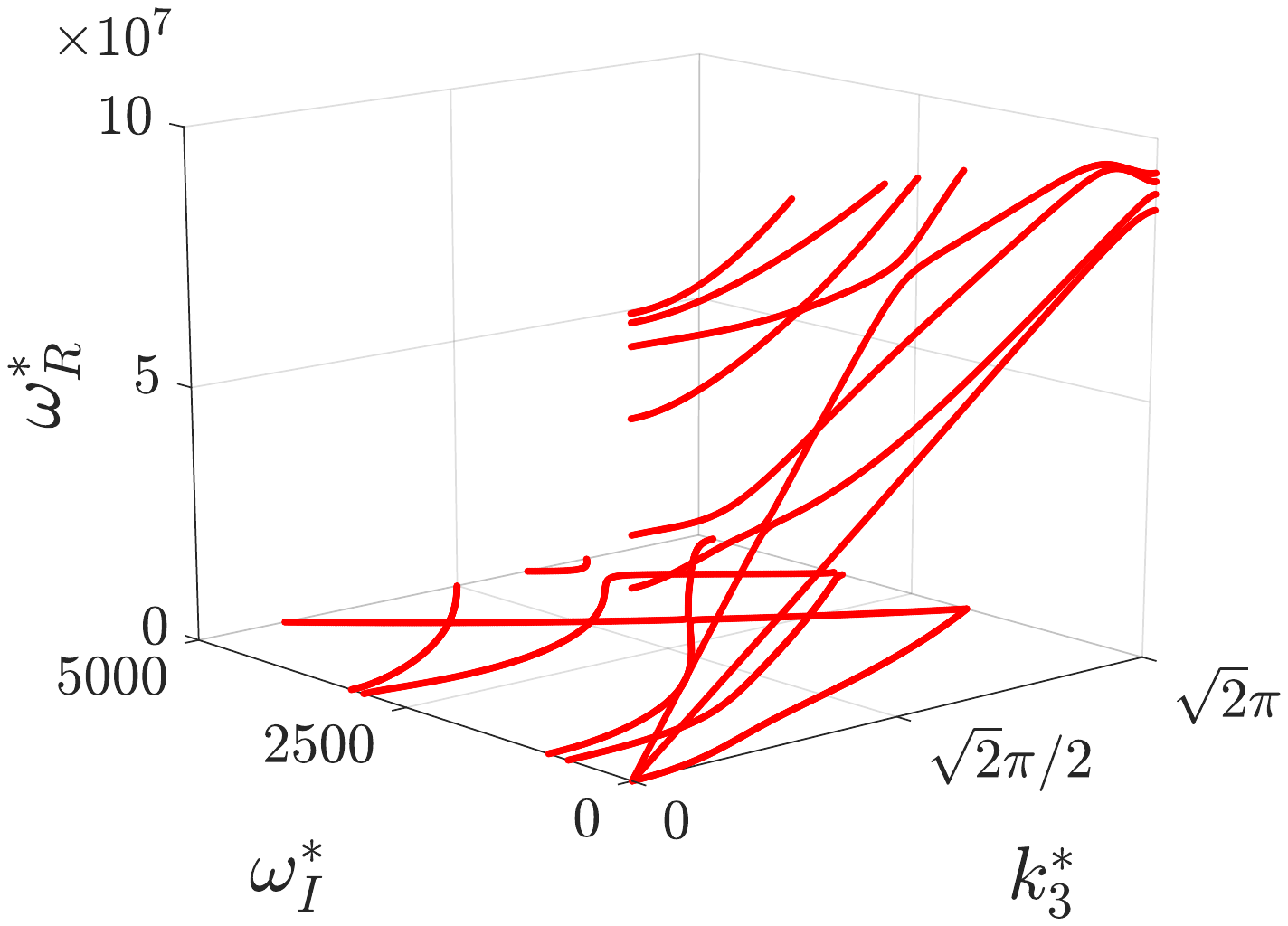}}
\subfigure[]
{\includegraphics[scale=0.6,trim={4cm 9.5cm 2.5cm 9.5cm}]{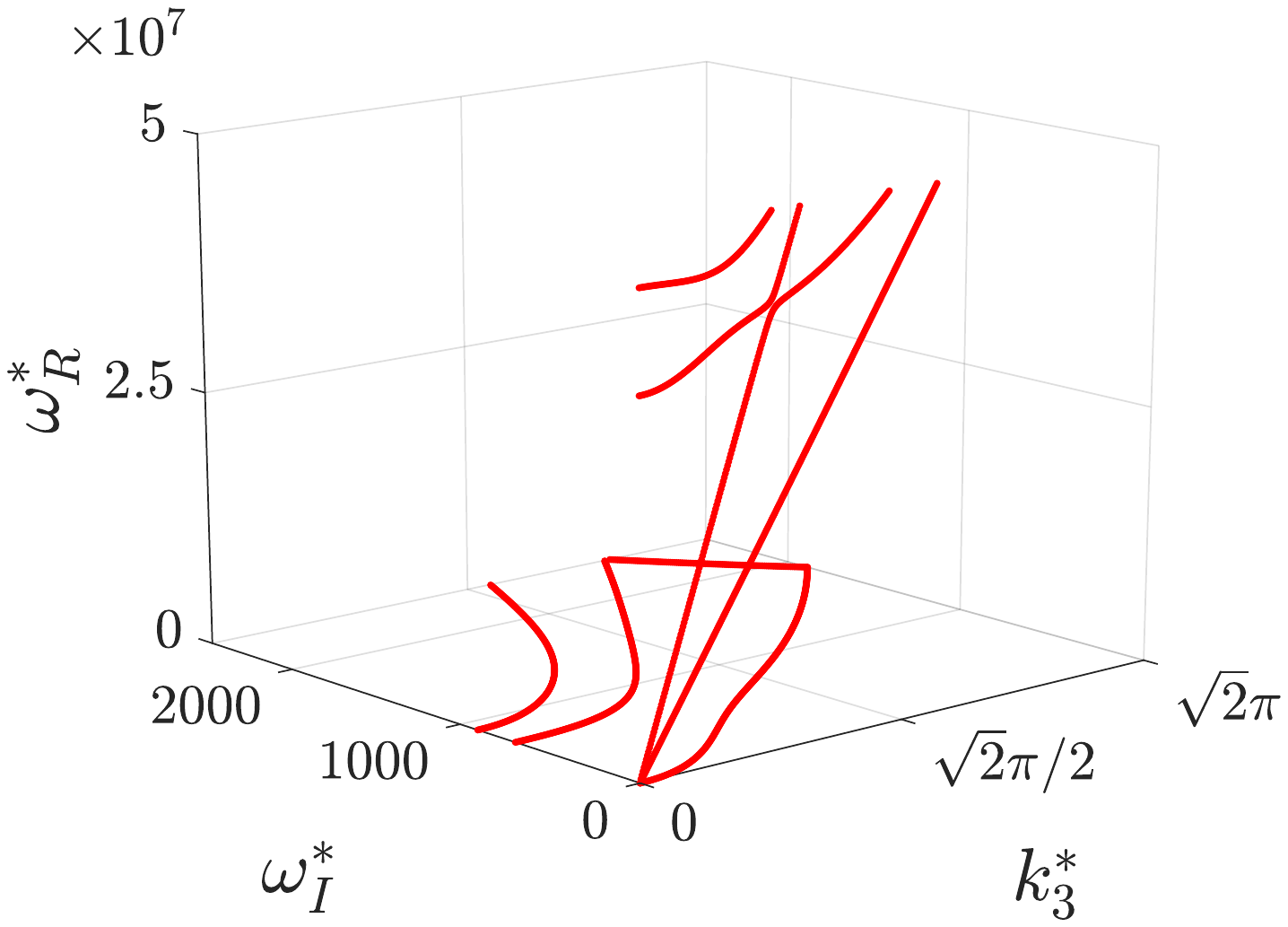}}\\
\subfigure[]
{\includegraphics[scale=0.6,trim={4cm 9.5cm 2.5cm 9.5cm}]{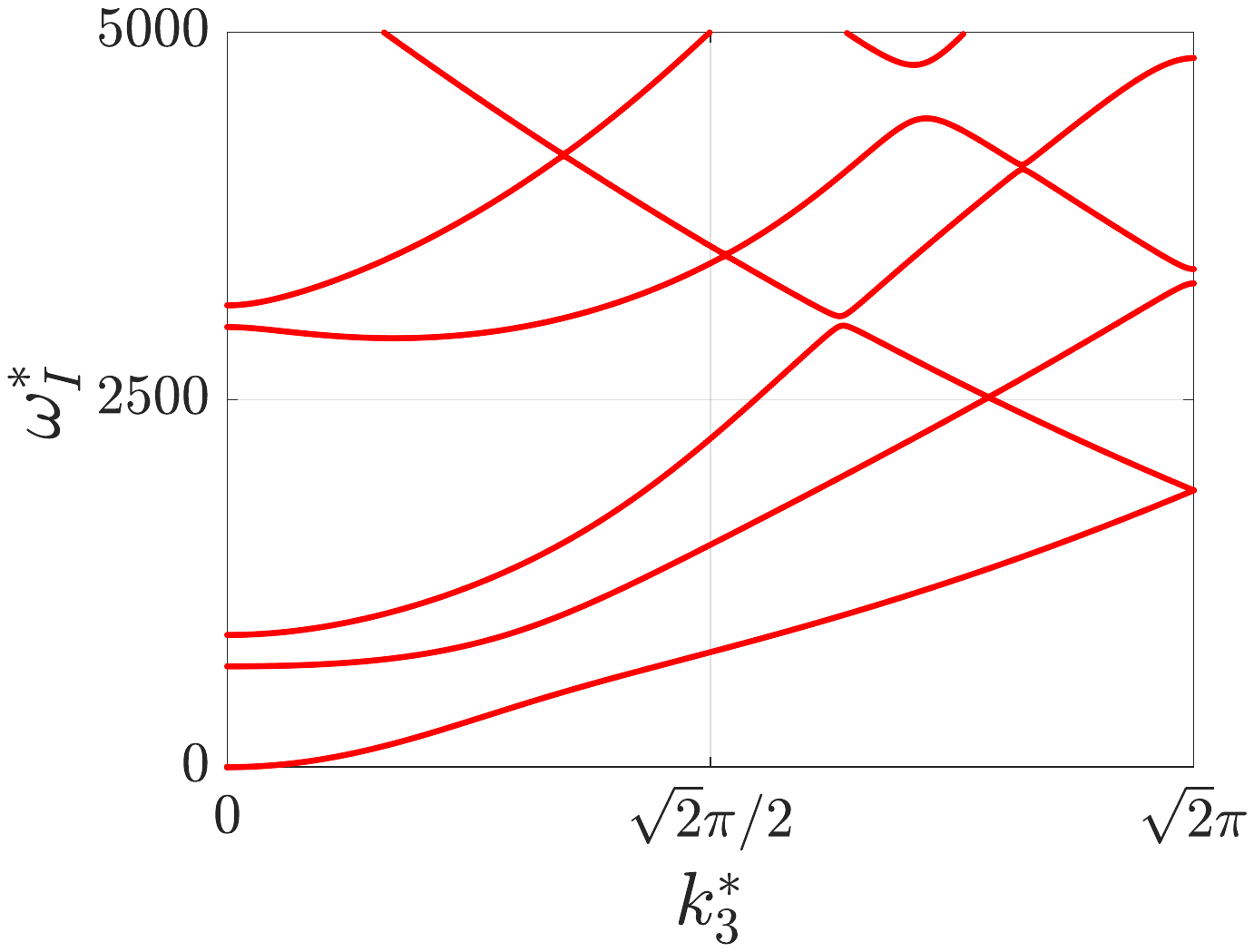}}
\subfigure[]
{\includegraphics[scale=0.6,trim={4cm 9.5cm 2.5cm 9.5cm}]{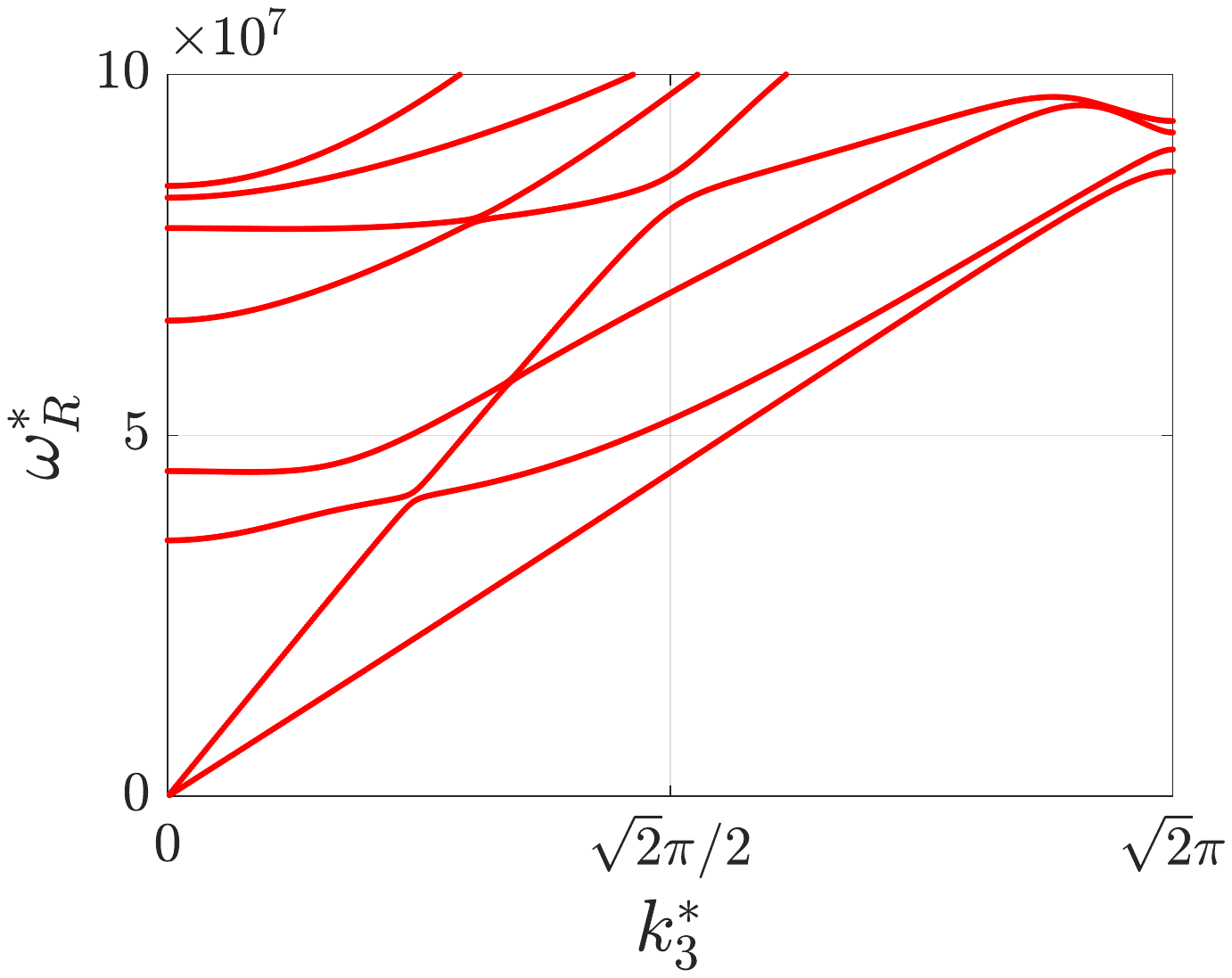}}
\caption{Complex Floquet-Bloch spectrum of the thermo-elastic SOFC-like material related to waves with unit vector of propagation $\textbf{m}_3^*$. (a) Complex frequencies $\omega$=$\omega_R^*$+$I$ $\omega_I^*$ versus $k_3^*$; (b) zoomed view of the lowest frequency branches of Figure \ref{fig:k1}(a); (c) view of the dispersive curves associated with temporal damping modes; (d) projection view on the plane $\omega_R^*$-$k_3^*$.}\label{fig:k12}
\end{figure}

 In the case with thermo-mechanical coupling, considering the first unit vector of propagation $\textbf{m}_1^*$, in Figure \ref{fig:k1} the regularized dispersive curves for $Q=4$, adopting a regular discretization of the variable $k_1$ with 500 points, are shown. In particular, Figure  \ref{fig:k1}(a) reports the complex spectrum where both the dimensionless real part $\omega_R^*=\omega_R/\omega_{ref}$, with $\omega_{ref}=1$ rad/sec, and imaginary part $\omega_I^*=\omega_I/\omega_{ref}$ of the complex angular frequency are plotted against the dimensionless wave number $k_1^*$.   A high spectral density is observed in the low frequency range, characterized both by acoustic and gathered optical branches. $Temporal$ $damping$ $modes$ are detected in the plane $\omega_R^*=0$, while for $\omega_R^*>0$ $mixed$ $modes$ appear, characterized in general by non-vanishing values of  both $\omega_R^*$ and $\omega_I^*$. It is, however, noted that as $\omega_R^*$ decreases the mixed modes tend toward $propagation$ $modes$ since $\omega_I^*$ 
 vanishes. In Figure  \ref{fig:k1}(b) a zoomed view of the 
 acoustic and first four optical branches is reported. Moreover, in Figure  \ref{fig:k1}(c) the temporal damping modes are shown separately. Several points of crossing
between acoustic and optical branches and also veering phenomena, i.e. repulsion between two branches, are evident.
Finally, in Figure \ref{fig:k1}(d) a projection view on the plane $\omega_R^*$-$k_1^*$ is reported, where the high 
spectral density prevents the presence of partial band-gaps 
associated with waves with unit vector of propagation $\textbf{m}_1^*$.\\
It is observed that the imaginary parts of the complex generalized eigenvalues turn out to be always non-negative, i.e. only damped waves are detected in the material.  Moreover, the spectrum results to be anti-symmetric with both respect to the plane  $\omega_I^*$-$k_1^*$, and to the plane
$\omega_R^*$-$\omega_I^*$ characterizing the progressive and regressive waves propagating in the periodic medium. By exploiting these symmetry condition, the only octant with positive values of $k_1^*$, $\omega_R^*$, $\omega_I^*$  is plotted in  Figure \ref{fig:k1}(a)-(b). Moreover, the real parts of the complex
generalized eigenvalues are typically of some orders of magnitude larger than the corresponding imaginary parts,
thus confirming the need of a very high precision enabled by the Avanpix-Multiprecision Computing Toolbox.\\
\begin{figure}[t!]
\centering
\subfigure[]
{\includegraphics[scale=0.6,trim={4cm 9.5cm 2.5cm 9.5cm}]{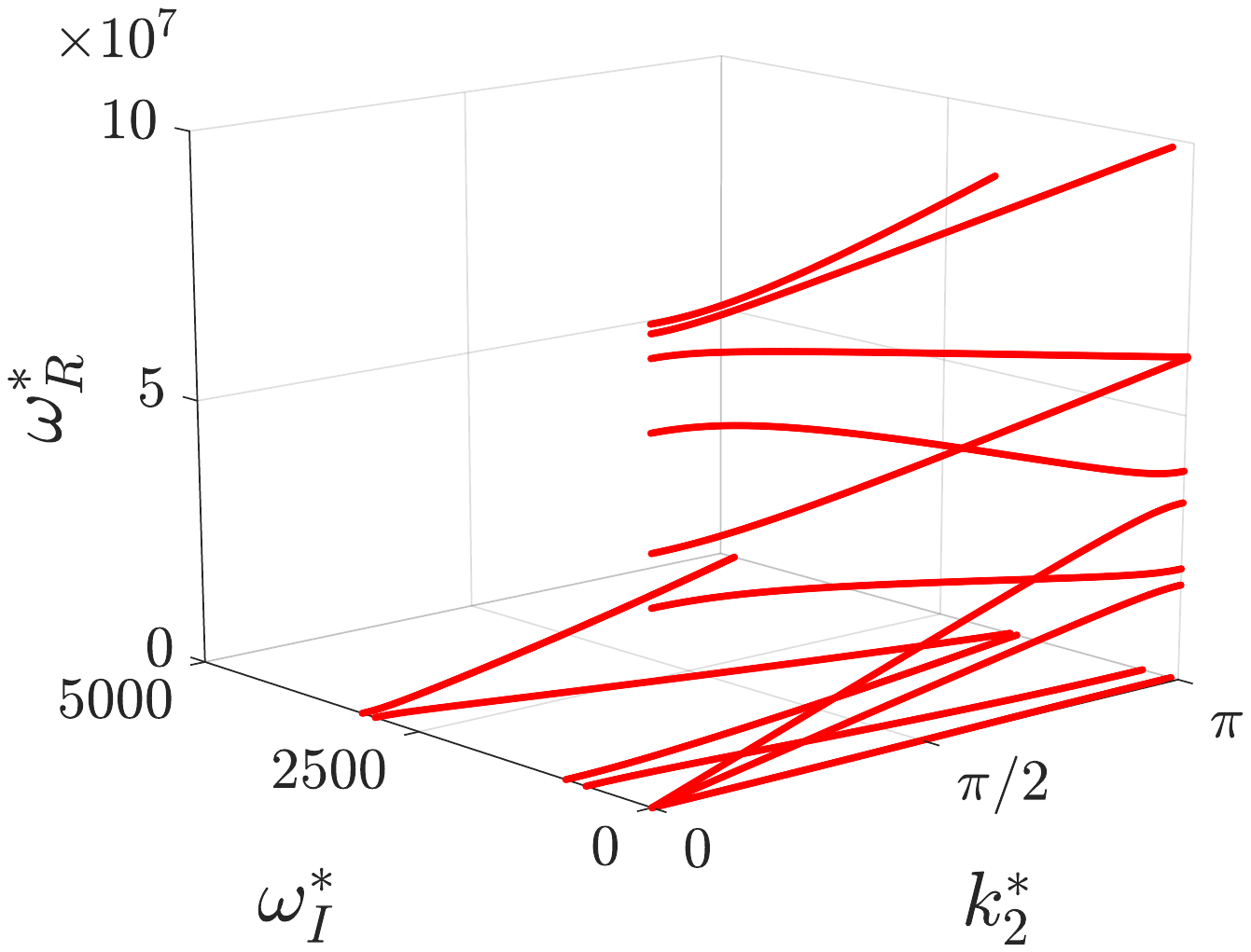}}
\subfigure[]
{\includegraphics[scale=0.6,trim={4cm 9.5cm 2.5cm 9.5cm}]{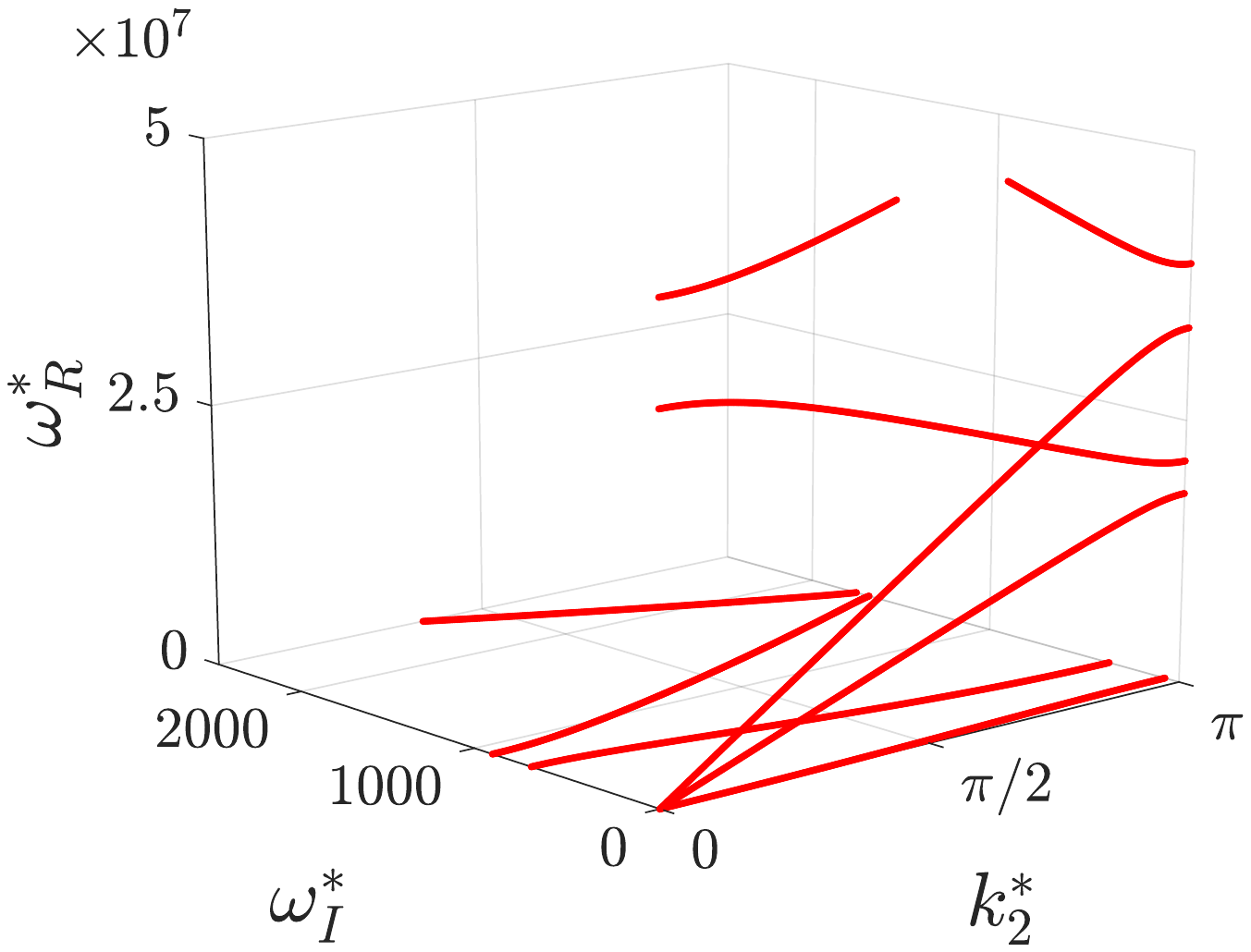}}\\
\subfigure[]
{\includegraphics[scale=0.6,trim={4cm 9.5cm 2.5cm 9.5cm}]{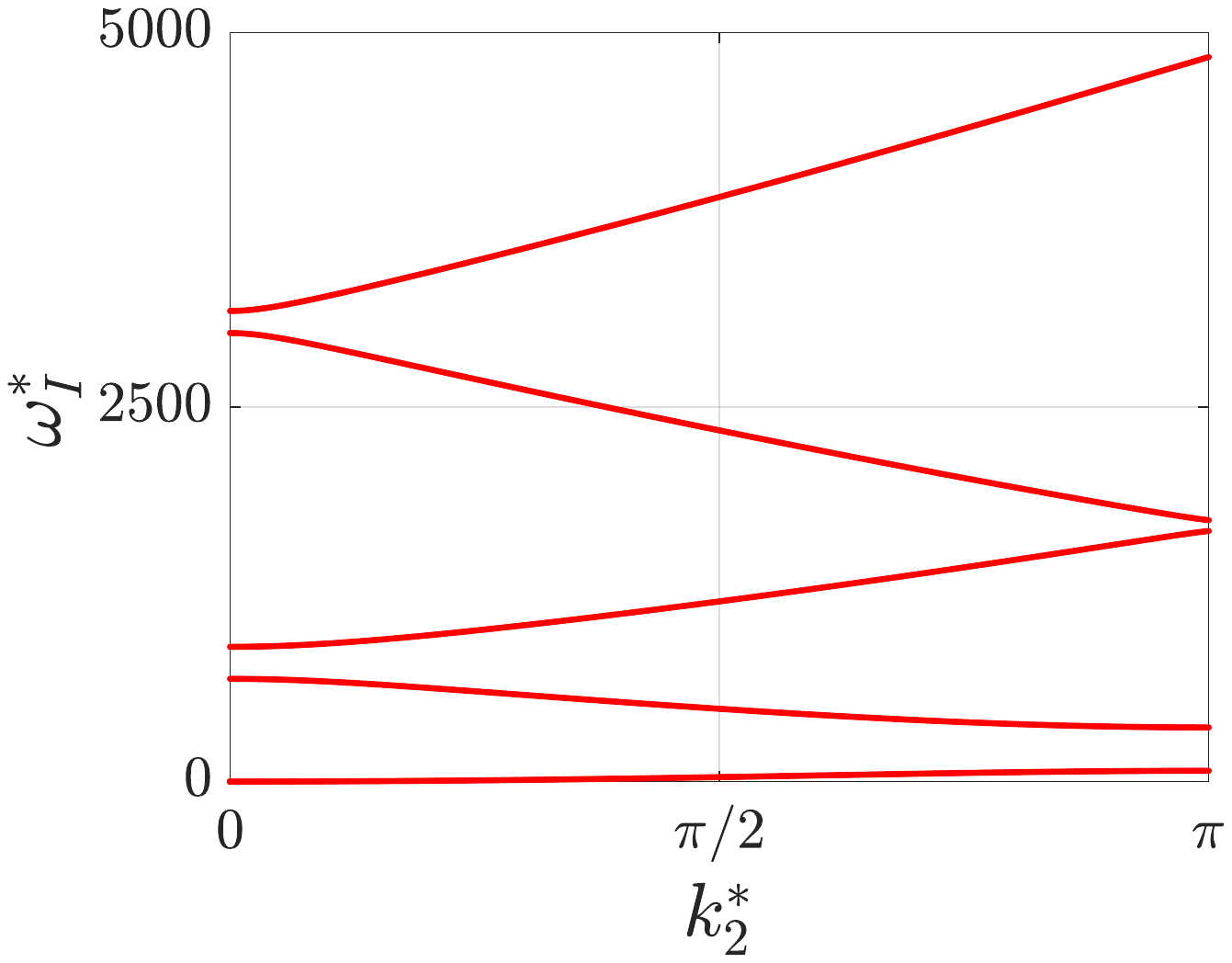}}
\subfigure[]
{\includegraphics[scale=0.6,trim={4cm 9.5cm 2.5cm 9.5cm}]{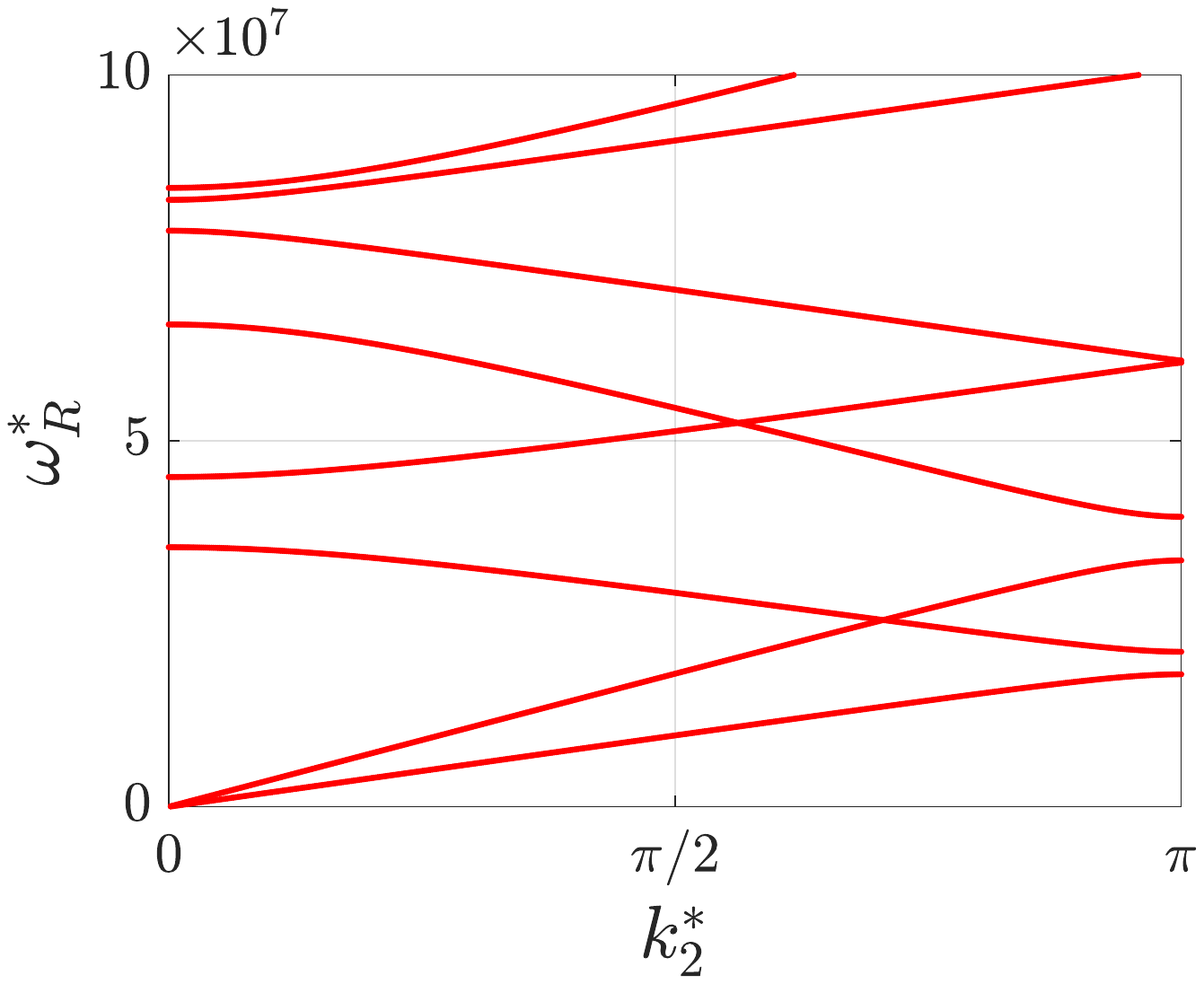}}
\caption{Complex Floquet-Bloch spectrum of the thermo-elastic SOFC-like material related to waves with unit vector of propagation $\textbf{m}_2^*$. (a) Complex frequencies $\omega$=$\omega_R^*$+$I$ $\omega_I^*$ versus $k_3^*$; (b) zoomed view of the lowest frequency branches of Figure \ref{fig:k12}(a); (c) view of the dispersive curves associated with temporal damping modes; (d) projection view on the plane $\omega_R^*$-$k_3^*$.}\label{fig:k2}
\end{figure}
Analogously, in Figure \ref{fig:k2} the complex Floquet-Bloch spectrum associated with the unit vector of propagation $\textbf{m}_2^*$ is investigated. In particular, in Figure \ref{fig:k2}(a) $\omega_R^*$ and  $\omega_I^*$ are plotted versus the dimensionless wave number $k_2^*$. Again in Figure \ref{fig:k2}(b) a zoomed view of the lowest frequency branches of Figure \ref{fig:k2}(a) is shown. Figure \ref{fig:k2}(c) illustrates the temporal damping modes in the plane $\omega_R^*=0$.  Finally, in Figure \ref{fig:k2}(d) the projection view on the plane $\omega_R^*$-$k_2^*$ is shown. Two partial band gaps are herein detected in the low frequencies range.\\
In Figure \ref{fig:k12}, finally, the unit vector of propagation   $\textbf{m}_3^*$ is taken into account. In this case $k_3^*$ is  the wave number in the direction of $\textbf{m}_3^*$. Herein, Figure \ref{fig:k12}(a) shows $\omega_R^*$ and  $\omega_I^*$ versus the dimensionless wave number $k_3^*$, while in Figure \ref{fig:k12}(b) a zoomed view of the lowest frequency branches is reported. In Figure \ref{fig:k12}(c) various crossing points, together with two veering points are evident in  the temporal damping modes. Due to the high spectral density, no partial band gaps are detected in the projection view on the plane $\omega_R^*$-$k_3^*$ , as shown in Figure \ref{fig:k12}(d).\\
\begin{figure}[hbtp]
\centering
\subfigure[]
{\includegraphics[scale=0.6,trim={4cm 9.5cm 2.5cm 9.5cm}]{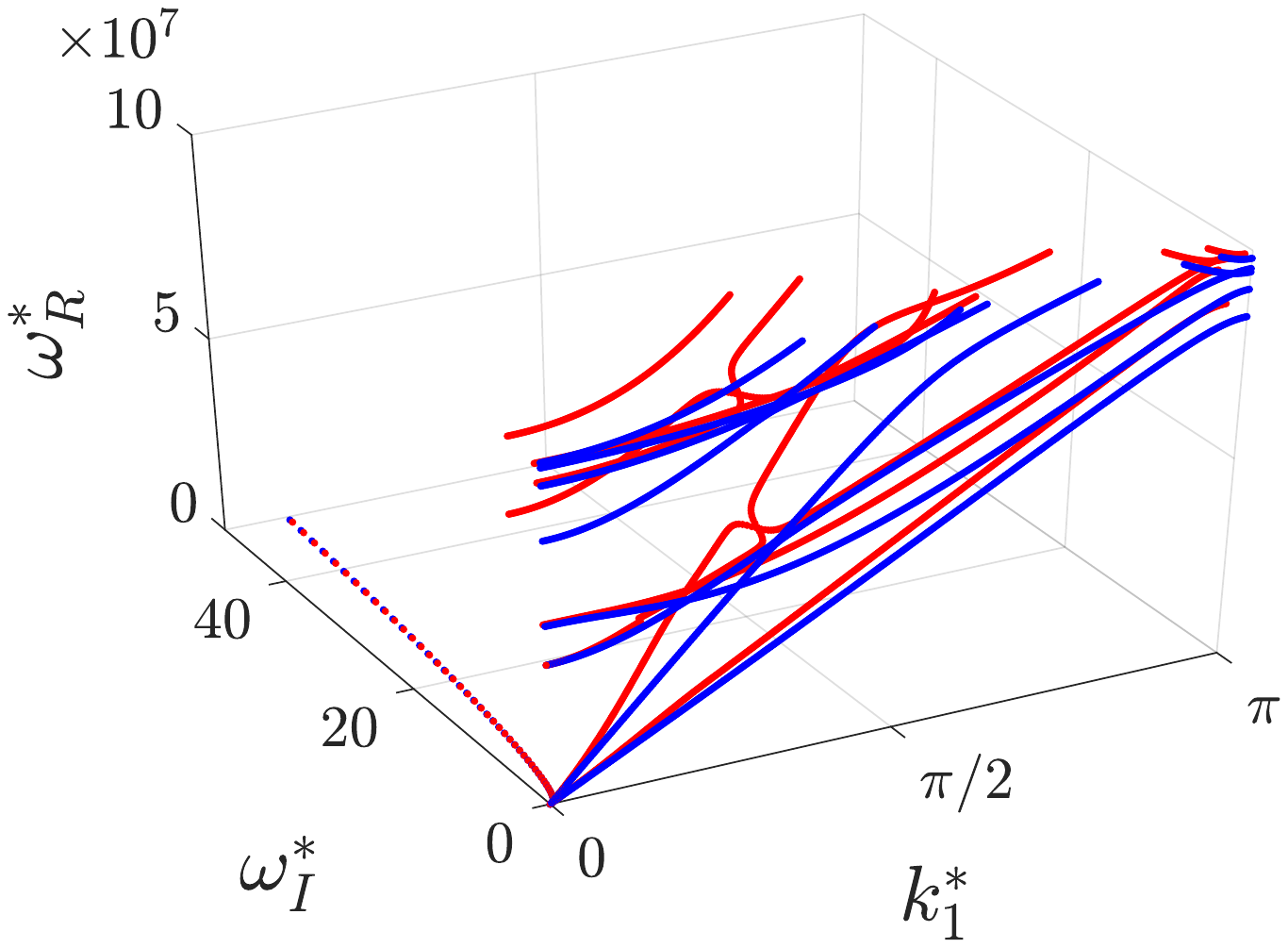}}
\subfigure[]
{\includegraphics[scale=0.6,trim={4cm 9.5cm 2.5cm 9.5cm}]{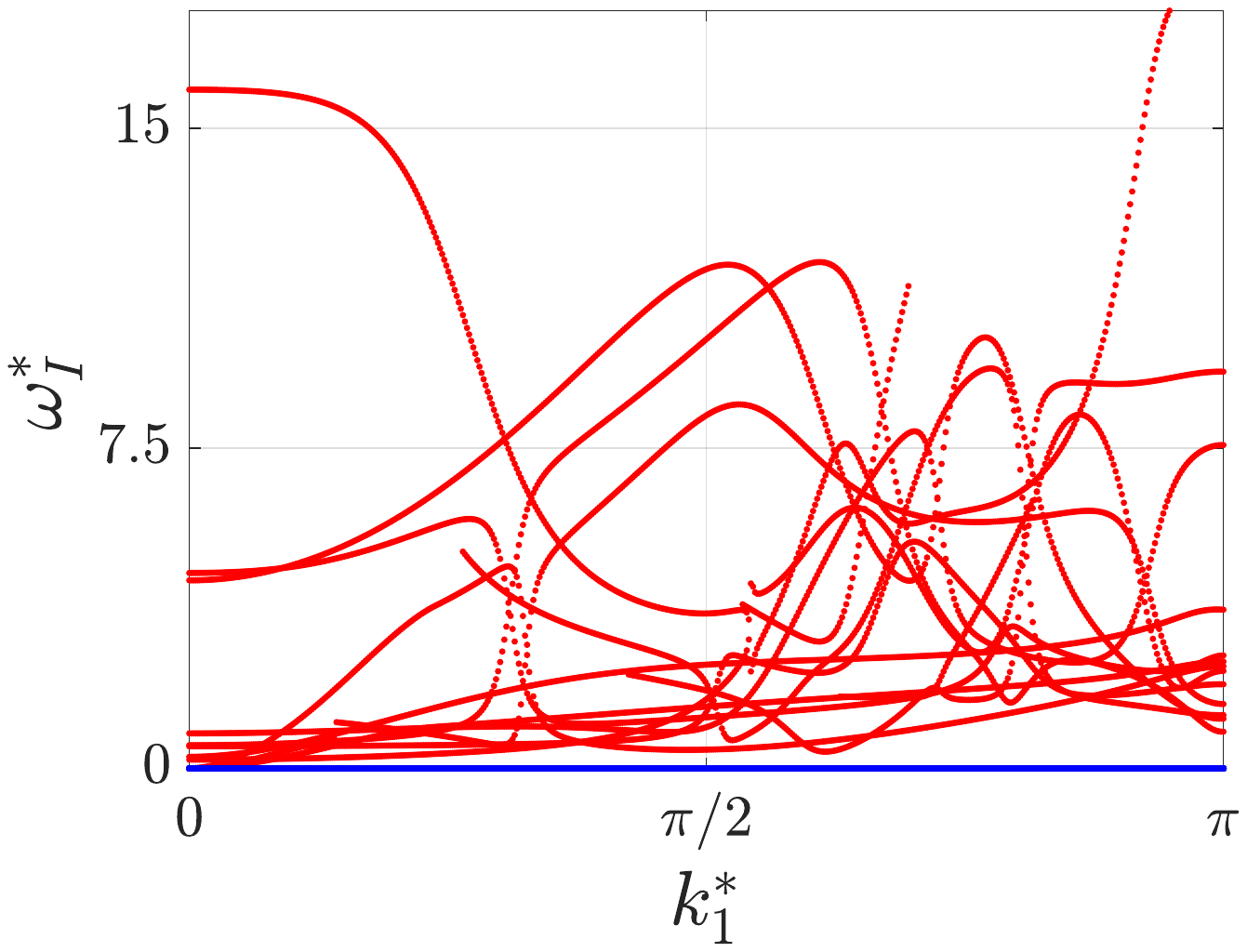}}
\caption{Comparison of the complex Floquet Bloch spectra, of SOFC-like material,  between the case with thermo-mechanical coupling (red curves) and without coupling (blue curves) with unit vector of propagation $\textbf{m}_1^*$. (a) Complex frequencies $\omega$=$\omega_R^*$+$I$ $\omega_I^*$ versus $k_1^*$; (b) projection view on the plane $\omega_I^*$-$k_1^*$
.}\label{fig:k1comparisons}
\end{figure}
\begin{figure}[ht]
\centering
\subfigure[]
{\includegraphics[scale=0.6,trim={4cm 9.5cm 2.5cm 9.5cm}]{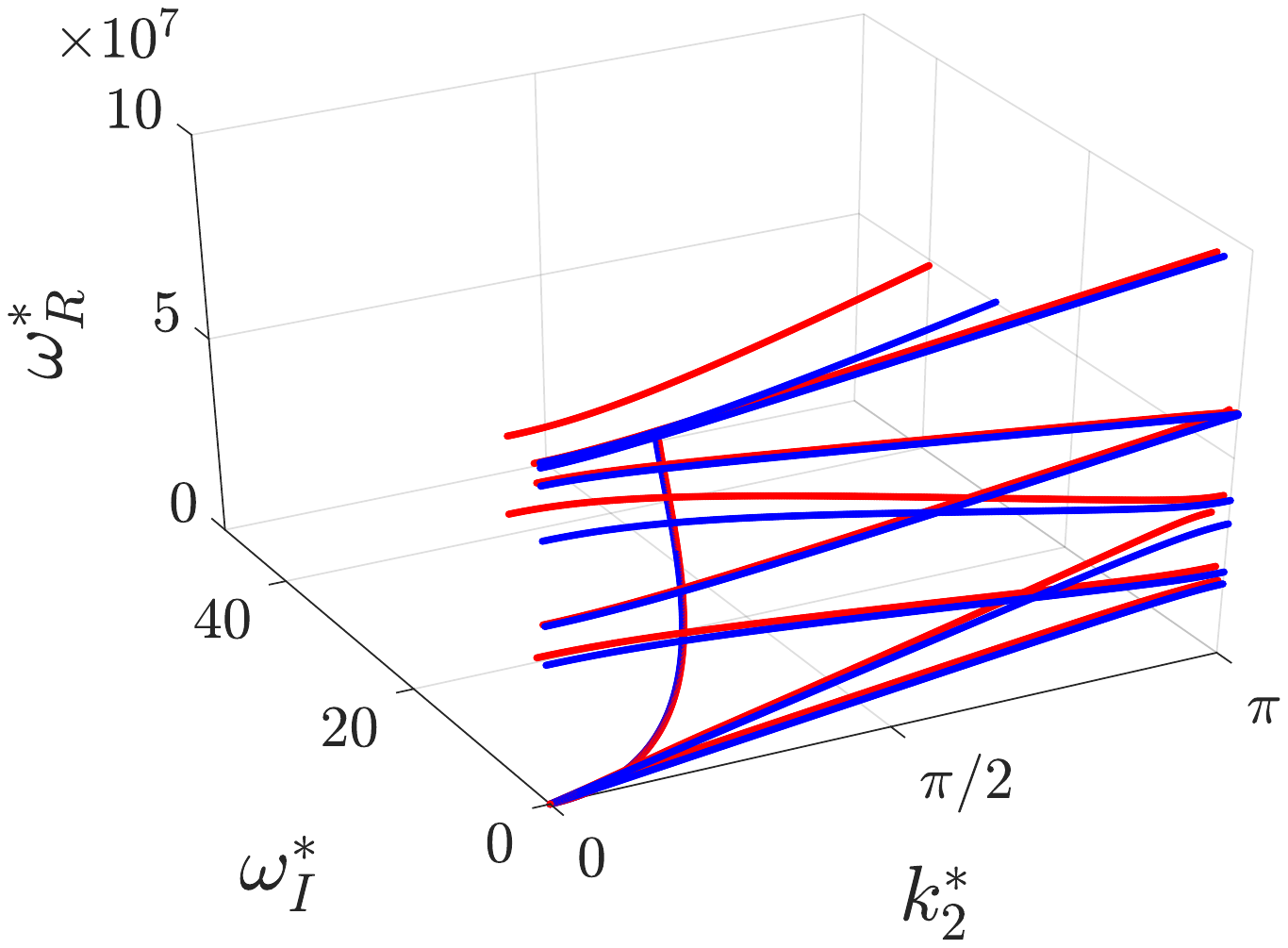}}
\subfigure[]
{\includegraphics[scale=0.6,trim={4cm 9.5cm 2.5cm 9.5cm}]{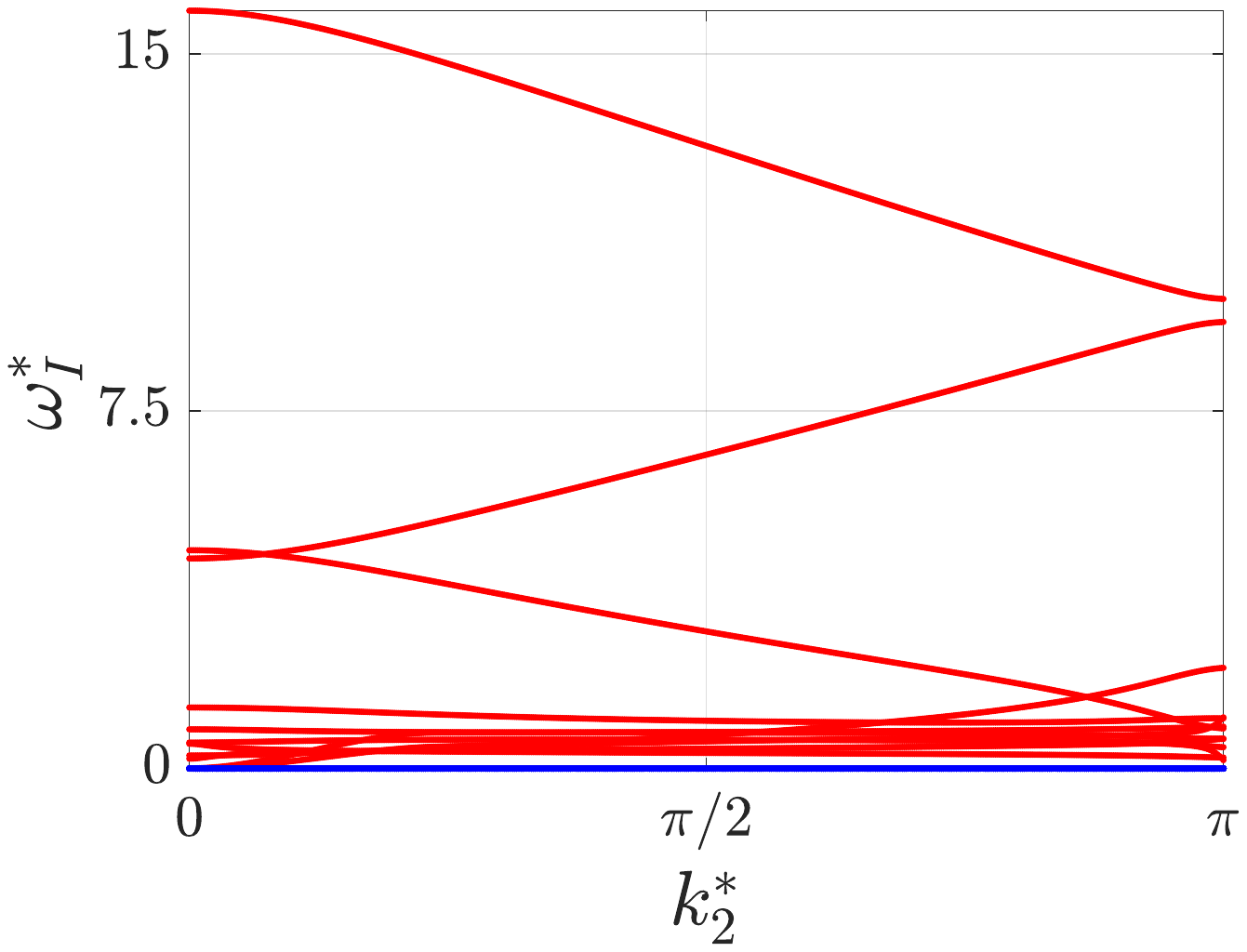}}
\caption{Comparison of the complex Floquet Bloch spectra, of SOFC-like material,  between the case with thermo-mechanical coupling (red curves) and without coupling (blue curves) with unit vector of propagation $\textbf{m}_2^*$. (a) Complex frequencies $\omega$=$\omega_R^*$+$I$ $\omega_I^*$ versus $k_2^*$; (b) projection view on the plane $\omega_I^*$-$k_2^*$.}\label{fig:k2comparisons}
\end{figure}
\begin{figure}[ht]
\centering
\subfigure[]
{\includegraphics[scale=0.6,trim={4cm 9.5cm 2.5cm 9.5cm}]{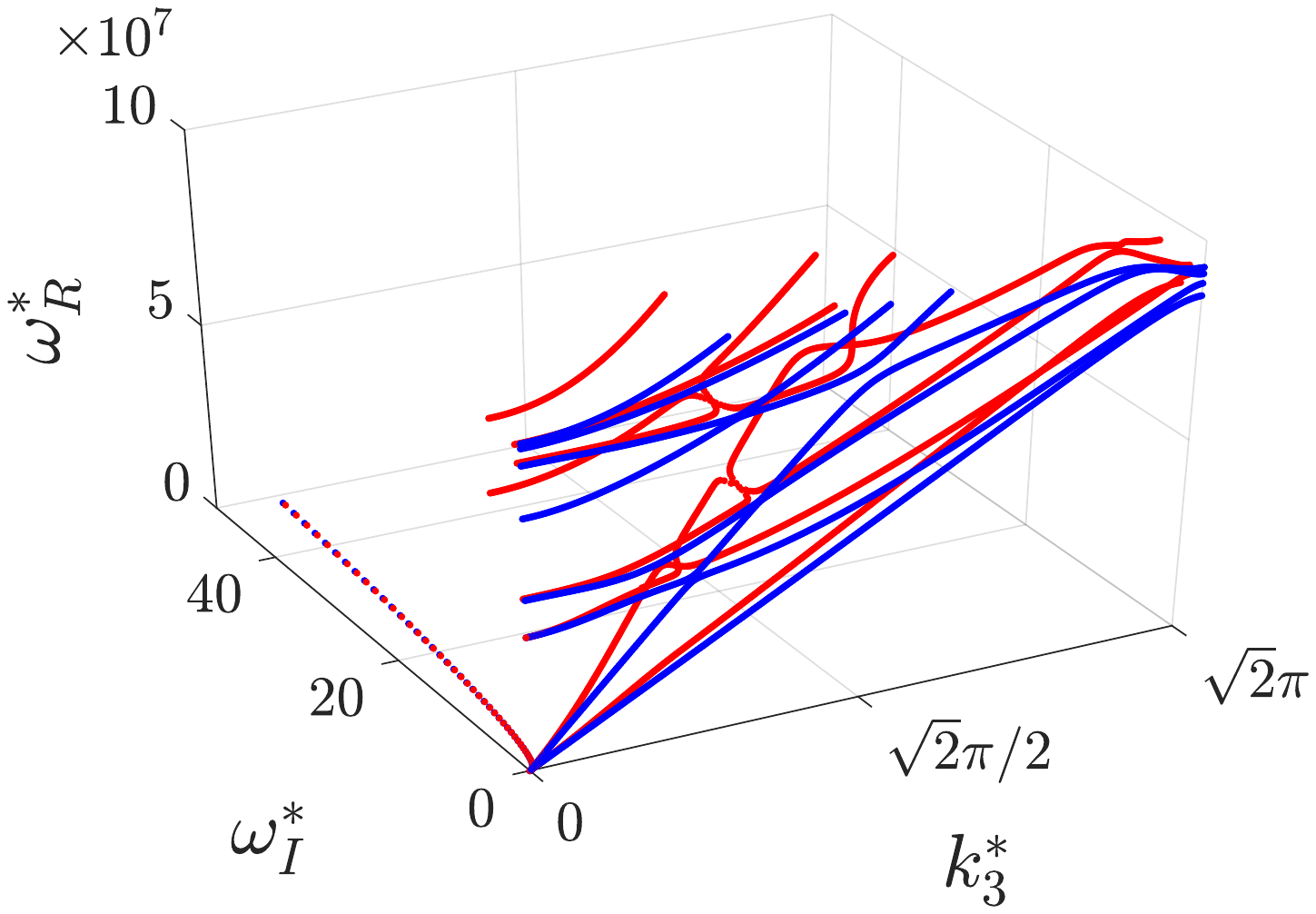}}
\subfigure[]
{\includegraphics[scale=0.6,trim={4cm 9.5cm 2.5cm 9.5cm}]{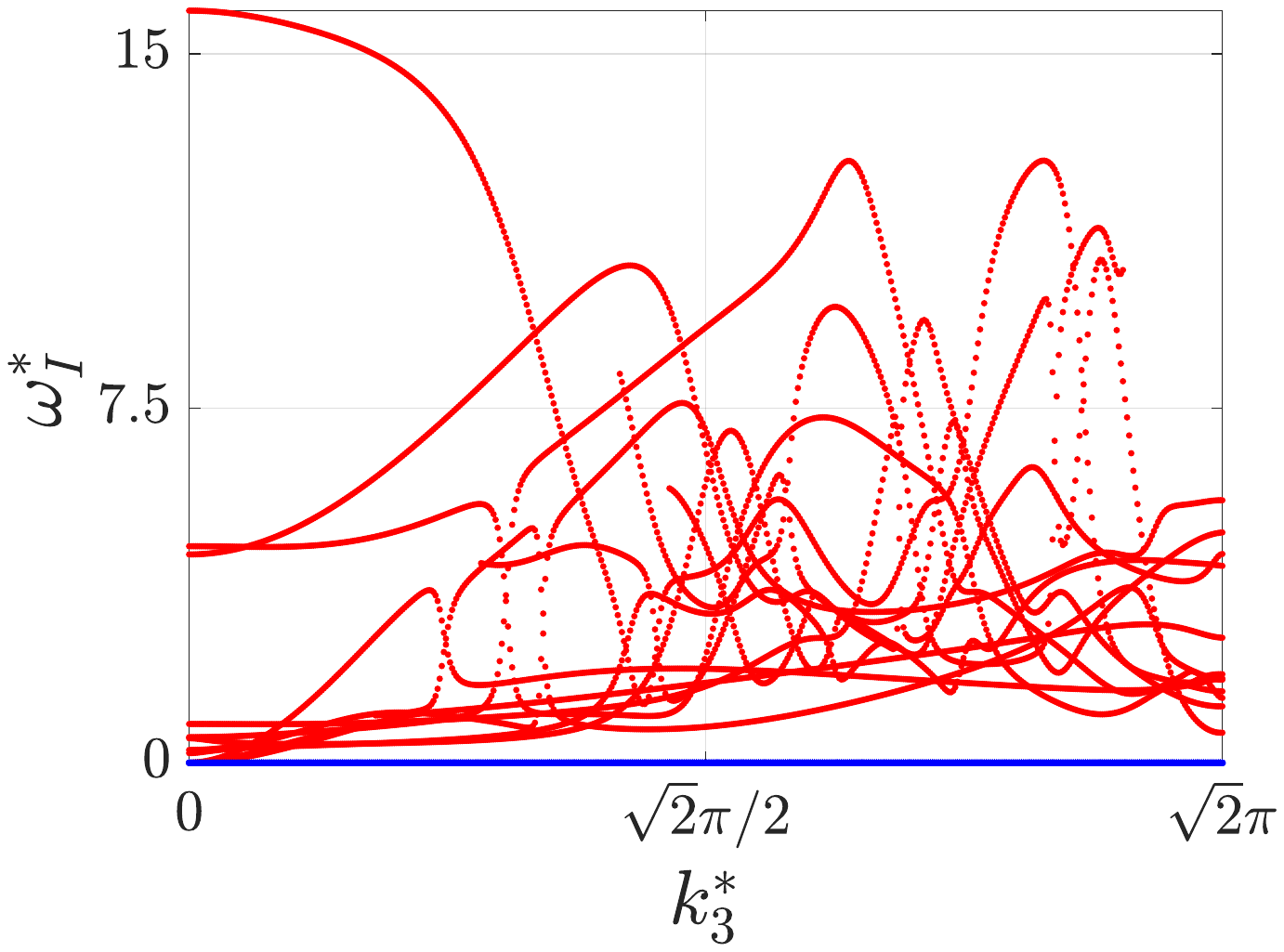}}
\caption{Comparison of the complex Floquet Bloch spectra, of SOFC-like material,  between the case with thermo-mechanical coupling (red curves) and without coupling (blue curves) with unit vector of propagation $\textbf{m}_3^*$. (a) Complex frequencies $\omega$=$\omega_R^*$+$I$ $\omega_I^*$ versus $k_3^*$; (b) projection view on the plane $\omega_I^*$-$k_3^*$.}\label{fig:k12comparisons}
\end{figure}
A final investigation concerns the comparison of Floquet-Bloch spectra in the cases with and without thermo-mechanical coupling,  depending on whether the components of the tensor $\boldsymbol{\alpha}$ are non-vanishing or vanishing, respectively. Therefore, in Figures \ref{fig:k1comparisons}, \ref{fig:k2comparisons}, and \ref{fig:k12comparisons} the red curves  are referred to the case  with thermo-mechanical coupling, while the blue ones to the case without coupling.
In particular, it is possible to emphasize that a qualitatively different behaviour is shown in the two considered cases. In the case with coupling, indeed, mixed modes are detected, while when the coupling is neglected only pure propagation modes appear. Moreover, the blue and red curves tend to increasingly depart from each other as $\omega_R^*$ increases. In Figures \ref{fig:k1comparisons}(a) the mixed modes (red curves) and the propagation modes (blue curves) in a low frequency range are shown, considering the unit vector of propagation $\textbf{m}_1^*$. Figures \ref{fig:k1comparisons}(b) a projection view on the plane $\omega_I^*$-$k_1^*$ of the complex spectrum is shown with $0 \leq \omega_R^* \leq 1.5 \cdot 10^8$. It stands to reason that while mixed modes are characterized by non-vanishing values of $\omega_I^*$, the propagation modes are defined for $\omega_I^*=0$.
Analogous considerations apply when  the unit vector of propagation $\textbf{m}_2^*$, Figure \ref{fig:k2comparisons}, and  the unit vector of propagation $\textbf{m}_3^*$, Figure \ref{fig:k12comparisons}, are considered. In both Figure \ref{fig:k2comparisons}(b) and Figure \ref{fig:k12comparisons}(b) the projection view is taken with $0 \leq \omega_R^* \leq 1.5 \cdot 10^8$.

\clearpage
\section{Final Remarks}\label{S:FR}
Thermo-diffusive phenomena involving SOFC-like periodic materials have been investigated in the dynamic regime.
Adopting a micromechanical approach, the dispersive waves propagation within the periodic medium has been analysed taking into account coupling effects between thermal mechanical and diffusive phenomena. To this aim, a generalization of the Floquet-Bloch theory has been adopted in order to determine the complex band structure of such materials. The Christoffel equations has been obtained from the governing equations of the first order thermo-diffusive medium. By adopting the Floquet-Bloch decomposition and exploiting proper integral transforms, an infinite algebraic linear system, involving complex frequencies and wave vectors, has been determined. Such infinite system has been truncated  taking into account a limited number of equations, which number has been obtained via a convergence analysis. In order to avoid the well-known Gibbs phenomenon, a regularization technique has been utilized. Finally, the complex angular frequencies, corresponding to a given wave vector, are found by solving the finite sequence of eigenvalue problems resulting from a discretization of the wave vector space. \\
The numerical investigations concern a periodic cell mimicking a SOFC device, in which the diffusive phenomena are neglected.
Considering the thermo-mechanical coupling, the complex dispersion curves in the first Brillouin zone, examined in the dimensionless space, are reported for three unit vectors of propagation. 
In general, a high spectral density is observed in the low-frequency range, characterized both by acoustic and gathered optical branches. Both  temporal damping modes and mixed modes feature the band structure of the periodic material. In particular, the mixed modes tend toward propagation modes as the real part of the complex frequency decreases.
Concerning the temporal damping modes, several points of crossing
between acoustic and optical branches and also veering phenomena are found. Partial band gaps only characterize waves propagating orthogonally to the material layers, while no partial band gaps are associated with the other two considered unit vectors of propagation.\\
Finally, in order to emphasize the effects of coupling phenomena, also the uncoupled problem is investigated. Qualitatively different behaviours are observed. Indeed, it results that in the absence of coupling phenomena the mixed modes tend to become propagation modes, and only the real part of the complex frequencies is different from zero.

\bibliographystyle{apalike}

\begin{thebibliography}{}

\bibitem[{\AA}berg and Gudmundson, 2000]{aaberg2000micromechanical}
{\AA}berg, M. and Gudmundson, P. (2000).
\newblock Micromechanical modeling of transient waves from matrix cracking and
  fiber fracture in laminated beams.
\newblock {\em International Journal of Solids and Structures},
  37(30):4083--4102.

\bibitem[Aboudi, 2001]{aboudi2001micromechanical}
Aboudi, J. (2001).
\newblock Micromechanical analysis of fully coupled
  electro-magneto-thermo-elastic multiphase composites.
\newblock {\em Smart materials and structures}, 10(5):867.

\bibitem[Aboudi et~al., 2001]{aboudi2001linear}
Aboudi, J., Pindera, M.-J., and Arnold, S. (2001).
\newblock Linear thermoelastic higher-order theory for periodic multiphase
  materials.
\newblock {\em Journal of Applied Mechanics}, 68(5):697--707.

\bibitem[Addessi et~al., 2013]{Addessi2013}
Addessi, D., De~Bellis, M.~L., and Sacco, E. (2013).
\newblock Micromechanical analysis of heterogeneous materials subjected to
  overall cosserat strains.
\newblock {\em Mechanics Research Communications}, 54:27 -- 34.

\bibitem[Addessi et~al., 2016]{Addessi2016}
Addessi, D., De~Bellis, M.~L., and Sacco, E. (2016).
\newblock A micromechanical approach for the cosserat modeling of composites.
\newblock {\em Meccanica}, 51(3):569--592.

\bibitem[Bacca et~al., 2013a]{Bacca2013a}
Bacca, M., Bigoni, D., Corso, F.~D., and Veber, D. (2013a).
\newblock Mindlin second-gradient elastic properties from dilute two-phase
  cauchy-elastic composites. part i: Closed form expression for the effective
  higher-order constitutive tensor.
\newblock {\em International Journal of Solids and Structures}, 50(24):4010 --
  4019.

\bibitem[Bacca et~al., 2013b]{Bacca2013b}
Bacca, M., Bigoni, D., Corso, F.~D., and Veber, D. (2013b).
\newblock Mindlin second-gradient elastic properties from dilute two-phase
  cauchy-elastic composites part ii: Higher-order constitutive properties and
  application cases.
\newblock {\em International Journal of Solids and Structures}, 50(24):4020 --
  4029.

\bibitem[Bacigalupo, 2014]{Bacigalupo2014}
Bacigalupo, A. (2014).
\newblock Second-order homogenization of periodic materials based on asymptotic
  approximation of the strain energy: formulation and validity limits.
\newblock {\em Meccanica}, 49(6):1407--1425.

\bibitem[Bacigalupo and De~Bellis, 2015]{BacigalupoDeBellis2015}
Bacigalupo, A. and De~Bellis, M.~L. (2015).
\newblock Auxetic anti-tetrachiral materials: equivalent elastic properties and
  frequency band-gaps.
\newblock {\em Composite Structures}, 131:530--544.

\bibitem[Bacigalupo and Gambarotta, 2013]{Bacigalupo2013}
Bacigalupo, A. and Gambarotta, L. (2013).
\newblock Multi-scale strain-localization analysis of a layered strip with
  debonding interfaces.
\newblock {\em Int J Solids Struct}, 50:2061--2077.

\bibitem[Bacigalupo et~al., 2014]{BACIGALUPO201415017}
Bacigalupo, A., Morini, L., and Piccolroaz, A. (2014).
\newblock Effective elastic properties of planar sofcs: A non-local dynamic
  homogenization approach.
\newblock {\em International Journal of Hydrogen Energy}, 39(27):15017 --
  15030.

\bibitem[Bacigalupo et~al., 2016a]{BacigalupoMorini2016}
Bacigalupo, A., Morini, L., and Piccolroaz, A. (2016a).
\newblock Multiscale asymptotic homogenization analysis of thermo-diffusive
  composite materials.
\newblock {\em International Journal of Solids and Structures}, 85-86:15--33.

\bibitem[Bacigalupo et~al., 2016b]{bacigalupo2016overall}
Bacigalupo, A., Morini, L., and Piccolroaz, A. (2016b).
\newblock Overall thermomechanical properties of layered materials for energy
  devices applications.
\newblock {\em Composite Structures}, 157:366--385.

\bibitem[Bacigalupo et~al., 2017]{BACIGALUPO2017}
Bacigalupo, A., Paggi, M., Corso, F.~D., and Bigoni, D. (2017).
\newblock Identification of higher-order continua equivalent to a cauchy
  elastic composite.
\newblock {\em Mechanics Research Communications}.

\bibitem[Bigoni and Drugan, 2007]{Bigoni2007}
Bigoni, D. and Drugan, W. (2007).
\newblock Analytical derivation of cosserat moduli via homogenization of
  heterogeneous elastic materials.
\newblock {\em Journal of Applied Mechanics}, 74(4):741--753.

\bibitem[Bove and Ubertini, 2008]{BoveUbertini2008}
Bove, R. and Ubertini, S. (2008).
\newblock {\em Modeling solid oxide fuel cells: methods, procedures and
  techniques}.
\newblock Springer.

\bibitem[Chen and Fish, 2001]{chen2001dispersive}
Chen, W. and Fish, J. (2001).
\newblock A dispersive model for wave propagation in periodic heterogeneous
  media based on homogenization with multiple spatial and temporal scales.
\newblock {\em Journal of applied mechanics}, 68(2):153--161.

\bibitem[De~Bellis and Addessi, 2011]{DeBellis-Addessi11}
De~Bellis, M.~L. and Addessi, D. (2011).
\newblock A {C}osserat based multi--scale model for masonry structures.
\newblock {\em International Journal for Multiscale Computational Engineering},
  9(5):543--563.

\bibitem[De~Bellis and Bacigalupo, 2017]{debellis2017auxetic}
De~Bellis, M.~L. and Bacigalupo, A. (2017).
\newblock Auxetic behavior and acoustic properties of microstructured
  piezoelectric strain sensors.
\newblock {\em Smart Materials and Structures}, 26(8):085037.

\bibitem[Deraemaeker and Nasser, 2010]{DERAEMAEKER20103272}
Deraemaeker, A. and Nasser, H. (2010).
\newblock Numerical evaluation of the equivalent properties of macro fiber
  composite (mfc) transducers using periodic homogenization.
\newblock {\em International Journal of Solids and Structures}, 47(24):3272 --
  3285.

\bibitem[Dev et~al., 2014a]{Devetal2014}
Dev, B., Walter, M.~E., Arkenberg, G.~B., and Swartz, S.~L. (2014a).
\newblock Mechanical and thermal characterization of a ceramic/glass composite
  seal for solid oxide fuel cells.
\newblock {\em Journal of Power Sources}, 245(1).
\newblock Article number: 958e66.

\bibitem[Dev et~al., 2014b]{dev2014mechanical}
Dev, B., Walter, M.~E., Arkenberg, G.~B., and Swartz, S.~L. (2014b).
\newblock Mechanical and thermal characterization of a ceramic/glass composite
  seal for solid oxide fuel cells.
\newblock {\em Journal of Power Sources}, 245:958--966.

\bibitem[Diouf and Pode, 2015]{DIOUF2015375}
Diouf, B. and Pode, R. (2015).
\newblock Potential of lithium-ion batteries in renewable energy.
\newblock {\em Renewable Energy}, 76:375 -- 380.

\bibitem[Ellis et~al., 2011]{Ellisetal2012}
Ellis, B.~L., Kaitlin, T., and Nazar, L.~F. (2011).
\newblock New composite materials for lithium-ion batteries.
\newblock {\em Electrochimica Acta}, 84(1):145--154.

\bibitem[Fantoni et~al., 2017]{fantoni2017multi}
Fantoni, F., Bacigalupo, A., and Paggi, M. (2017).
\newblock Multi-field asymptotic homogenization of thermo-piezoelectric
  materials with periodic microstructure.
\newblock {\em International Journal of Solids and Structures}, 120:31--56.

\bibitem[Francfort, 1983]{francfort1983homogenization}
Francfort, G.~A. (1983).
\newblock Homogenization and linear thermoelasticity.
\newblock {\em SIAM Journal on Mathematical Analysis}, 14(4):696--708.

\bibitem[Freund et~al., 2014]{FREUND201456}
Freund, J., Karakoç, A., and Sjölund, J. (2014).
\newblock Computational homogenization of regular cellular material according
  to classical elasticity.
\newblock {\em Mechanics of Materials}, 78:56 -- 65.

\bibitem[Galka et~al., 1996]{galka1996some}
Galka, A., Telega, J., and Wojnar, R. (1996).
\newblock Some computational aspects of homogenization of thermopiezoelectric
  composites.
\newblock {\em Comput. Assist. Mech. Eng. Sci}, 3(2):133--154.

\bibitem[Geers et~al., 2010]{Geers2010}
Geers, M., Kouznetsova, V., and Brekelmans, W. (2010).
\newblock Multi-scale computational homogenization: Trends and challenges.
\newblock {\em Journal of Computational and Applied Mathematics}, 234(7):2175
  -- 2182.

\bibitem[Golub and van Loan, 1996]{GolvanLoa1996}
Golub, G.~H. and van Loan, C.~F. (1996).
\newblock {\em Matrix computations}.
\newblock Johns Hopkins University Press.

\bibitem[Hajimolana et~al., 2011]{Hajetal2011}
Hajimolana, S.~A., Hussain, M.~A., Wan~Daud, W.~M.~A., Soroush, M., and
  Shamiri, A. (2011).
\newblock Mathematical modeling of solid oxide fuel cells: a review.
\newblock {\em Renewable \& Sustainable Energy Reviews}, 15(4).
\newblock Article number: 1893e917.

\bibitem[Hawwa and Nayfeh, 1995]{hawwa1995general}
Hawwa, M.~A. and Nayfeh, A.~H. (1995).
\newblock The general problem of thermoelastic waves in anisotropic
  periodically laminated composites.
\newblock {\em Composites Engineering}, 5(12):1499--1517.

\bibitem[Ignaczak and Ostoja-Starzewski, 2010]{ignaczak2010thermoelasticity}
Ignaczak, J. and Ostoja-Starzewski, M. (2010).
\newblock {\em Thermoelasticity with finite wave speeds}.
\newblock Oxford University Press.

\bibitem[Jerri, 1998]{Jerri1998}
Jerri, A.~E. (1998).
\newblock {\em The {G}ibbs phenomenon in {F}ourier analysis, splines and
  wavelet approximations}.
\newblock Springer.

\bibitem[Johnson and Qu, 2008]{johnson2008effective}
Johnson, J. and Qu, J. (2008).
\newblock Effective modulus and coefficient of thermal expansion of ni--ysz
  porous cermets.
\newblock {\em Journal of Power Sources}, 181(1):85--92.

\bibitem[Kakaç et~al., 2007]{KAKAC2007761}
Kakaç, S., Pramuanjaroenkij, A., and Zhou, X.~Y. (2007).
\newblock A review of numerical modeling of solid oxide fuel cells.
\newblock {\em International Journal of Hydrogen Energy}, 32(7):761 -- 786.
\newblock Fuel Cells.

\bibitem[Kanout{\'e} et~al., 2009]{kanoute2009multiscale}
Kanout{\'e}, P., Boso, D., Chaboche, J., and Schrefler, B. (2009).
\newblock Multiscale methods for composites: a review.
\newblock {\em Archives of Computational Methods in Engineering}, 16(1):31--75.

\bibitem[Lord and Shulman, 1967]{lord1967generalized}
Lord, H.~W. and Shulman, Y. (1967).
\newblock A generalized dynamical theory of thermoelasticity.
\newblock {\em Journal of the Mechanics and Physics of Solids}, 15(5):299--309.

\bibitem[Nakajo et~al., 2012]{Naketal2012}
Nakajo, A., Kuebler, J., Faes, A., Vogt, U.~F., Schindler, H.~J., Chiang,
  L.~K., Modena, S., herle J., V., and Hocker, T. (2012).
\newblock Compilation of mechanical properties for the structural analysis of
  solid oxide fuel cell stacks. constitutive materials of anode-supported
  cells.
\newblock {\em Ceramics International}, 38(5):3907--3927.

\bibitem[Nowacki, 1974a]{nowacki1974dynamical1}
Nowacki, W. (1974a).
\newblock Dynamical problems of thermo diffusion in solids {I}.
\newblock {\em Bulletin of Polish academy of Science and Technology},
  22(1):43--51.

\bibitem[Nowacki, 1974b]{nowacki1974dynamical2}
Nowacki, W. (1974b).
\newblock Dynamical problems of thermo diffusion in solids {II}.
\newblock {\em Bulletin of Polish academy of Science and Technology},
  22(3):129--135.

\bibitem[Nowacki, 1974c]{nowacki1974dynamical3}
Nowacki, W. (1974c).
\newblock Dynamical problems of thermo diffusion in solids {III}.
\newblock {\em Bulletin of Polish academy of Science and Technology},
  22(4):161--170.

\bibitem[Nowacki, 1976]{NOWACKI1976261}
Nowacki, W. (1976).
\newblock Dynamic problems of diffusion in solids.
\newblock {\em Engineering Fracture Mechanics}, 8(1):261 -- 266.

\bibitem[Olesiak, 1994]{Olesiak1994}
Olesiak, Z.~S. (1994).
\newblock Stresses in coated matrices caused by thermodiffusion.
\newblock {\em Materials Science}, 29(6):622--632.

\bibitem[Pitakthapanaphong and Busso, 2005]{PitBus2005}
Pitakthapanaphong, S. and Busso, E.~P. (2005).
\newblock Finite element analysis of the fracture behaviour of multi layered
  systems used in solid oxide fuel cell applications.
\newblock {\em Modelling and Simulation in Materials Science and Engineering},
  13(4).
\newblock Article number: 531e40.

\bibitem[Richardson et~al., 2012]{Ricetal2012}
Richardson, G., Denuault, G., and Please, C.~P. (2012).
\newblock Multiscale modelling and analysis of lithium-ion battery charge and
  discharge.
\newblock {\em Journal of Engineering Mathematics}, 72(1):41--72.

\bibitem[Salvadori et~al., 2014]{salvadori2014computational}
Salvadori, A., Bosco, E., and Grazioli, D. (2014).
\newblock A computational homogenization approach for li-ion battery cells:
  Part 1--formulation.
\newblock {\em Journal of the Mechanics and Physics of Solids}, 65:114--137.

\bibitem[Sharma and Sharma, 2009]{sharma2009modelling}
Sharma, J. and Sharma, R. (2009).
\newblock Modelling of thermoelastic rayleigh waves in a solid underlying a
  fluid layer with varying temperature.
\newblock {\em Applied Mathematical Modelling}, 33(3):1683--1695.

\bibitem[Sherief et~al., 2004]{SHERIEF2004591}
Sherief, H., Hamza, F.~A., and Saleh, H.~A. (2004).
\newblock The theory of generalized thermoelastic diffusion.
\newblock {\em International Journal of Engineering Science}, 42(5):591 -- 608.

\bibitem[Sridhar et~al., 2018]{SRIDHAR2018414}
Sridhar, A., Kouznetsova, V., and Geers, M. (2018).
\newblock A general multiscale framework for the emergent effective
  elastodynamics of metamaterials.
\newblock {\em Journal of the Mechanics and Physics of Solids}, 111:414 -- 433.

\bibitem[Suiker et~al., 2001]{suiker2001micro}
Suiker, A., De~Borst, R., and Chang, C. (2001).
\newblock Micro-mechanical modelling of granular material. part 2: Plane wave
  propagation in infinite media.
\newblock {\em Acta Mechanica}, 149(1-4):181--200.

\bibitem[Van~der Pol and Bremmer, 1950]{VanBre1947}
Van~der Pol, B. and Bremmer, H. (1950).
\newblock {\em Operational calculus based on the two-sided {L}aplace integral}.
\newblock Cambridge University Press.

\bibitem[Ward, 1981]{Ward1981}
Ward, R.~C. (1981).
\newblock Balancing the generalized eigenvalue problem.
\newblock {\em Journal on Scientific and Statistical Computing}, 2(2):141--152.

\bibitem[Weinberger, 1965]{Wei1965}
Weinberger, H.~F. (1965).
\newblock {\em A first course in partial differential equations with complex
  variables and transform methods}.
\newblock Blaisdell Publishing Company.

\bibitem[Zhang et~al., 2007]{zhang2007thermo}
Zhang, H., Zhang, S., Bi, J.~Y., and Schrefler, B. (2007).
\newblock Thermo-mechanical analysis of periodic multiphase materials by a
  multiscale asymptotic homogenization approach.
\newblock {\em International Journal for Numerical Methods in Engineering},
  69(1):87--113.

\end{thebibliography}


\section*{Appendix A: Explicit form of the governing equations in the transformed space }
By exploiting the convolution products, Equations (\ref{eq:ChrFourier1})-(\ref{eq:ChrFourier3}) can be written in the following form
\begin{eqnarray}
  && I {r_j} \int_{\mathbb{R}^2} \mathcal{F}_{\mathbf{x}} \left[ {C_{ijhk}} \right](\mathbf{r}-\mathbf{q})\, \left(I q_k \int_{\mathbb{R}^2}\mathcal{F}_{\mathbf{x}} \left[ {{{\tilde u}_h}} \right] (\mathbf{q}-\mathbf{w}) \mathcal{F}_{\mathbf{x}} \left[ e^{I\left( {{\mathbf{k}} \cdot {\mathbf{x}}} \right)} \right] (\mathbf{w}) \, d{\mathbf{w}} \right) \, d{\mathbf{q}} \nonumber \\
	&& - I{r_j} \int_{\mathbb{R}^2} \mathcal{F}_{\mathbf{x}} \left[ {\alpha _{ij}} \right](\mathbf{r}-\mathbf{q})\, \left(\int_{\mathbb{R}^2}\mathcal{F}_{\mathbf{x}} \left[ {\tilde \vartheta } \right] (\mathbf{q}-\mathbf{w}) \mathcal{F}_{\mathbf{x}} \left[ e^{I\left( {{\mathbf{k}} \cdot {\mathbf{x}}} \right)} \right] (\mathbf{w}) \, d{\mathbf{w}} \right) \, d{\mathbf{q}} \nonumber \\
	&& - I {r_j} \int_{\mathbb{R}^2} \mathcal{F}_{\mathbf{x}} \left[ {\beta_{ij}} \right](\mathbf{r}-\mathbf{q})\, \left(\int_{\mathbb{R}^2}\mathcal{F}_{\mathbf{x}} \left[ {\tilde \eta } \right] (\mathbf{q}-\mathbf{w}) \mathcal{F}_{\mathbf{x}} \left[ e^{I\left( {{\mathbf{k}} \cdot {\mathbf{x}}} \right)} \right] (\mathbf{w}) \, d{\mathbf{w}} \right) \, d{\mathbf{q}} \nonumber \\
	&& + \omega^2 \int_{\mathbb{R}^2} \mathcal{F}_{\mathbf{x}} \left[ {\rho} \right](\mathbf{r}-\mathbf{q})\, \left(\int_{\mathbb{R}^2}\mathcal{F}_{\mathbf{x}} \left[ {{{\tilde u}_i}} \right] (\mathbf{q}-\mathbf{w}) \mathcal{F}_{\mathbf{x}} \left[ e^{I\left( {{\mathbf{k}} \cdot {\mathbf{x}}} \right)} \right] (\mathbf{w}) \, d{\mathbf{w}} \right) \, d{\mathbf{q}} = 0 \,,\label{eq:ChrFourier1mod}\\[5pt]
&& I {r_i} \int_{\mathbb{R}^2} \mathcal{F}_{\mathbf{x}} \left[ {K_{ij}} \right](\mathbf{r}-\mathbf{q})\, \left(I q_j \int_{\mathbb{R}^2}\mathcal{F}_{\mathbf{x}} \left[ {{{\tilde \vartheta}}} \right] (\mathbf{q}-\mathbf{w}) \mathcal{F}_{\mathbf{x}} \left[ e^{I\left( {{\mathbf{k}} \cdot {\mathbf{x}}} \right)} \right] (\mathbf{w}) \, d{\mathbf{w}} \right) \, d{\mathbf{q}} \nonumber \\
	&& - I \omega \int_{\mathbb{R}^2} \mathcal{F}_{\mathbf{x}} \left[ {\alpha_{ij}} \right](\mathbf{r}-\mathbf{q})\, \left(I q_j \int_{\mathbb{R}^2}\mathcal{F}_{\mathbf{x}} \left[ {{{\tilde u}_i}} \right] (\mathbf{q}-\mathbf{w}) \mathcal{F}_{\mathbf{x}} \left[ e^{I\left( {{\mathbf{k}} \cdot {\mathbf{x}}} \right)} \right] (\mathbf{w}) \, d{\mathbf{w}} \right) \, d{\mathbf{q}} \nonumber \\ \nonumber \\
	&&  - I \omega \int_{\mathbb{R}^2} \mathcal{F}_{\mathbf{x}} \left[ {\Psi} \right](\mathbf{r}-\mathbf{q})\, \left(\int_{\mathbb{R}^2}\mathcal{F}_{\mathbf{x}} \left[ {{\tilde \eta}} \right] (\mathbf{q}-\mathbf{w}) \mathcal{F}_{\mathbf{x}} \left[ e^{I\left( {{\mathbf{k}} \cdot {\mathbf{x}}} \right)} \right] (\mathbf{w}) \, d{\mathbf{w}} \right) \, d{\mathbf{q}} \nonumber \\
	&& - I \omega \int_{\mathbb{R}^2} \mathcal{F}_{\mathbf{x}} \left[ {p} \right](\mathbf{r}-\mathbf{q})\, \left(\int_{\mathbb{R}^2}\mathcal{F}_{\mathbf{x}} \left[ {{\tilde \vartheta}} \right] (\mathbf{q}-\mathbf{w}) \mathcal{F}_{\mathbf{x}} \left[ e^{I\left( {{\mathbf{k}} \cdot {\mathbf{x}}} \right)} \right] (\mathbf{w}) \, d{\mathbf{w}} \right) \, d{\mathbf{q}} = 0 \,,\label{eq:ChrFourier2mod}\\[5pt]
	&& I {r_i} \int_{\mathbb{R}^2} \mathcal{F}_{\mathbf{x}} \left[ {D_{ij}} \right](\mathbf{r}-\mathbf{q})\, \left(I q_j \int_{\mathbb{R}^2}\mathcal{F}_{\mathbf{x}} \left[ {{{\tilde \eta}}} \right] (\mathbf{q}-\mathbf{w}) \mathcal{F}_{\mathbf{x}} \left[ e^{I\left( {{\mathbf{k}} \cdot {\mathbf{x}}} \right)} \right] (\mathbf{w}) \, d{\mathbf{w}} \right) \, d{\mathbf{q}} \nonumber \\
	&& - I \omega \int_{\mathbb{R}^2} \mathcal{F}_{\mathbf{x}} \left[ {\beta_{ij}} \right](\mathbf{r}-\mathbf{q})\, \left(I q_j \int_{\mathbb{R}^2}\mathcal{F}_{\mathbf{x}} \left[ {{{\tilde u}_i}} \right] (\mathbf{q}-\mathbf{w}) \mathcal{F}_{\mathbf{x}} \left[ e^{I\left( {{\mathbf{k}} \cdot {\mathbf{x}}} \right)} \right] (\mathbf{w}) \, d{\mathbf{w}} \right) \, d{\mathbf{q}} \nonumber \\ 
	&&  - I \omega \int_{\mathbb{R}^2} \mathcal{F}_{\mathbf{x}} \left[ {\Psi} \right](\mathbf{r}-\mathbf{q})\, \left(\int_{\mathbb{R}^2}\mathcal{F}_{\mathbf{x}} \left[ {{\tilde \vartheta}} \right] (\mathbf{q}-\mathbf{w}) \mathcal{F}_{\mathbf{x}} \left[ e^{I\left( {{\mathbf{k}} \cdot {\mathbf{x}}} \right)} \right] (\mathbf{w}) \, d{\mathbf{w}} \right) \, d{\mathbf{q}} \nonumber \\
	&& - I \omega \int_{\mathbb{R}^2} \mathcal{F}_{\mathbf{x}} \left[ {c} \right](\mathbf{r}-\mathbf{q})\, \left(\int_{\mathbb{R}^2}\mathcal{F}_{\mathbf{x}} \left[ {{\tilde \eta}} \right] (\mathbf{q}-\mathbf{w}) \mathcal{F}_{\mathbf{x}} \left[ e^{I\left( {{\mathbf{k}} \cdot {\mathbf{x}}} \right)} \right] (\mathbf{w}) \, d{\mathbf{w}} \right) \, d{\mathbf{q}} = 0 \,.\label{eq:ChrFourier3mod}
  \end{eqnarray}
with $\textbf{r}, \textbf{q}, \textbf{w} \in \textbf{B}$. Note that in Equations (\ref{eq:ChrFourier1mod}), (\ref{eq:ChrFourier2mod}), (\ref{eq:ChrFourier3mod}) the  Fourier transform
of the exponential function , i.e. $\mathcal{F}_{\mathbf{x}} \left[ e^{I\left( {{\mathbf{k}} \cdot {\mathbf{x}}} \right)} \right]$, results as 
\begin{equation}
\mathcal{F}_{\mathbf{x}} \left[ e^{I\left( {{\mathbf{k}} \cdot {\mathbf{x}}} \right)} \right](\mathbf{w})=4 \pi^2\delta(\mathbf{w}-\mathbf{k}),
\end{equation}
where $\delta$ is the Dirac-delta generalized function centered at $\mathbf{w}=\mathbf{k}$. It follows that the convolution product between the Fourier transform $\mathcal{F}_{\mathbf{x}} \left[ g(\mathbf{x}) \right]$, with $g(\mathbf{x})$ a generic function,  and $\mathcal{F}_{\mathbf{x}} \left[ e^{I\left( {{\mathbf{k}} \cdot {\mathbf{x}}} \right)} \right]$
can be written as
\begin{align}\label{eq:simple_comvolution}
& \int_{\mathbb{R}^2} \mathcal{F}_{\mathbf{x}} \left[ g(\mathbf{x}) \right] (\mathbf{q}-\mathbf{w}) \mathcal{F}_{\mathbf{x}} \left[ e^{I\left( {{\mathbf{k}} \cdot {\mathbf{x}}} \right)} \right] (\mathbf{w}) \, d{\mathbf{w}} = \nonumber\\
&\int_{\mathbb{R}^2} \mathcal{F}_{\mathbf{x}} \left[ g(\mathbf{x}) \right] (\mathbf{q}-\mathbf{w}) 4 \pi^2 \delta (\mathbf{w}-\mathbf{k}) \, d{\mathbf{w}} = 4 \pi^2 \mathcal{F}_{\mathbf{x}} \left[ g(\mathbf{x}) \right] (\mathbf{q}-\mathbf{k})\,.
\end{align}
Finally, by applying (\ref{eq:simple_comvolution}) to (\ref{eq:ChrFourier1mod})-(\ref{eq:ChrFourier3mod}), one obtains (\ref{eq:conv1})-(\ref{eq:conv1var2}). 

\section*{Appendix B: Linear operators involved in the generalized eigenvalue problems}

\subsection*{Appendix B.1:{ Vector} ${\textbf{z}}$, {and operators} ${\textbf{\textit{A}}}$, ${\textbf{\textit{B}}}$, {and} ${\textbf{\textit{C}}}$}
In order to express Equations (\ref{eq:conv2})-(\ref{eq:conv2var2}) in compact form,
the  linear operators $\textbf{\textit{A}}$, $\textbf{\textit{B}}$, and $\textbf{\textit{C}}$ are herein defined, in terms of their actions on the argument $\textbf{z}$, expressed in the following form
\begin{equation}
\textbf{z}={\rm col}({\mathbf{\tilde u}}_1, {\mathbf{\tilde u}}_2, {\bm{\tilde \vartheta}}, {\bm{\tilde \eta}}) \in (l_2(\mathbb{Z}^2))^4\,,
\end{equation}
where ${\mathbf{\tilde u}}_1$, ${\mathbf{\tilde u}}_2$, $\bm{\tilde \vartheta}$,  ${\bm{\tilde \eta}}$ are vectors collecting, respectively, the Fourier coefficients ${\tilde u}_1^{\bar{q}_1 \,\bar{q}_2}$, ${\tilde u}_2^{\bar{q}_1 \,\bar{q}_2}$, ${\tilde \vartheta}^{\bar{q}_1 \,\bar{q}_2}$,  ${\tilde \eta}^{\bar{q}_1 \,\bar{q}_2}$, the ${\rm col}$ operator stacks its vector arguments column-wise into a single column vector, $l_2(\mathbb{Z}^2)$ denotes the space of square-summable sequences with two integer indices, and $(l_2(\mathbb{Z}^2))^4$ stands for $l_2(\mathbb{Z}^2) \times l_2(\mathbb{Z}^2) \times l_2(\mathbb{Z}^2) \times l_2(\mathbb{Z}^2)$.\\
 \noindent The first operator $\textbf{\textit{A}}: (l_2(\mathbb{Z}^2))^4 \to (l_2(\mathbb{Z}^2))^4$ is  defined equation-wise as follows 
\begin{eqnarray}
\left[\textbf{\textit{A}} \,{\rm col}({\mathbf{\tilde u}}_1, {\mathbf{\tilde u}}_2, {\bm{\tilde \vartheta}}, {\bm{\tilde \eta}})\right]^{\bar{r}_1\,\, \bar{r}_2 \,\, 1}&=&\sum_{\bar{\mathbf{q}} \in {\mathbb{Z}^2}} {\rho}^{\bar{r}_1-\bar{q}_1\,\,\bar{r}_2-\bar{q}_2} {\tilde u}_1^{\bar{q}_1 \bar{q}_2}\,,
\end{eqnarray}
\begin{eqnarray}
\left[\textbf{\textit{A}} \,{\rm col}({\mathbf{\tilde u}}_1, {\mathbf{\tilde u}}_2, {\bm{\tilde \vartheta}}, {\bm{\tilde \eta}})\right]^{\bar{r}_1\,\, \bar{r}_2 \,\, 2}&=&\sum_{\bar{\mathbf{q}} \in {\mathbb{Z}^2}} {\rho}^{\bar{r}_1-\bar{q}_1\,\,\bar{r}_2-\bar{q}_2} {\tilde u}_2^{\bar{q}_1 \bar{q}_2}\,,
\end{eqnarray}
\begin{eqnarray}
\left[\textbf{\textit{A}} \,{\rm col}({\mathbf{\tilde u}}_1, {\mathbf{\tilde u}}_2, {\bm{\tilde \vartheta}}, {\bm{\tilde \eta}})\right]^{\bar{r}_1\,\, \bar{r}_2 \,\, 3}&=&0\,,
\end{eqnarray}
\begin{eqnarray}
\left[\textbf{\textit{A}} \,{\rm col}({\mathbf{\tilde u}}_1, {\mathbf{\tilde u}}_2, {\bm{\tilde \vartheta}}, {\bm{\tilde \eta}})\right]^{\bar{r}_1\,\, \bar{r}_2 \,\, 4}&=&0\,.
\end{eqnarray}
\noindent The second operator
$\textbf{\textit{B}}: (l_2(\mathbb{Z}^2))^4 \to (l_2(\mathbb{Z}^2))^4$ is  defined equation-wise as follows
\begin{eqnarray}
\left[\textbf{\textit{B}} \,{\rm col}({\mathbf{\tilde u}}_1, {\mathbf{\tilde u}}_2, {\bm{\tilde \vartheta}}, {\bm{\tilde \eta}})\right]^{\bar{r}_1\,\, \bar{r}_2 \,\, 1}&=&0\,,
\end{eqnarray}
\begin{eqnarray}
\left[\textbf{\textit{B}}  \,{\rm col}({\mathbf{\tilde u}}_1, {\mathbf{\tilde u}}_2, {\bm{\tilde \vartheta}}, {\bm{\tilde \eta}})\right]^{\bar{r}_1\,\, \bar{r}_2 \,\, 2}&=&0\,,
\end{eqnarray}
\begin{eqnarray}
\left[\textbf{\textit{B}}  \,{\rm col}({\mathbf{\tilde u}}_1, {\mathbf{\tilde u}}_2, {\bm{\tilde \vartheta}}, {\bm{\tilde \eta}})\right]^{\bar{r}_1\,\, \bar{r}_2 \,\, 3}&=& \sum_{\bar{\mathbf{q}} \in {\mathbb{Z}^2}} \left(\frac{2 \pi \bar{q}_j}{d_j}+k_j\right)  \alpha_{ij}^{\bar{r}_1-\bar{q}_1\,\,\bar{r}_2-\bar{q}_2} {\tilde u}_i^{\bar{q}_1 \bar{q}_2}\nonumber \\
	&& -  I \sum_{\bar{\mathbf{q}} \in {\mathbb{Z}^2}} \psi^{\bar{r}_1-\bar{q}_1\,\,\bar{r}_2-\bar{q}_2} {\tilde \eta}^{\bar{q}_1 \bar{q}_2} \nonumber \\
	&& - I \sum_{\bar{\mathbf{q}} \in {\mathbb{Z}^2}} {p}^{\bar{r}_1-\bar{q}_1\,\,\bar{r}_2-\bar{q}_2} {\tilde \vartheta}^{\bar{q}_1 \bar{q}_2}\,,
\end{eqnarray}
\begin{eqnarray}
\left[\textbf{\textit{B}}  \,{\rm col}({\mathbf{\tilde u}}_1, {\mathbf{\tilde u}}_2, {\bm{\tilde \vartheta}}, {\bm{\tilde \eta}})\right]^{\bar{r}_1\,\, \bar{r}_2 \,\, 4}&=& \sum_{\bar{\mathbf{q}} \in {\mathbb{Z}^2}} \left(\frac{2 \pi \bar{q}_j}{d_j}+k_j\right) \beta_{ij}^{\bar{r}_1-\bar{q}_1\,\,\bar{r}_2-\bar{q}_2} {\tilde u}_i^{\bar{q}_1 \bar{q}_2}\nonumber \\
	&& -  I \sum_{\bar{\mathbf{q}} \in {\mathbb{Z}^2}} \psi^{\bar{r}_1-\bar{q}_1\,\,\bar{r}_2-\bar{q}_2} {\tilde \vartheta}^{\bar{q}_1 \bar{q}_2} \nonumber \\
	&& - I \sum_{\bar{\mathbf{q}} \in {\mathbb{Z}^2}} {c}^{\bar{r}_1-\bar{q}_1\,\,\bar{r}_2-\bar{q}_2} {\tilde \eta}^{\bar{q}_1 \bar{q}_2}\,.
\end{eqnarray}
\noindent The third operator
$\textbf{\textit{C}}: (l_2(\mathbb{Z}^2))^4 \to (l_2(\mathbb{Z}^2))^4$ is  defined equation-wise as follows
\begin{eqnarray}
\left[\textbf{\textit{C}}  \,{\rm col}({\mathbf{\tilde u}}_1, {\mathbf{\tilde u}}_2, {\bm{\tilde \vartheta}}, {\bm{\tilde \eta}})\right]^{\bar{r}_1\,\, \bar{r}_2 \,\, 1}&=&- \left(\frac{2 \pi \bar{r}_j}{d_j}+k_j\right) \sum_{\bar{\mathbf{q}} \in {\mathbb{Z}^2}} \left(\frac{2 \pi {\color{black} \bar{q}_k}}{d_k} +k_k \right) C_{ijhk}^{\bar{r}_1-\bar{q}_1\,\,\bar{r}_2-\bar{q}_2} {\tilde u}_1^{\bar{q}_1 \bar{q}_2} \nonumber \\
&& - I \left(\frac{2 \pi \bar{r}_j}{d_j} +k_j \right) \sum_{\bar{\mathbf{q}} \in {\mathbb{Z}^2}} \alpha_{ij}^{\bar{r}_1-\bar{q}_1\,\,\bar{r}_2-\bar{q}_2} {\tilde \vartheta}^{\bar{q}_1 \bar{q}_2}\nonumber \\
	&& - I \left(\frac{2 \pi \bar{r}_j}{d_j} +k_j \right) \sum_{\bar{\mathbf{q}} \in {\mathbb{Z}^2}} \beta_{ij}^{\bar{r}_1-\bar{q}_1\,\,\bar{r}_2-\bar{q}_2} {\tilde \eta}^{\bar{q}_1 \bar{q}_2}\,,
\end{eqnarray}
\begin{eqnarray}
\left[\textbf{\textit{C}} \,{\rm col}({\mathbf{\tilde u}}_1, {\mathbf{\tilde u}}_2, {\bm{\tilde \vartheta}}, {\bm{\tilde \eta}})\right]^{\bar{r}_1\,\, \bar{r}_2 \,\, 2}&=&- \left(\frac{2 \pi \bar{r}_j}{d_j}+k_j\right) \sum_{\bar{\mathbf{q}} \in {\mathbb{Z}^2}} \left(\frac{2 \pi {\color{black} \bar{q}_k}}{d_k} +k_k \right) C_{ijhk}^{\bar{r}_1-\bar{q}_1\,\,\bar{r}_2-\bar{q}_2} {\tilde u}_2^{\bar{q}_1 \bar{q}_2} \,,\nonumber \\
&& - I \left(\frac{2 \pi \bar{r}_j}{d_j} +k_j \right) \sum_{\bar{\mathbf{q}} \in {\mathbb{Z}^2}} \alpha_{ij}^{\bar{r}_1-\bar{q}_1\,\,\bar{r}_2-\bar{q}_2} {\tilde \vartheta}^{\bar{q}_1 \bar{q}_2}\nonumber \\
	&& - I \left(\frac{2 \pi \bar{r}_j}{d_j} +k_j \right) \sum_{\bar{\mathbf{q}} \in {\mathbb{Z}^2}} \beta_{ij}^{\bar{r}_1-\bar{q}_1\,\,\bar{r}_2-\bar{q}_2} {\tilde \eta}^{\bar{q}_1 \bar{q}_2}\,,
\end{eqnarray}
\begin{eqnarray}
\left[\textbf{\textit{C}} \,{\rm col}({\mathbf{\tilde u}}_1, {\mathbf{\tilde u}}_2, {\bm{\tilde \vartheta}}, {\bm{\tilde \eta}})\right]^{\bar{r}_1\,\, \bar{r}_2 \,\, 3}&=&- \left(\frac{2 \pi \bar{r}_i}{d_i}+k_i\right) \sum_{\bar{\mathbf{q}} \in {\mathbb{Z}^2}} \left(\frac{2 \pi {\color{black} \bar{q}_j}}{d_j}+k_j\right) K_{ij}^{\bar{r}_1-\bar{q}_1\,\,\bar{r}_2-\bar{q}_2} {\tilde \vartheta}^{\bar{q}_1 \bar{q}_2}\,,
\end{eqnarray}	
\begin{eqnarray}
\left[\textbf{\textit{C}} \,{\rm col}({\mathbf{\tilde u}}_1, {\mathbf{\tilde u}}_2, {\bm{\tilde \vartheta}}, {\bm{\tilde \eta}})\right]^{\bar{r}_1\,\, \bar{r}_2 \,\, 4}&=&- \left(\frac{2 \pi \bar{r}_i}{d_i}+k_i\right) \sum_{\bar{\mathbf{q}} \in {\mathbb{Z}^2}}\left(\frac{2 \pi {\color{black} \bar{q}_j}}{d_j}+k_j\right) D_{ij}^{\bar{r}_1-\bar{q}_1\,\,\bar{r}_2-\bar{q}_2} {\tilde \eta}^{\bar{q}_1 \bar{q}_2}\,.
\end{eqnarray}

\subsection*{Appendix B.2: Vector ${\textbf{z}'}$, and operators ${\textbf{\textit{A}}'}$, ${\textbf{\textit{B}}'}$, and ${\textbf{\textit{C}}'}$}
\noindent Moreover, the operators $\textbf{\textit{A}}'$ and $\textbf{\textit{B}}'$  appearing in Equation (\ref{eq:generalized2}) are now defined, in terms of their actions on the argument
\begin{equation}
\textbf{z}' = {\rm col}(\textbf{z}_1, \textbf{z}_2) \in (l_2(\mathbb{Z}^2))^4 \times (l_2(\mathbb{Z}^2))^4\,.
\end{equation}
\noindent These operators are expressed, alternatively, in terms of the operators $\textbf{\textit{A}}$ or $\textbf{\textit{B}}$.\\
\noindent The first operator 
\begin{eqnarray}
\textbf{\textit{A}}'&:& (l_2(\mathbb{Z}^2))^4 \times (l_2(\mathbb{Z}^2))^4 \to (l_2(\mathbb{Z}^2))^4 \times (l_2(\mathbb{Z}^2))^4
\end{eqnarray}
is defined as follows
\begin{equation}\label{eq:operator1bis}
\textbf{\textit{A}}' \, \textbf{z} ={\rm col} \left(\textbf{\textit{A}} \, \textbf{z}_1 + \textbf{\textit{0}} \, \textbf{z}_2 ,
\textbf{\textit{0}} \, \textbf{z}_1 + \textbf{\textit{I}} \, \textbf{z}_2 \right),
\end{equation}
where $\textbf{\textit{0}}$ and $\textbf{\textit{I}}$  denote the zero and identity infinite-dimensional linear operators from $(l_2(\mathbb{Z}^2))^4$ to $(l_2(\mathbb{Z}^2))^4$, respectively.\\
\noindent The second operator 
\begin{eqnarray}
\textbf{\textit{B}}'&:& (l_2(\mathbb{Z}^2))^4 \times (l_2(\mathbb{Z}^2))^4 \to (l_2(\mathbb{Z}^2))^4 \times (l_2(\mathbb{Z}^2))^4
\end{eqnarray}
is defined as follows
\begin{equation}\label{eq:operator2bis}
\textbf{\textit{B}}' \, \textbf{z} = {\rm col} \left(
\textbf{\textit{B}} \, \textbf{z}_1 + \textbf{\textit{C}} \, \textbf{z}_2, 
-\textbf{\textit{I}} \, \textbf{z}_1 + \textbf{\textit{0}} \, \textbf{z}_2 \right)\,,
\end{equation}
\noindent By applying the operators $\textbf{\textit{A}}'$ and $\textbf{\textit{B}}'$ to the vector $\textbf{z}'$ defined as
\begin{equation}\label{eq:operatorzbis}
\textbf{z}'= {\rm col} \left(\textbf{z}_1, \textbf{z}_2 \right)=
{\rm col} \left(\omega \textbf{z}, 
\textbf{z} \right)
\end{equation}
one gets
\begin{eqnarray}
\left(\omega \textbf{\textit{A}}'+\textbf{\textit{B}}'\right) \, \mathbf{z}'&=&
{\rm col} \left(\omega \, \textbf{\textit{A}} \, \omega \mathbf{z} + \omega \, \textbf{\textit{0}} \, \mathbf{z} + \textbf{\textit{B}} \, \omega \mathbf{z} + \textbf{\textit{C}} \, \mathbf{z},
\omega \, \textbf{\textit{0}} \, \omega \mathbf{z} + \omega \, \textbf{\textit{I}} \, \mathbf{z} - \textbf{\textit{I}} \, \omega \mathbf{z} + \textbf{\textit{0}} \, \mathbf{z} \right) \nonumber \\
&=&
{\rm col} \left(\left(\omega^2 \textbf{\textit{A}} + \omega \textbf{\textit{B}} + \textbf{\textit{C}}\right) \, \mathbf{z}, 
{\mathbf{0}}\right),
\end{eqnarray}
which shows the equivalence between Equations (\ref{eq:generalized}) and  (\ref{eq:generalized2}).\\

\subsection*{Appendix B.3: Vector ${\mathbf{z}^{(f, \,\,reg)}}$ and $\mathbf{z}'^{(f,\, reg)}$, and operators ${\textbf{\textit{A}}^{(f,\,\,reg)}}$, ${\textbf{\textit{B}}'^{(f,\,\,reg)}}$, and ${\textbf{\textit{C}}'^{(f,\,\,reg)}}$}
\vspace{2mm}
\noindent Concerning the finite-dimensional generalized quadratic eigenvalue problem in Equation (\ref{eq:generalized3}), the vector $\mathbf{z}^{(f, \,\,reg)}$  results

 \begin{equation}
\mathbf{z}^{(f, \,\,reg)} = {\rm col}({\mathbf{\tilde u}}_1^{(f, \,\,reg)}, {\mathbf{\tilde u}}_2^{(f, \,\,reg)}, {\bm{\tilde \vartheta}}^{(f, \,\,reg)}, {\bm{\tilde \eta}}^{(f, \,\,reg)}) \in (\mathbb{R}^{(2Q +1)^2})^4\,,
\end{equation}
\noindent where $(\mathbb{R}^{(2Q +1)^2})^4$ stands for $\mathbb{R}^{(2Q +1)^2} \times \mathbb{R}^{(2Q +1)^2} \times \mathbb{R}^{(2Q +1)^2} \times \mathbb{R}^{(2Q +1)^2}$.\\
\noindent The operators ${\textbf{\textit{A}}^{(f,\,\,reg)}}$, ${\textbf{\textit{B}}^{(f,\,\,reg)}}$, ${\textbf{\textit{C}}^{(f,\,\,reg)}}: (\mathbb{R}^{(2Q +1)^2})^4 \to (\mathbb{R}^{(2Q +1)^2})^4$ are defined in a similar way as ${\textbf{\textit{A}}}$, ${\textbf{\textit{B}}}$, and ${\textbf{\textit{C}}}$, by restricting $\bar{\mathbf{r}}$ and $\bar{\mathbf{q}}$ to $\{-Q,\ldots,Q\}^2$.\\ 
\noindent Analogously to the infinite dimensional case, the operators
$\textbf{\textit{A}}'^{(f,\,\,reg)}$ and $\textbf{\textit{B}}'^{(f,\,\,reg)}$ in Equation (\ref{eq:generalized4}) can be defined, after introducing the argument 
\begin{equation}
\mathbf{z}'^{(f,\, reg)}={\rm col} \left(\omega \mathbf{z}^{(f,\, reg)}, \mathbf{z}^{(f,\, reg)}\right) \in (\mathbb{R}^{(2Q +1)^2})^4 \times (\mathbb{R}^{(2Q +1)^2})^4.
\end{equation}
\noindent The first operator
$\textbf{\textit{A}}'^{(f,\,\,reg)}: (\mathbb{R}^{(2Q +1)^2})^4 \times (\mathbb{R}^{(2Q +1)^2})^4 \to (\mathbb{R}^{(2Q +1)^2})^4 \times (\mathbb{R}^{(2Q +1)^2})^4$ is  defined equation-wise as follows
\begin{equation}
\textbf{\textit{A}}'^{(f,\,\,reg)} \, \mathbf{z}'^{(f,\, reg)} = {\rm col}
\left(\textbf{\textit{A}}^{(f,\,\,reg)} \omega \mathbf{z}^{(f,\, reg)} + \textbf{\textit{0}}^{(f)} \, \mathbf{z}^{(f,\, reg)}, 
\textbf{\textit{0}}^{(f)} \omega \mathbf{z}^{(f,\, reg)} + \textbf{\textit{I}}^{(f)} \, \mathbf{z}^{(f,\, reg)} \right),
\end{equation}
\noindent where $\textbf{\textit{0}}^{(f)}$ and $\textbf{\textit{I}}^{(f)}$ denote the zero and identity operators from $(\mathbb{R}^{(2Q +1)^2})^4$ to $(\mathbb{R}^{(2Q +1)^2})^4$, respectively.
\noindent The second operator
$\textbf{\textit{B}}'^{(f,\,\,reg)}: (\mathbb{R}^{(2Q +1)^2})^4 \times (\mathbb{R}^{(2Q +1)^2})^4 \to (\mathbb{R}^{(2Q +1)^2})^4 \times (\mathbb{R}^{(2Q +1)^2})^4$ is  defined equation-wise as follows
\begin{equation}
\textbf{\textit{B}}'^{(f,\, reg)} \, \mathbf{z}^{(f,\, reg)}= {\rm col}
\left(\textbf{\textit{B}}^{(f,\,\,reg)} \, \omega \mathbf{z}^{(f,\, reg)} + \textbf{\textit{C}}^{(f,\,\,reg)}  \mathbf{z}^{(f,\, reg)}, 
-\textbf{\textit{I}}^{(f)} \omega \mathbf{z}^{(f,\, reg)} + \textbf{\textit{0}}^{(f)} \omega \mathbf{z}^{(f,\, reg)} \right).
\end{equation}

\end{document}